\documentclass[aps,prb,floatfix,amsmath,amssymb,preprint,eqsecnum,nofootinbib]{revtex4-1}
\usepackage{graphicx}
\usepackage{dcolumn}
\usepackage{bm}
\usepackage{epstopdf}
\usepackage{color}

\usepackage{setspace} 
\newcommand{\beq}{\begin{equation}}
\newcommand{\eeq}{\end{equation}}
\newcommand{\ba}{\begin{array}{ccc}}
\newcommand{\ea}{\end{array}}
\newcommand{\nn}{\nonumber}
 \renewcommand{\d}{\partial}
\def\bea{\begin{eqnarray}}
\def\eea{\end{eqnarray}}

\def\prl{\parallel}

\def\<{\langle}
\def\>{\rangle}
\usepackage{amsmath}
\usepackage{amssymb}
\begin{document}

\title{Quantum phase transitions of metals in two spatial dimensions:\\ II. Spin density wave order}

\author{Max A. Metlitski}
\affiliation{Department of Physics, Harvard University, Cambridge MA
02138}

\author{Subir Sachdev}
\affiliation{Department of Physics, Harvard University, Cambridge MA
02138}

\date{\today \\
\vspace{1.6in}}
\begin{abstract}
We present a field-theoretic renormalization group analysis of Abanov and Chubukov's model of the spin
density wave transition in two dimensional metals. We identify the independent 
field scale and coupling constant renormalizations in a local field theory, and argue that the damping constant of 
spin density wave fluctuations tracks the renormalization of the local couplings. The divergences at two-loop
order overdetermine the renormalization constants, and are shown to be consistent with our renormalization scheme.
We describe the physical consequences of our renormalization group equations, including the breakdown of Fermi liquid
behavior near the ``hot spots'' on the Fermi surface. In particular, we find that the dynamical critical exponent
$z$ receives corrections to its mean-field value $z = 2$. At higher orders in the loop expansion, 
we find infrared singularities similar to those found by S.-S.~Lee
for the problem of a Fermi surface coupled to a gauge field. A treatment of these singularities implies that an expansion in $1/N$,
(where $N$ is the number of fermion flavors) fails for the present problem. 
We also discuss the renormalization of the
pairing vertex, and find an enhancement which scales as logarithm-squared of the energy scale.
A similar enhancement is also found for a modulated bond order which is locally an Ising-nematic order.
\end{abstract}

\maketitle

\section{Introduction}
\label{sec:intro}

There is little doubt that the quantum transition involving the onset of spin density wave (SDW) order
in a metal is of vital importance to the properties of a variety of correlated electron metals. This is amply
illustrated by some recent experimental studies.
In the cuprates, Daou {\em et al.\/} \cite{nlsco1} argued that the Fermi surface change associated with
such a transition was the key in understanding the physics of the strange metal. In the pnictide
superconductors, experiments \cite{joerg0,joerg1,joerg2} have explored the interesting coupling
between the onsets of SDW order and superconductivity. In CeRhIn$_5$ (and other `115' compounds),
Knebel {\em et al.} \cite{knebel} have described the suppression of the SDW order by pressure,
and the associated enhancement of superconductivity.

The theory of Hertz \cite{hertz,monod,vojtarev} has formed much of the basis of the study of the spin density wave transition
in the literature. The central step of this theory is the derivation of an effective action
for the spin density wave order parameter, after integrating out all the low energy excitations near the Fermi
surface. A conventional renormalization group (RG) is then applied to this effective action, and this can be extended to high
order using standard
field-theoretic techniques \cite{HertzMillisGamma}. However, it has long been clear that the full integration of 
the Fermi surface excitations is potentially dangerous, because the Fermi surface structure undergoes a singular
renormalization from the SDW fluctuations.

Important advances were subsequently 
made in the work of Abanov and Chubukov \cite{ChubukovShort0,ChubukovShort}.
They argued that the Hertz analysis was essentially correct in spatial dimension $d=3$, but that it broke
down seriously in $d=2$. They proposed an alternative low energy field theory for $d=2$, involving the bosonic
SDW order parameter and fermions along arcs of the Fermi surface; the arcs are located near Fermi surface ``hot spots''
which are directly connected by SDW ordering wavevector.
They also presented a RG study of this field theory, and found
interesting renormalizations of the Fermi velocities at the arcs.

This paper will present a re-examination of the model of Abanov and Chubukov, using a field-theoretic RG
method. We will begin in Section~\ref{sec:model} by introducing the low energy field theory
for the SDW transition in two dimensional metals, and reviewing the Abanov-Chubukov argument
for the breakdown of the Hertz theory. 
Section~\ref{naive} will define the independent renormalization constants using the structure of the local field theory,
and determine their values using the divergences in a $1/N$ expansion (where $N$ is the number of fermion flavors)
to two loop order. 
Actually, the two-loop divergences overdetermine the renormalization constants, but we will
find a consistent solution: this is a significant
check on the consistency of our renormalization procedure. While our renormalizations
of the Fermi velocities agree with those of Abanov and Chubukov, we find significant differences in the other
renormalizations, and associated physical consequences. 
At two-loop order, the ratio of the velocities scales logarithmically to zero (as specified by Eq.~(\ref{alphaflow})), 
and consequently we are able to compute RG-improved results for a variety of physical observables
(which differ from previous results \cite{ChubukovShort0,ChubukovShort}):
\begin{itemize}
\item The non-Fermi liquid behavior at the hotspot is controlled by the fermion
self energy given by Eq.~(\ref{Sigmaimp}).
\item Moving away from the hot spot, we find that Fermi liquid
behavior is restored, but the quasiparticle residue and the Fermi velocity vary strongly as a function of the
momentum ($p_{\parallel}$) along the Fermi surface: these are given in Eq.~(\ref{Zimp}).
\item The bosonic SDW spectrum does not obey dynamic scaling with $z=2$, but instead obeys the `super power-law'
form in Eq.~(\ref{dynimp}), and the amplitude of the spectrum scales as in Eq.~(\ref{Dimp}).
\end{itemize}
Going beyond two-loops, we also explored the consequences of a strong-coupling fixed point at which the velocity
ratio and other couplings reach finite fixed-point values. Here the boson and fermion Green's functions obey
the scaling forms in Eqs.~(\ref{zeta}-\ref{Gfixed}), and the non-Fermi liquid behavior at the hotspot
is specified in Eq.~(\ref{Ghsscal}). Moving away from the hotspot, we have the Fermi liquid form
in Eq.~(\ref{GFSscal}), with the Fermi velocity and quasiparticle residue given by Eq.~(\ref{Zvscal}).

In Section~\ref{sec:genus}, we describe the structure of the field theory at higher loop order. 
Similar to the effects pointed out recently
by S.-S.~Lee \cite{SSLee} for the problem of a Fermi surface coupled to a gauge field, we find that there are infrared
singularities which lead to a breakdown in the naive counting of powers of $1/N$. However, unlike in the problem
of a gauge field coupled to a single patch of the Fermi surface \cite{SSLee}, we find that the higher order diagrams
cannot be organized into an expansion in terms of the genus of a surface associated with the graph. Rather, diagrams that scale 
as increasingly higher powers of $N$ are generated upon increasing the number of loops.

In Section~\ref{sec:pairing}, we consider the onset of pairing near the SDW transition, a question 
examined
previously by Abanov, Chubukov, Finkel'stein, and 
Schmalian \cite{ChubukovFinkelstein,ChubukovSchmalian,ChubukovLong}.
Like them, we find that the corrections to the $d$-wave pairing vertex are enhanced relative to the naive
counting of powers of $1/N$. However, we also find an enhancement factor which scales as the logarithm-squared
of the energy scale: this is the result in Eq.~(\ref{Vfin}). We will discuss the interpretation of this log-squared term in Section~\ref{sec:pairing}.

In Section~\ref{sec:cdw} we show that a similar log-squared enhancement is present for
the vertex of a bond order which is locally an Ising-nematic order; this order parameter is illustrated
in Figs.~\ref{fig:density} and~\ref{fig:density2}.
 The unexpected
 similarity between this order, and the pairing vertex, is a consequence of emergent SU(2) pseudospin
symmetries of the continuum theory of the SDW transition, with independent pseudospin rotations on 
different pairs of hot spots. One of the pseudospin rotations
is the particle-hole transformation, and the other pseudospin symmetries will be described more completely 
in Section~\ref{sec:model}.

\section{Low energy field theory}
\label{sec:model}

We will study the generic 
phase transition between a Fermi liquid and a SDW state in two spatial dimensions, and our discussion also
easily generalizes to charge density wave order. The wavevector of the density wave order is $\vec{Q}$, and 
we assume that there exist points on the Fermi surface connected by $\vec{Q}$; these points are known as hot spots.
We assume further that the Fermi velocities at a pair of hot spots connected by $\vec{Q}$ are not parallel to each other;
this avoids the case of `nested Fermi surfaces', which we will not treat here.

A particular realization of the above situation is provided by the case of SDW ordering on the square
lattice at wavevector $\vec{Q} = (\pi, \pi)$. We also take a Fermi surface appropriate for the cuprates, generated
by a tight-binding model with first and second neighbor hopping. We will restrict all our subsequent discussion
to this case for simplicity.

\begin{figure}[t]
\begin{center}
\includegraphics[width=3.3in]{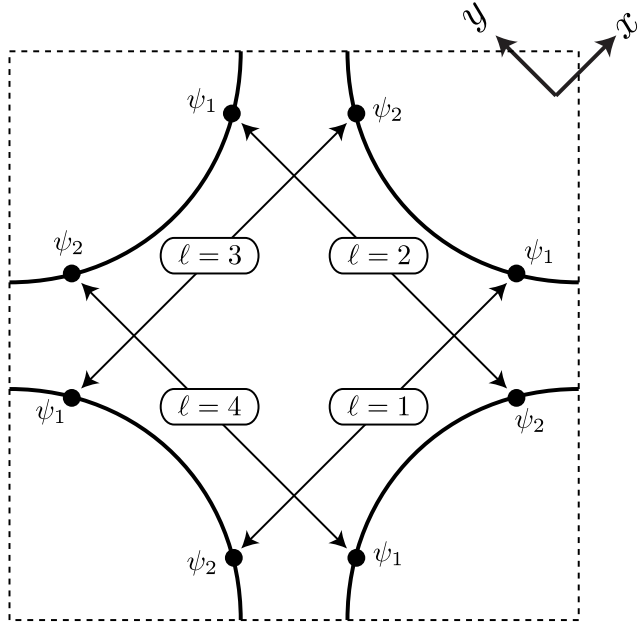}
\caption{Square lattice Brillouin zone showing the Fermi surface appropriate to the
cuprates. The filled circles are the hot spots connected by the SDW wavevector $\vec{Q} = (\pi,\pi)$.
The locations of the continuum fermion fields $\psi_1^\ell$ and $\psi_2^\ell$ is indicated.}
\label{fig:hotspots}
\end{center}
\end{figure}

\begin{figure}[t]
\begin{center}
\includegraphics[width=3.3in]{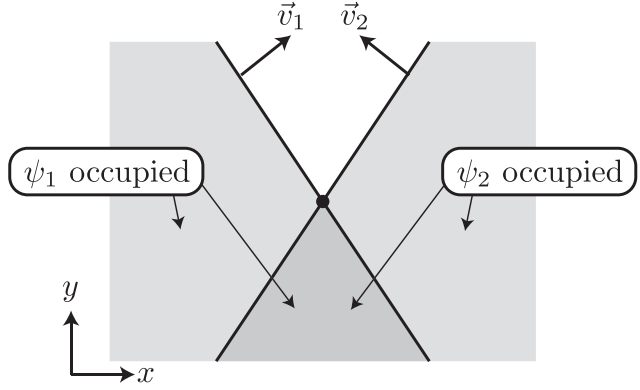}
\caption{Configuration of the $\ell=1$ pair of hot spots, with the momenta
of the fermion fields measured from the common hot spot at $\vec{k}=0$, indicated by the filled circle. The Fermi
velocities $\vec{v}_{1,2}$ of the $\psi_{1,2}$ fermions are indicated.}
\label{fig:fermions}
\end{center}
\end{figure}

At wavevector $\vec{Q}= (\pi, \pi)$ the SDW ordering is collinear, and so
is described by a three component real field $\phi^a$, $a=x,y,z$.
There are $n = 4$ pairs of hot spots, as shown in Fig.~\ref{fig:hotspots}.
We introduce fermion fields $(\psi^{\ell}_{1\sigma}, \psi^{\ell}_{2\sigma})$, $\ell = 1 ... n$, $\sigma = \uparrow \downarrow$ for each pair of hot spots. Lattice rotations map the pairs of hot spots into each other, acting cyclically on the index $\ell$. Moreover, the two hot spots within each pair are related by a reflection across a lattice diagonal. It will be useful to promote each field $\psi$ to have $N$-flavors with an eye to performing a $1/N$ expansion. (Note that in Ref. \onlinecite{ChubukovLong}, the total number of hot spots $2 n N$ is denoted as $N$.) The flavor index is suppressed in all the expressions. The low energy effective theory is given by the Lagrangian,
\bea L &=&   \frac{N}{2 c^2} (\d_{\tau} \vec{\phi})^2 + \frac{N}{2}(\nabla \vec{\phi})^2 + \frac{N r}{2} \vec{\phi}^2 + \frac{N u}{4} (\vec{\phi}^2)^2\nn\\ &+&
\psi^{\dagger\ell}_1 (\d_{\tau} - i \vec{v}^{\ell}_1 \cdot \nabla) \psi^{\ell}_1 + \psi^{\dagger\ell}_2 (\d_{\tau} - i \vec{v}^{\ell}_2 \cdot \nabla) \psi^{\ell}_2 \nn\\ &+& \lambda \phi^a \left(\psi^{\dagger\ell}_{1\sigma} \tau^a_{\sigma \sigma'} \psi^{\ell}_{2\sigma'} + \psi^{\dagger\ell}_{2\sigma} \tau^a_{\sigma \sigma'} \psi^{\ell}_{1\sigma'}\right)\label{L}\eea
The first line in Eq.~(\ref{L}) is the usual O(3) model for the SDW order parameter, the second line is the fermion kinetic energy and the third line is the interaction between the SDW order parameter and the fermions at the hot spots. Here, we have linearized the fermion dispersion near the hot spots and $\vec{v}^{\ell}$ are the corresponding Fermi velocities. It is convenient to choose coordinate axes along directions $\hat{x} = \frac{1}{\sqrt{2}}(1,1)$ and $\hat{y} = \frac{1}{\sqrt{2}} (-1,1)$, so that 
\begin{equation}
\vec{v}^{\ell = 1}_1 = (v_x, v_y)~~,~~\vec{v}^{\ell = 1}_2 = (-v_x, v_y);
\end{equation}
these Fermi velocities are indicated in Fig.~\ref{fig:fermions}.
The other Fermi velocities are related by rotations, $\vec{v}^{\ell} = (R_{\pi/2})^{\ell-1} \vec{v}^{\ell = 1}$. 

We choose the coefficient $\lambda$ of the fermion-SDW interaction to be of $\mathcal{O}(1)$ in $N$. 
As a result, the coefficients in the first line of Eq.~(\ref{L}) are all scaled by $N$ as this factor will automatically appear upon integrating out the high-momentum/frequency modes of the fermion fields. 
\begin{figure}[t]
\centering
 \includegraphics[width=2.2in]{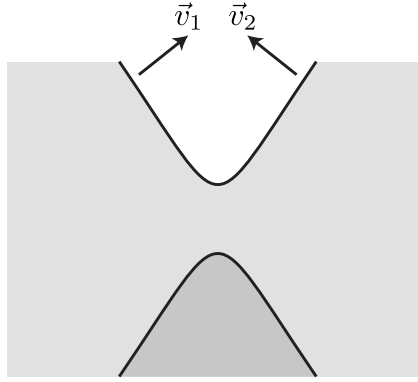}
 \caption{Modification of the Fermi surfaces in Fig.~\ref{fig:fermions} by SDW order
with $\langle \phi \rangle
 \neq 0$. The full lines are the Fermi surfaces, and the white, light shaded, and
dark shaded regions denote
 momenta where 0, 1, and 2 of the bands are occupied. The upper and lower lines are
 boundaries of hole and electron pockets respectively.}
 \label{fig:sdwfs2}
\end{figure}

Before proceeding with the analysis of the theory (\ref{L}), let us note its symmetries. Besides the microscopic translation, point-group, spin-rotation and time-reversal symmetries, the low energy theory possesses a set of four emergent $SU(2)$ pseudospin symmetries associated with 
particle-hole transformations. Let us introduce a four-component spinor,
\beq \Psi^{\ell}_i = \left(\begin{array}{c} \psi^{\ell}_i\\ i \tau^2 \psi^{\dagger \ell}_i\end{array}\right)\label{Psi}\eeq
We will denote the particle-hole indices in the four-component spinor by $\alpha, \beta$. The spinor (\ref{Psi}) satisfies the hermiticity condition,
\beq i \tau^2 \left(\begin{array}{cc} 0 & -1\\ 1 & 0\end{array}\right) \Psi^{\ell}_i = \Psi^{*\ell}_i \label{Herm}\eeq
Then, the fermion part of the Lagrangian (\ref{L}) can be rewritten as,
\beq L_\psi = \frac{1}{2} \Psi^{\dagger\ell}_1 (\d_{\tau} - i \vec{v}^{\ell}_1 \cdot \nabla) \Psi^{\ell}_1 + \frac{1}{2}\Psi^{\dagger\ell}_2 (\d_{\tau} - i \vec{v}^{\ell}_2 \cdot \nabla) \Psi^{\ell}_2  + \frac{1}{2}\lambda \vec{\phi} \cdot \left(\Psi^{\dagger\ell}_{1} \vec{\tau} \Psi^{\ell}_{2} + \Psi^{\dagger\ell}_{2} \vec{\tau} \Psi^{\ell}_{1}\right)\label{LPsi}\eeq
Now the Lagrangian (\ref{LPsi}) and the condition (\ref{Herm}) are manifestly invariant under,
\beq SU(2)_{\ell}:\,\, \Psi^{\ell}_i \to U_{\ell} \Psi^{\ell}_i \label{SU2}\eeq
with $U_{\ell}$ - $SU(2)$ matrices. We note that the diagonal subgroup of (\ref{SU2}) is associated with independent conservation of the fermion number at each hot spot pair. The symmetry (\ref{SU2}) is a consequence of linearization of the fermion spectrum near the hot spots and is broken by higher order terms in the dispersion. The diagonal subgroup noted above is preserved by higher order terms in the dispersion, but is broken by four-fermi interactions, which map fermion pairs from opposite hot spots into each other. Both symmetry breaking effects are irrelevant in the scaling limit discussed below.

The pseudospin symmetry (\ref{SU2}) constrains the form of the fermion Green's function to be,
\beq - \langle \Psi^{\ell}_{i \alpha \sigma} \Psi^{m \dagger}_{j \beta \sigma'}\rangle = \delta^{\ell m} \delta_{ij}\delta_{\alpha \beta} \delta_{\sigma \sigma'}  G^{\ell}_i(x-x')\eeq
which implies,
\beq G^{\ell}_i(x-x') = - G^{\ell}_i(x'-x) \eeq
The corresponding  expression in momentum space, $G^{\ell}_i(k) = - G^{\ell}_i(-k)$, implies that the location of hot spots in the Brillouin zone is not renormalized by the spin wave fluctuations in the low energy theory.

Another important manifestation of the particle-hole symmetry is the equality of any Feynman graphs, which are related by a reversal of a fermion loop direction.

\subsection{The Hertz action}
\label{sec:hertz}

The Hertz action is derived by working in the metallic phase, and 
integrating out the fermions in Eq.~(\ref{L}), leaving an  effective theory for $\phi$ alone. 
In particular, the one-loop self-energy of the field $\phi$ is evaluated in Appendix~\ref{app:rpa},
and is given by
\beq \Pi^{0}(\omega, \vec{q}) = \Pi^{0}(\omega = 0, \vec{q} = 0) + N \gamma |\omega| + ..., \quad \gamma = \frac{n \lambda^2}{2 \pi v_x v_y}\label{Pi} \eeq
The presence of the non-analytic term $|\omega|$ is due to the fact that the density of particle-hole pairs with momentum $\vec{Q}$ and energy $\omega$ scales as $\omega$. As usual, the constant piece $\Pi^0(q = 0)$ is eliminated by tuning the coefficient $r$. The ellipses in Eq.~(\ref{Pi}) denote terms analytic in $\omega$ and $\vec{q}$, starting with $\omega^2$ and $\vec{q}^2$. These terms formally disappear when we take the cut-off of the effective theory (\ref{L}) to infinity. Thus, the quadratic part of the effective action for the field $\phi$ reads 
\beq S_2 = \frac{N}{2} \int \frac{d\omega d^2k }{(2 \pi)^3}  \phi^a(-k,-\omega) \left(\gamma |\omega| + \frac{1}{c^2} \omega^2 + \vec{k}^2 + r \right) \phi^a(k, \omega)\eeq
At sufficiently low energies, the analytic term $\omega^2$ in the boson self-energy coming from the bare action, Eq.~(\ref{L}), can be neglected compared to the dynamically generated $|\omega|$ term. Thus, at low energies the propagation of collective spin excitations becomes diffusive, due to the damping by the fermions at the hot spots. 

Hertz \cite{hertz} proceeds by neglecting all the quartic and higher order self-interactions of the field $\phi$, which are generated when the fermions are eliminated. This is justified if such interactions are local, as one can then absorb them into operators, which are polynomial in the order parameter and its derivatives (the simplest of which is just the operator $(\vec{\phi}^2)^2$). The theory then reduces to,
\beq S_H = \frac{N}{2} \int \frac{d\omega d^2k }{(2 \pi)^3}  \phi^a(-k,-\omega) \left(\gamma |\omega| + \vec{k}^2 + r \right) \phi^a(k, \omega) + \frac{N u}{4} \int d\tau d^2 x (\vec{\phi}^2)^2 \label{SHertz}\eeq
The quadratic part of the action (\ref{SHertz}) is invariant under scaling with the dynamical critical exponent $z = 2$,
\beq \vec{k} \to s \vec{k}, \quad \omega \to s^2 \omega, \quad \phi(\vec{x}, \vec{\tau}) \to s \phi(s \vec{x}, s^2 \tau)\label{scalz2}\eeq 
Thus the theory is effectively $d + z = 4$ dimensional and the quartic coupling $u$ is marginal by power-counting in $d=2$.

At one loop order, the flow of $u$ follows easily from the conventional momentum shell RG \cite{Millis}
\begin{equation}
\frac{du}{d \ell} = - \frac{11}{2 \pi^2 N \gamma} u^2, \label{hertzflow}
\end{equation}
where $s=e^{-\ell}$ is the renormalization scale. Thus $u$ is marginally irrelevant, and flows to the Gaussian fixed point with $u=0$
in the infrared.
This stability of the Gaussian fixed point has formed the basis of much of the subsequent 
work \cite{vojtarev,HertzMillisGamma,Millis} on the Hertz theory.

\subsection{Breakdown of the Hertz theory}
\label{sec:breakdown}

The analysis in Section~\ref{sec:hertz} is valid only under the assumption that the fermion-induced quartic and 
higher order couplings of the field $\phi$ can be neglected. In fact, as observed in 
Refs.~\onlinecite{ChubukovLong,ChubukovShort}, this assumption is not justified in spatial dimension $d = 2$. 
Indeed, as shown in Ref. \onlinecite{ChubukovLong}, the fermion-induced four-point vertex is given by,
\beq \Gamma^{a_1 a_2 a_3 a_4}_4(q_1, q_2, q_3, q_4) = \lambda^4 f^{a_1 a_2 a_3 a_4}(q_1, q_2, q_3, q_4) + \mathrm{permutations \, of \, 2, 3, 4} \label{Gamma4}\eeq
\beq f^{a_1 a_2 a_3 a_4}(q_1, q_2, q_3, q_4) = \sum_{\ell} \frac{N(\delta^{a_1 a_2} \delta^{a_3 a_4} - \delta^{a_1 a_3} \delta^{a_2 a_4} + \delta^{a_1 a_4} \delta^{a_2 a_3}) (|\omega_1| - |\omega_2| + |\omega_3| - |\omega_4|)}{2 \pi v_x v_y (i (\omega_2 + \omega_3) - \vec{v}^{\ell}_1 \cdot 
(\vec{q}_2 + \vec{q}_3))(i (\omega_1 + \omega_2) - \vec{v}^{\ell}_2 \cdot 
(\vec{q}_1 + \vec{q}_2))}\label{f}\eeq
We see that the vertex (\ref{Gamma4}) is highly non-local. Moreover, under the $z=2$ scaling (\ref{scalz2}), we can neglect the frequency dependence in the denominators of Eq.~(\ref{f}), obtaining $\Gamma_4 \sim |\omega|/\vec{q}^2 \sim \mathcal{O}(1)$, which produces a marginal interaction. Similarly, one can show that all the higher order fermion-induced vertices behave as $\Gamma_{2n} \sim |\omega|/|\vec{q}|^{2n - 2} \sim |\vec{q}|^{4 - 2n}$, which is again marginal under (\ref{scalz2}) when combined with the scaling of the field-strength. 
Thus, the Hertz-Millis theory has an infinite number of non-local marginal perturbations and the 
standard action (\ref{SHertz}) is incomplete.

\subsection{RG interpretation}
\label{sec:RGinterp}

An RG interpretation of the results of Section~\ref{sec:breakdown} follows by performing 
a scaling analysis directly on the spin-fermion model (\ref{L}). As before, we will scale the boson fields according to Eq.~(\ref{scalz2}). Correspondingly, it is natural to scale the fermion momenta towards the hot spots,
\beq \psi^{\ell}_{12} (\vec{x}, \tau) \to s^{3/2} \psi^{\ell}_{12} (s \vec{x}, s^2 \tau) \label{scalferm} \eeq
Here the field-strength rescaling has been chosen to preserve the spatial gradient terms in the fermion action. We now see that the boson-fermion coupling $\lambda$ in (\ref{L}) is {\em marginal\/} under the field scalings in Eqs.~(\ref{scalz2})
and (\ref{scalferm}); a similar analysis
in $d=3$ would show that $\lambda$ is irrelevant.

The marginality of $\lambda$, and the infinite number of marginal couplings in Section~\ref{sec:breakdown}
indicate that all subsequent RG should be performed direction on the spin-fermion model (\ref{L}).
Further, with the scalings as in (\ref{scalz2}) and (\ref{scalferm}), we should not expand in powers of $\lambda$,
but rather analyze the theory at a {\em fixed} boson-fermion ``Yukawa'' coupling. A similar strategy was followed in 
Refs.~\onlinecite{eunah,yejin} for the Ising-nematic transition in a $d$-wave superconductor.

An important consequence of the scalings (\ref{scalz2}) and (\ref{scalferm}) on (\ref{L}) is that both 
the boson kinetic term $(\d_{\tau} \phi)^2$ and the fermion kinetic term $\psi^{\dagger} \d_{\tau} \psi$ 
are irrelevant. We may safely drop the boson kinetic energy. However, the fermion kinetic energy must be retained - otherwise, the theory does not possess any dynamics. We will return to this point shortly. 
Let us now rescale the fermion fields $\psi =  \tilde{\psi} /\sqrt{\lambda}$ to eliminate the marginal coupling $\lambda$. 
We define, $\eta = 1/\lambda$  and $\vec{\tilde{v}} = \vec{v}/\lambda$ . Note that $\tilde{v}$ has the unusual dimensions of $[\omega]^{1/2}/[k]$. We drop the tildes in what follows. Then,
\bea L &=&  \frac{N}{2}(\nabla \vec{\phi})^2 + \frac{N r}{2} \vec{\phi}^2 + \frac{N u}{4} (\vec{\phi}^2)^2\nn\\ &+&
\psi^{\dagger\ell}_1 (\eta \d_{\tau} - i \vec{v}^{\ell}_1 \cdot \nabla) \psi^{\ell}_1 + \psi^{\dagger\ell}_2 (\eta \d_{\tau} - i \vec{v}^{\ell}_2 \cdot \nabla) \psi^{\ell}_2 \nn \\ &+&  \phi^a \left(\psi^{\dagger\ell}_{1\sigma} \tau^a_{\sigma \sigma'} \psi^{\ell}_{2\sigma'} + \psi^{\dagger\ell}_{2\sigma} \tau^a_{\sigma \sigma'} \psi^{\ell}_{1\sigma'}\right)
\label{Lmarg}\eea
As already remarked, the coupling constant $\eta$ is irrelevant. Thus, we take the limit $\eta \to 0^+$ in all our calculations. In practice, $\eta$ gives the prescription for integrating over the poles of the fermion propagator. We will work with the action (\ref{Lmarg}) for the rest of this paper. At criticality it is characterized by two dimensionless constants,
\beq \alpha = \frac{v_y}{v_x}~~,~~  \tilde{u} = \frac{u}{\gamma} \label{dimless} \eeq
and a dimensionful constant $\gamma$, Eq. (\ref{Pi}),
\beq \gamma = \frac{n}{2 \pi v_x v_y}. \label{defgamma}
\eeq 
Thus, in the critical regime, the theory (\ref{Lmarg}) does not possess an expansion in any coupling constant.

\section{Field-theoretic RG}
\label{naive}

We begin by discussing the general renormalization structure of (\ref{Lmarg}). In the absence of a coupling constant, we will use the RPA based scaling (\ref{scalz2}) and (\ref{scalferm}) as the starting point of our analysis. Naively, one expects that this scaling is also obeyed by the $N = \infty$ limit of the theory and that corrections to it can be calculated in a systematic expansion in $1/N$. Indeed, the usual arguments would indicate that at $N = \infty$, the boson self-energy is given by the RPA bubble in Fig.~\ref{FigPi}, Eq.~(\ref{Pi}),
(see the Appendix~\ref{app:rpa} for details of the calculation). 
\begin{figure}[t]
\begin{center}
\includegraphics[width=2in]{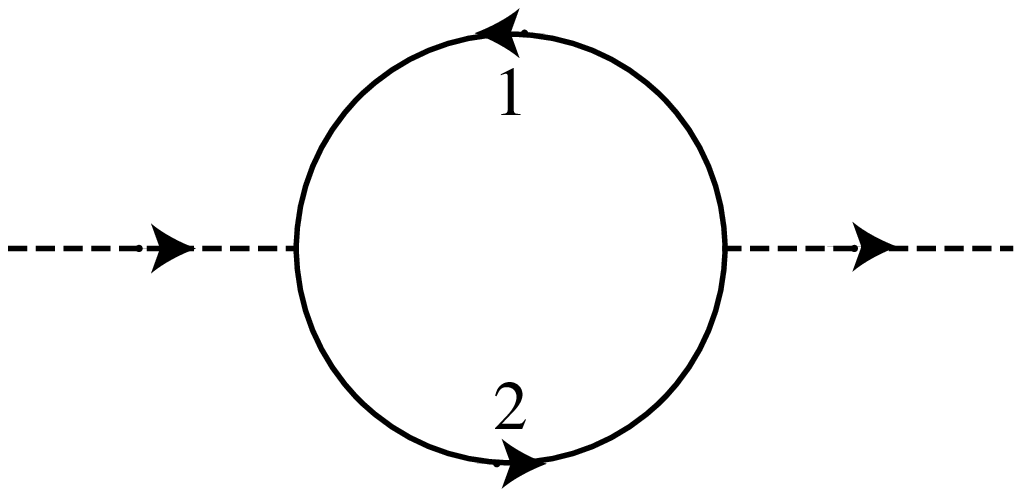}\\
\includegraphics[width=2in]{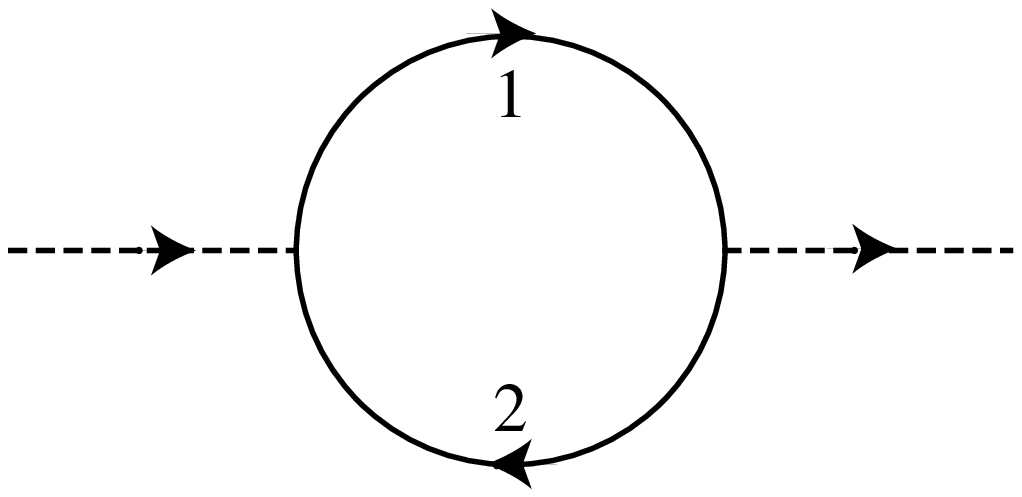}
\caption{The boson self-energy at $N = \infty$. The full lines represent the $\psi_{1,2}$ fermions, and the dashed lines represent the boson $\phi^a$.}
\label{FigPi}
\end{center}
\end{figure}
Hence, the bosonic propagator
\beq \langle \phi^a(x) \phi^b(x')\rangle = \delta^{ab}  D(x-x')\eeq
at $N = \infty$ takes the form,
\beq D(x) = \frac{1}{N} \int \frac{d\omega d^2 q}{(2 \pi)^3} \frac{1}{\gamma |\omega| + \vec{q}^2+r} e^{- i \omega \tau + i \vec{q} \vec{x}} \label{bprop} \eeq
which respects the scaling (\ref{scalz2}). On the other hand, the fermion propagator
\beq -\langle \psi^{\ell}_{i\sigma}(x) \psi^{\dagger m}_{j\sigma'}(x')\rangle = \delta^{\ell m} \delta_{ij} \delta_{\sigma \sigma'}  G^{\ell}_i(x-x')\nn\eeq
at $N = \infty$ is given by its free value,
\beq \quad G^{\ell}_{i}(x) = \int \frac{d\omega d^2 k}{(2 \pi)^3} \frac{1}{i \eta \omega  - \vec{v}^{\ell}_i \cdot \vec{k}} e^{- i \omega \tau + i \vec{k} \cdot \vec{x}}\label{fprop}\eeq
Applying scaling (\ref{scalferm}) to this propagator indicates $\eta$ scales to zero; we will eventually take this limit, but need a non-zero $\eta$ for now to properly define the fermion loop integrals.

As we will see later in Section~\ref{sec:genus}, the $N = \infty$ limit in the present theory turns out to be much more subtle and is not given by the simple forms in Eqs.~(\ref{bprop}),(\ref{fprop}). Moreover, the anomalous dimensions in this limit are not expected to be parametrically small. Nevertheless, we can reasonably expect that the RG structure presented here remains valid, even though we are not
able to accurately compute higher loop corrections to the renormalization constants. In addition, the difficulties with the $1/N$ expansion appear only at high loop order, which enables us to check the consistency of our approach to the order discussed below.

With the above remarks in mind, we are ready to discuss the renormalization of the theory in Eq.~(\ref{Lmarg}). The theory contains five operators that are 
marginal by power counting at $z=2$, and not related by symmetry. Two of these are eliminated by field-strength renormalizations,
\beq \phi = Z^{1/2}_{\phi} \phi_r, \quad \psi = Z^{1/2}_{\psi} \psi_r\eeq
As is conventional, we can fix $Z_\phi$ by demanding that the coefficient of $(\nabla \phi)^2$ remains invariant.
For fermion field, it is convenient to allow both velocities to flow, and so we renormalize these as
\beq v_x = Z^x_{v} v^r_x,\quad v_y = Z^y_{v} v^r_y.\eeq
The fermion spatial gradient terms are then not available to fix $Z_\psi$, and we cannot use
the fermion temporal gradient term because its coefficient $\eta$ scales to zero.
Instead we demand the invariance of the boson-fermion coupling term to fix
the fermion field strength renormalization; it is thus consistent to use a unit
coefficient for this term, as we have done in Eq.~(\ref{Lmarg}).
The quartic boson coupling renormalizes
\beq \quad \tilde{u} = \frac{Z_u Z^x_{v} Z^y_v}{ Z^2_{\phi}} \tilde{u}_r. \label{Zu} \eeq
It is also useful to track the renormalization of the dimensionless velocity ratio $\alpha$ in 
Eq.~(\ref{dimless})
\beq \alpha = \frac{Z^y_v}{Z^x_v} \alpha_r. \eeq
All the renormalization factors $Z$ depend only on $N$, $\alpha_r$, $\tilde u_r$ and the ratio $\mu/\Lambda$,
where $\mu$ is a renormalization scale and $\Lambda$ is a UV cutoff.

An important point is that the damping parameter $\gamma$ appearing in the boson propagator
does not have an independent renormalization constant. It is not a coupling in a local field theory,
and only appears in certain correlation functions as a measure of the strength of the particle-hole continuum, as 
determined by Eq.~(\ref{defgamma}). This implies that when we consider the renormalization
of the boson propagator, the renormalization of the parameter $\gamma$ should track the the renormalizations
of the velocties $v_{x,y}$ obtained from the renormalization of the fermion propagator; in other words,
the renormalization of $\gamma$ is
\beq \gamma = \frac{1}{Z^x_v Z^y_v} \gamma_r  \label{gammren} .\eeq
This tight coupling between
the boson and fermion sectors is a key feature of the theory (\ref{Lmarg}), and a primary reason for strong
coupling physics in $d=2$.

The theory (\ref{Lmarg}) contains two relevant perturbations. One of these is the usual $\vec{\phi}^2$ operator, whose coefficient
renormalizes as,
\beq r = \frac{Z_r}{Z_\phi} r_r \eeq
Here, $r$ always denotes the deviation from the critical point. The other relevant perturbation, whose discussion we have omitted thus far, 
is the chemical potential,
\beq \delta L = - \mu \psi^{\ell\dagger}_{i \sigma}\psi^{\ell}_{i \sigma}\eeq
However, this perturbation is redundant, as it can be absorbed into a shift of hot spot location. Moreover, as already observed
in section \ref{sec:model}, the location of the hot spots is not renormalized in the low-energy theory, which implies that there is no mixing between
the two relevant operators. This is unlike the situation for the Ising-nematic transition in a metal studied in Ref. \onlinecite{max1}, where such mixing leads to
a nontrivial shift of the Fermi surface as a function of deviation $r$ from the critical point.

Introducing the renormalized one-particle irreducible correlation functions of $n_f$ fermion and $n_b$ boson fields
\beq \Gamma^{n_f,n_b}_r = Z^{n_f/2}_\psi Z^{n_b/2}_\phi \Gamma^{n_f,n_b}\eeq
we can write down the renormalization group equations,
\beq \left(\mu \frac{\d}{\d \mu} + \beta_\alpha \frac{\d}{\d \alpha_r} + \beta_u \frac{\d}{\d \tilde{u}_r} + \eta_\gamma \gamma_r \frac{\d}{\d \gamma_r} - \eta_2 r_r \frac{\d}{\d r_r} - \frac{n_b \eta_\phi}{2} - \frac{n_f \eta_\psi}{2}\right) \Gamma^{n_b, n_f}_r(\{p\},\alpha_r,\tilde{u}_r,\gamma_r, r_r, \mu) = 0 \label{RGGamma}\eeq
Here, the $\beta$-functions and anomalous dimensions are functions of $\alpha_r$ and $\tilde{u}_r$ given by,
\bea \beta_{\alpha} &=& \mu \frac{\d \alpha_r}{\d \mu}\Big|_{\alpha, \tilde{u},  \Lambda},\quad \beta_u = \mu \frac{\d \tilde{u}_r}{\d \mu}\Big|_{\alpha, \tilde{u}, \Lambda}, \quad \eta_\gamma = \frac{1}{\gamma_r} \mu\frac{\d\gamma_r}{\d \mu} \Big|_{\alpha, \tilde{u}, \gamma, \Lambda},\\
\eta_\phi &=& \mu\frac{\d}{\d \mu} \log Z_\phi \Big|_{\alpha, \tilde{u}, \Lambda},\quad \eta_\psi = \mu\frac{\d}{\d \mu} \log Z_\psi \Big|_{\alpha, \tilde{u}, \Lambda},\quad
\eta_2 = \mu\frac{\d}{\d \mu}\log \frac{Z_r}{Z_\phi} \Big|_{\alpha, \tilde{u}, \Lambda} \eea
Using dimensional analysis,
\beq \Gamma^{n_b, n_f}_r(\{\omega\},\{\vec{p}\},\alpha_r,\tilde{u}_r,\gamma_r, r_r, \mu) = \gamma^{n_b/2 + n_f/4 -1}_r \mu^{4 - n_b - 3 n_f/2} f^{n_b,n_f}\left(\left\{\frac{\gamma_r \omega}{\mu^2}\right\}, \left\{\frac{\vec{p}}{\mu}\right\}, \alpha_r, \tilde{u}_r, \frac{r_r}{\mu^2}\right)
\eeq
Now, solving the RG equation (\ref{RGGamma}),
\bea f^{n_b,n_f}(\{\hat{\omega}\}, \{\hat{p}\},\alpha_r, \tilde{u}_r, \hat{r}) &=& s^{4-3 n_f/2 - n_b} {Z_\phi(s)}^{-n_b/2} {Z_\psi(s)}^{-n_f/2} {Z_\gamma(s)}^{n_b/2+n_f/4-1}\nn\\&\times&f^{n_b,n_f}(s^{-2} Z_\gamma(s) \{\hat{\omega}\}, s^{-1} \{\hat{p}\}, \alpha_r(s), \tilde{u}_r(s), Z_r(s) \hat{r})\label{RGsol}\eea
with
\bea s \frac{d \alpha_r}{d s} &=& \beta_\alpha(\alpha_r(s), \tilde{u}_r(s)), \quad\alpha_r(1) = \alpha_r, \quad \quad s \frac{d \tilde{u}_r}{d s} = \beta_u(\alpha_r(s), \tilde{u}_r(s)), \quad \tilde{u}_r(1) = \tilde{u}_r\nn\\
Z_\phi(s) &=& \exp\left(\int_1^s \frac{ds'}{s'} \eta_\phi(\alpha_r(s'), \tilde{u}_r(s'))\right), \quad Z_\psi(s) = \exp\left(\int_1^s \frac{ds'}{s'} \eta_\psi(\alpha_r(s'), \tilde{u}_r(s'))\right)\nn\\
Z_\gamma(s) &=& \exp\left(\int_1^s \frac{ds'}{s'} \eta_\gamma(\alpha_r(s'), \tilde{u}_r(s'))\right),\quad Z_r(s) = \exp\left(-\int_1^s \frac{ds'}{s'} \eta_2(\alpha_r(s'), \tilde{u}_r(s'))\right) \nn\\\eea

Now, let us construct the scaling forms of the correlation functions assuming that the couplings $\alpha_r$, $\tilde{u}_r$ have a stable fixed point. Actually, as we will see below, this assumption is not supported by explicit calculations of low loop contributions to the $\beta$-functions and anomalous dimensions. However, as already remarked, higher loop diagrams, which are naively suppressed by powers of $1/N$, actually scale as progressively higher powers of $N$ and might modify the RG flow significantly. Thus, the fixed-point form of the correlation functions satisfies,
\beq f(s^{2-\eta_\gamma} \{\hat{\omega}\}, s \{\hat{p}\}, s^{2+\eta_2} \hat{r}) = s^{4-\eta_\gamma - 
(3+\eta_\psi - \eta_\gamma/2)n_f/2 - (2 + \eta_\phi - \eta_\gamma)n_b/2} f(\{\hat{\omega}\},  \{\hat{p}\},  \hat{r}) \eeq
Hence, typical frequencies and momenta are related by $\omega \sim |\vec{p}|^z$, with the dynamical critical exponent $z$ being given by,
\beq z = 2-\eta_\gamma\label{zeta}\eeq
Moreover, the correlation length $\xi$ away from the critical point scales as $\xi \sim r^{-\nu}$ with
\beq \nu =  \frac{1}{2 + \eta_2}\eeq
Specializing to boson and fermion two-point functions,
\bea D^{-1}(\omega, \vec{p}) &\sim& \xi^{-(2 - \eta_\phi)} K(\omega \xi^z, \vec{p} \xi) \quad \stackrel{\xi \to \infty}{\to} \quad |\vec{p}|^{2-\eta_\phi} \tilde{K}(\omega/|\vec{p}|^z, \hat{p})\label{Dfixed}\\
G^{-1}(\omega, \vec{p}) &\sim& \xi^{-(z/2-\eta_\psi)} L(\omega \xi^z, \vec{p} \xi) \quad \stackrel{\xi \to \infty}{\to} \quad |\vec{p}|^{z/2-\eta_\psi} \tilde{L}(\omega/|\vec{p}|^z,\hat{p}) \label{Gfixed}\eea
Here, the expressions on the right give the correlation functions at the critical point to which we confine our attention from here on. From Eq.~(\ref{Gfixed}) we may infer the fate of the Fermi surface at the critical point. We expect that as $\xi \to \infty$ the Fermi-surface remains sharply defined. Close to the hot spots, the Fermi surfaces of fermions $\psi_1$ and $\psi_2$ will evolve into straight lines with a fixed angle between them. At the hot spot, the fermion self-energy takes the form,
\beq G^{-1}(\omega, \vec{p} = 0) \sim \omega^{1/2 - \eta_\psi/z} \label{Ghsscal}\eeq
which is generally non Fermi-liquid like. On the other hand, away from the hot spot, if we define $p_\perp$ as the distance to the Fermi surface and $p_\parallel$ as the distance to the hot spot, for $p_\perp \ll p_\parallel$ and $\omega \ll p^z_\parallel$, we expect well-defined Landau quasi-particles,
\beq G(\omega, \vec{p}) \sim \frac{{\cal Z}}{ i \omega - v_F p_\perp}\label{GFSscal}\eeq
with the Fermi velocity $v$ and quasiparticle residue ${\cal Z}$ vanishing as we approach the hot spot as,
\beq v_F(p_\parallel) \sim p^{z-1}_\parallel, \quad {\cal Z}(p_\parallel) \sim p^{z/2 + \eta_\psi}_\parallel\label{Zvscal}\eeq

The remainder of this section will provide a computation of the 4 renormalization constants
$Z_\phi$, $Z_\psi$, $Z_v^x$, $Z_v^y$ to leading order in $1/N$. At this order, the constants
will depend only upon the dimensionless constant $\alpha$, and do not involve $u$. We discuss the renormalization of $u$
in Appendix~\ref{app:u}. Thus our considerations here will involve the RG flow only of the single
coupling $\alpha$, the ratio of the velocities, and a discussion of its physical implications. For completeness,
we will also compute the renormalization constant $Z_r$, which determines the scaling of the correlation length away from the 
critical point. This constant will depend upon both $\alpha$ and $u$ already at leading order in $1/N$.

As we will see below, the 4 renormalization constants will be overdetermined from the structure of the
$1/N$ corrections to the fermion self energy, the boson-fermion vertex, and the boson self energy.
Computations of these quantities are provided in the appendix, and we use the results here to 
compute the $Z$'s.

The first correction to the self-energy of the fermion $\psi^{\ell = 1}_1$ is given by Fig. \ref{FigSigma}, and computed
in Appendix~\ref{app:fself}.
\begin{figure}[t]
\begin{center}
\includegraphics[width=3in]{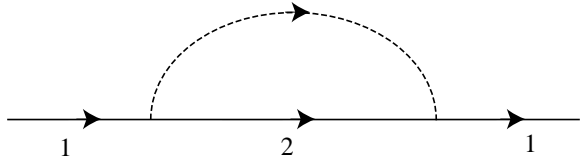}
\caption{The leading contribution to the fermion self-energy.}
\label{FigSigma}
\end{center}
\end{figure}
\beq \Sigma_1(\omega, \vec{p}) = - \frac{3}{2 \pi N |\vec{v}| \gamma}\left(i \mbox{sgn}(\omega) (\sqrt{\gamma |\omega| + (\hat{v}_2 \cdot \vec{p})^2} - |\hat{v}_2 \cdot \vec{p}|) + \frac{2}{\pi} \hat{v}_2 \cdot \vec{p} \log \frac{\Lambda}{|\hat{v}_2 \cdot \vec{p}|}\right)\label{Sigma}\eeq
Note that unless otherwise stated, we will discuss the $\ell = 1$ hot spot and drop the index $\ell$. 
We see that at the hot spot, $\vec{p} = 0$, the self-energy has a non-Fermi liquid form, \cite{millis2,ChubukovShort0}
\beq \Sigma(\vec{p} = 0) = - i \frac{3}{(2 \pi n)^{1/2} N} \left(\frac{1}{\alpha} + \alpha\right)^{-1/2} |\omega|^{1/2} \mbox{sgn} (\omega)
\label{Sigmak0} \eeq
This result is consistent with our scaling form (\ref{Ghsscal}); to this order the anomalous dimension $\eta_\psi = 0$. On the other hand, away from the hot spot, in the regime $\gamma |\omega| \ll (\hat{v}_2 \cdot \vec{p})^2$, the fermion propagator takes the Fermi-liquid form (\ref{GFSscal}).
To leading order, the Fermi surface is given by $\hat{v}_1 \cdot \vec{p} = 0$. The Fermi velocity and quasiparticle residue vanish with the distance $p_\parallel$ along the Fermi-surface to the hot spot as,
\beq v_F = \frac{4 n N}{3\gamma} p_\parallel, \quad {\cal Z} = \frac{4 N}{3} (2 \pi n)^{1/2} \gamma^{-1/2} \left(\frac{1}{\alpha} + \alpha\right)^{-1/2} p_\parallel \label{ZVoneloop}\eeq
consistent with the scaling form (\ref{Zvscal}) with mean-field exponents $z = 2$, $\eta_\psi = 0$.

The last term in Eq.~(\ref{Sigma}) contributes to the renormalization of $v_x, v_y$, and so constrains the 
renormalization constants by
\bea Z_\psi Z^x_v &=& 1 - \frac{6}{\pi n N} \frac{\alpha}{1+\alpha^2} \log(\Lambda/\mu) \label{Zx} \\
Z_\psi Z^y_v &=& 1 + \frac{6}{\pi n N} \frac{\alpha}{1+\alpha^2} \log(\Lambda/\mu)\label{Zy}\eea 

Next we consider the correction to the boson-fermion vertex,
\beq -\langle \psi_{2 \sigma}(p') \psi^{\dagger}_{1 \sigma'}(p) \phi^a(-q)\rangle_{1PI} = \tau^a_{\sigma \sigma'} \Gamma_{\phi \psi_2 \psi^{\dagger}_1}(p,q) (2 \pi)^3 \delta^3(p' - p - q)\eeq
This is given by Fig. \ref{FigVertex} and computed in Appendix~\ref{app:vertex}. 
\begin{figure}[t]
\begin{center}
\includegraphics*[width=1.6in]{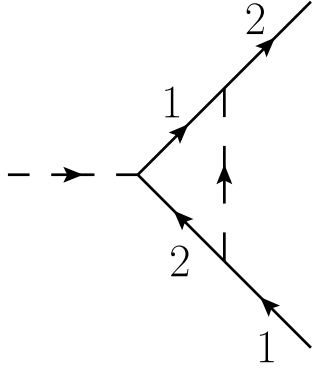}
\caption{The leading correction to the boson-fermion vertex.}
\label{FigVertex}
\end{center}
\end{figure}
We need only the
UV divergent part, which is
\beq \Gamma_{\phi \psi_2 \psi^{\dagger}_1}(p,q) = 1 + \frac{2}{\pi n N} \tan^{-1} \frac{1}{\alpha} \log \Lambda \label{Gamma} \eeq
Eq.~(\ref{Gamma}) constrains the renormalizations by
\beq Z^{1/2}_\phi Z_{\psi} = 1 - \frac{2}{\pi n N} \tan^{-1}\frac{1}{\alpha} \log (\Lambda/\mu)\label{ZGamma}\eeq

Finally, we consider the corrections to the boson two-point function, shown in Fig.~\ref{FigPiCorr}, and computed in 
Appendix~\ref{app:bself}. 
\begin{figure}[t]
\begin{center}
\includegraphics*[width=3in]{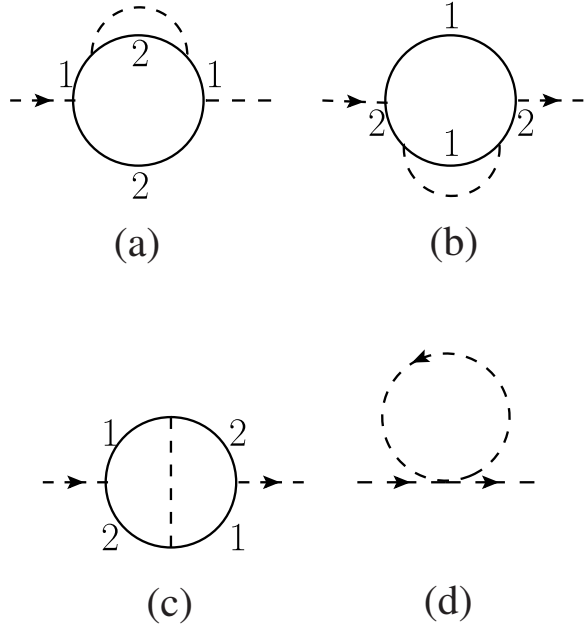}
\caption{The leading correction to the boson polarization. A sum over both directions of the fermion loop is implied.}
\label{FigPiCorr}
\end{center}
\end{figure}
These yield
\bea D^{-1}(\omega, \vec{q}) &=& N \gamma |\omega| \left[1 + \frac{4}{\pi n N} \tan^{-1} \frac{1}{\alpha} \log \Lambda\right] \nn \\ &+& N \vec{q}^2 \left[1 + \frac{2}{\pi n N} \left( \frac{1}{\alpha} - \alpha + \left(\frac{1}{\alpha^2} + \alpha^2\right) \tan^{-1} \frac{1}{\alpha}\right) \log \Lambda \right]\nn\\
&+& N r \left[1+\left(\frac{4}{\pi n N} \tan^{-1} \frac{1}{\alpha} - \frac{5}{2 \pi^2 N} \tilde{u}\right) \log \Lambda\right] \label{Dcorr}\eea
Note that both the frequency and momentum dependent parts of the boson propagator receive renormalization
corrections. As we discussed earlier, the corrections to the coefficient of $|\omega|$ should not be considered
as renormalizations of an independent coupling $\gamma$, but should rather track the renormalizations of the
fermion velocities. Consequently, from Eqs.~(\ref{gammren}) and (\ref{Dcorr}), we conclude that
\beq Z_{\phi} (Z^x_v Z^y_v)^{-1}  = 1 - \frac{4}{\pi n N} \tan^{-1} \frac{1}{\alpha}\log (\Lambda/\mu)\label{ZPiomega}\eeq
From the momentum dependent part of (\ref{Dcorr}) we immediately obtain the bosonic field strength renormalization,
\beq Z_\phi = 1 - \frac{2}{\pi n N} \left(\frac{1}{\alpha} - \alpha + \left(\frac{1}{\alpha^2} + \alpha^2\right) \tan^{-1}\frac{1}{\alpha}\right) \log (\Lambda/\mu) \label{Zphi} \eeq
while the $r$ dependent part of (\ref{Dcorr}) yields the renormalization constant $Z_r$,
\beq Z_r = 1- \left(\frac{4}{\pi n N} \tan^{-1}\frac{1}{\alpha} - \frac{5}{2 \pi^2 N} \tilde{u} \right) \log (\Lambda/\mu)\label{Zr}\eeq

We note that while our results for the fermion self-energy (\ref{Sigma}) and the vertex (\ref{Gamma}) are in agreement with Ref.~\onlinecite{ChubukovLong}, the expression for the boson two-point function Eq.~(\ref{Dcorr}) differs from that of Ref.~\onlinecite{ChubukovLong}. More precisely, the frequency dependent part of our $D^{-1}$ agrees with Ref.~\onlinecite{ChubukovLong}, while the momentum dependent part does not. As already noted, the renormalization of the frequency dependent part of $D^{-1}$ is constrained by that of the fermion self-energy and the vertex. On the other hand, the renormalization of the momentum dependent part is completely independent. The authors of Ref.~\onlinecite{ChubukovLong} found that both the frequency and the momentum parts are renormalized by the same factor, which would imply that the dynamical critical exponent $z = 2$ to this order. However, our calculations indicate that the two renormalizations are equal only at $\alpha = 1$ and, as we will see below, the dynamical critical exponent $z$ receives corrections already at the present order in $1/N$. 

We now have 5 equations for 4 renormalization constants: Eqs.~(\ref{Zx}), (\ref{Zy}), (\ref{ZGamma}), (\ref{ZPiomega}),
and (\ref{Zphi}). It is easily verified that they are consistent with each other. This is a strong check on our renormalization
procedure, and verifies the consistency of tying $\gamma$ to the velocities by Eq.~(\ref{defgamma}).
We can solve these equations to obtain
\bea \frac{Z^y_v}{Z^x_v} &=& 1 + \frac{12}{\pi n N} \frac{\alpha}{1+\alpha^2} \log(\Lambda/\mu) \nn \\
Z^x_v Z^y_v  &=& 1  - \frac{2}{\pi n N} \left(\frac{1}{\alpha} - \alpha\right)\left(1 + \left(\frac{1}{\alpha} - \alpha\right)\tan^{-1}\frac{1}{\alpha} \right)\log (\Lambda/\mu) \nn \\
Z_{\psi} &=& 1  + \frac{1}{\pi n N} \left(\frac{1}{\alpha} - \alpha\right)\left(1 + \left(\frac{1}{\alpha} - \alpha\right)\tan^{-1}\frac{1}{\alpha} \right)\log (\Lambda/\mu) \label{Zres} \eea
\subsection{RG flows}
\label{sec:rgflow}
The renormalization constants in Eq.~(\ref{Zres}) determine the flow of the dimensionless coupling $\alpha$
with the $\beta$-function
\beq \beta(\alpha_r) = \frac{12}{\pi n N} \frac{\alpha^2_r}{\alpha^2_r+1} \label{betaalpha}
\eeq
The $\beta$ function for the velocity anisotropy $\alpha$ has an infrared stable fixed point $\alpha = 0$ and an infrared unstable fixed point $\alpha = \infty$. Physically, both fixed points correspond to a nested Fermi surface. For $\alpha = 0$, the Fermi-velocities at the two hot spots are anti-parallel, while for $\alpha = \infty$ they are parallel. The flows to the two fixed points are logarithmic. In particular, near the infrared stable fixed point $\alpha = 0$,
\beq \alpha_r(s) = \frac{\alpha_r}{1 + \displaystyle \frac{12 \alpha_r}{\pi n N} \log(1/s)}\label{alphaflow}\eeq
Here we've assumed that the starting point of the flow $\alpha_r \ll 1$. Note that the logarithmic 
flow to $\alpha \rightarrow 0$ in the infrared, with vanishing velocity ratio, is similar to that
found recently in Ref.~\onlinecite{yejin} in a different physical context.

Let us now discuss the physics of the $\alpha=0$ fixed point. The renormalization constants
in (\ref{Zphi}),(\ref{Zr}), (\ref{Zres}) also determine the renormalization of the velocities, the anomalous dimensions of the bosons, fermions and of the $\phi^2$ operator. For the velocities, the ratio is already specified by $\alpha$, and it is convenient to take
$\gamma$ as the other independent combination of the velocities. We have therefore
\bea
\eta_\gamma &=&  \frac{2}{\pi n N} \left(\frac{1}{\alpha_r} - \alpha_r\right) \left(1 + \left(\frac{1}{\alpha_r} - \alpha_r\right) \tan^{-1} \frac{1}{\alpha_r}\right) \nn \\
\eta_{\phi} &=&  \frac{2}{\pi n N} \left(\frac{1}{\alpha_r} - \alpha_r + \left(\frac{1}{\alpha^2_r} + \alpha^2_r\right) \tan^{-1} \frac{1}{\alpha_r}\right) \nn 
\eea
\bea
\eta_{\psi} &=&  -\frac{1}{\pi n N} \left(\frac{1}{\alpha_r} - \alpha_r\right) \left(1 + \left(\frac{1}{\alpha_r} - \alpha_r\right) \tan^{-1} \frac{1}{\alpha_r}\right)\nn\\
\eta_2 &=& -\frac{2}{\pi n N} \left(\frac{1}{\alpha_r} - \alpha_r\right) \left(1 + \left(\frac{1}{\alpha_r} - \alpha_r\right) \tan^{-1} \frac{1}{\alpha_r}\right) - \frac{5}{2 \pi^2 N} \tilde{u}_r
\label{betares} 
\eea
Note that as can be seen from Eqs.~(\ref{RGsol}),(\ref{zeta}) the flow of the dimensionful constant $\gamma_r$ described by the exponent $\eta_\gamma$ is equivalent to an anomalous dynamical critical exponent $z$. Since $\eta_\gamma$ is non-zero, the dynamical behaviour of the theory deviates from the simple Hertz-Millis scaling with $z=2$.


 As $\alpha$ flows slowly to $0$, the critical exponents in Eq.~(\ref{betares}) slowly vary: 
\beq \eta_{\phi} \to \frac{1}{n N} \frac{1}{\alpha^2_r}, \quad \eta_{\psi} \to -\frac{1}{2 n N} \frac{1}{\alpha^2_r}, \quad
\eta_\gamma \to \frac{1}{n N} \frac{1}{\alpha^2_r},\quad\eta_2 \to -\frac{1}{n N} \frac{1}{\alpha^2_r}, \quad \alpha_r \to 0 \label{expalphazero}\eeq
Observe that the corrections to the critical exponents diverge as $\alpha_r \to 0$. Thus, for sufficiently small momenta the $1/N$ expansion breaks down. From Eq.~(\ref{expalphazero}) we see that this will happen when $\alpha \sim 1/\sqrt{N}$;
from Eq.~(\ref{alphaflow}), we can estimate that this occurs at a momentum scale $k \sim \exp(-N^{3/2})$. This is parametrically smaller than the scale $k \sim \exp(-N)$ at which the direct expansion in $1/N$ (without RG improvement) becomes invalid.

Despite the breakdown of the RG at the longest scales, there is an intermediate asymptotic regime, $1/\sqrt{N} \ll \alpha_r \ll 1$, where Eq.~(\ref{expalphazero}) remains valid, and we can integrate the RG equations and find interesting consequences for both the fermionic and bosonic spectra. 

\begin{figure}[t]
\centering
 \includegraphics[width=2.2in]{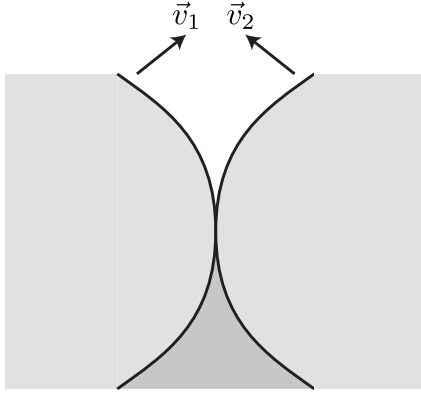}
 \caption{Modification of the Fermi surfaces in Fig.~\ref{fig:fermions} at the SDW
quantum critical point. 
 As in Figs.~\ref{fig:fermions} and~\ref{fig:sdwfs2}, the full lines are the Fermi
surfaces, and the white, light shaded, and dark shaded regions denote
 momenta where 0, 1, and 2 of the bands are occupied. The equation of one of the
Fermi surfaces
 is given in (\ref{FSlog}).}
 \label{fig:sdwfs3}
\end{figure}
For the fermions, the location of the $\psi_1$ Fermi surface is given at tree-level by $\hat{v}_1 \cdot \vec{p} = 0$,
or $p_y = - v_x p_x/v_y = - p_x / \alpha$. Evaluating $\alpha$ at $s = \mu/p_x$, we find the 
Fermi surface at
\begin{equation}
p_y = - \frac{12}{\pi n N} p_x \log (\mu/|p_x|)\label{FSlog}
\end{equation}
The resulting Fermi surface distorts from the shape shown in Fig.~\ref{fig:hotspots}
to that in Fig.~\ref{fig:sdwfs3}. We may also use RG to improve the one-loop result for the fermion self-energy (\ref{Sigma}). From Eq.~(\ref{RGsol}), the fermion self-energy at the hot spot is,
\beq \Sigma(\omega, \vec{p} = 0) \sim -i \exp\left(-\frac{3}{\pi^2 n^3 N^3} \log^3\frac{\mu^2}{\gamma_r |\omega|}\right) |\omega|^{1/2} \mbox{sgn}(\omega), \label{Sigmaimp}\eeq
Along the Fermi surface away from the hot spot, the quasiparticle residue and Fermi velocity behave as,
\beq v_F \sim \exp\left(\frac{48}{\pi^2 n^3 N^3} \log^3\frac{\mu}{p_\parallel}\right)p_\parallel, \quad {\cal Z} \sim \left(\log\frac{\mu}{p_\parallel}\right)^{-1/2} p_\parallel \label{Zimp}\eeq


The characteristic frequency of the bosonic spectrum is $\omega \sim \vec{q}^2 / \gamma_r$;
evaluating $\gamma_r$ at $s = \mu/|\vec{q}|$, we find that 
it scales with a `super power-law' of the momentum
\begin{equation}
\omega\sim \vec{q}^2 \exp \left( \frac{48}{\pi^2 n^3 N^3} \log^3 \frac{\mu}{|\vec{q}|} \right).
\label{dynimp}\end{equation}
From Eq.~(\ref{RGsol}) we also obtain the static and dynamic scaling of the bosonic propagator,
\bea D^{-1}(\omega, \vec{q} = 0) &\sim& |\omega|^{1-\frac{1}{nN}} \exp\left(\frac{6}{\pi^2 n^4 N^4} \log^3 \frac{\mu^2}{\gamma_r |\omega|}\right) \left(\log \frac{\mu^2}{\gamma_r |\omega|}\right)^{-1/3}\nn \\
D^{-1}(\omega = 0, \vec{q}) &\sim& |\vec{q}|^2 \exp\left(\frac{48}{\pi^2 n^3 N^3} \log^3 \frac{\mu}{|\vec{q}|} \right)\label{Dimp}\eea

Note that the unusual super-power law dependencies in Eqs.~(\ref{Sigmaimp}), (\ref{Zimp}),(\ref{dynimp}),(\ref{Dimp})  are consequences of the scaling of $\alpha_r \rightarrow 0$ in the
infrared and associated divergences of the anomalous dimensions.

\section{Counting powers of $N$}
\label{sec:genus}

As written in Eq.~(\ref{Lmarg}), our field theory offers a potentially simple way of organizing perturbation theory
in powers of $1/N$: each boson propagator comes with a power of $1/N$, each fermion loop yields a power of $N$,
and each $u$ interaction yields a factor $N$: we refer to this as the ``naive'' $1/N$ expansion, and it has been
the basis of our computations so far. 

However, because we have to take $\eta \rightarrow 0$ in the scaling limit, there is a danger that
some of the higher order diagrams will have a singular dependence on $\eta$. The fermion propagators
in such diagrams need to include self-energy corrections for the diagrams to be finite in the $\eta\rightarrow 0$ limit.
The price we will pay for this regularization is that the diagram will acquire additional powers of $N$,
and the naive counting of powers of $1/N$ will break down.

%
Recently, in the context of a theory of a Fermi surface interacting with a gauge field, S.-S.~Lee \cite{SSLee}
has given a procedure for identifying diagrams with a breakdown of naive $1/N$ counting,
and shown that the expansion in powers of $1/N$ is actually an expansion in the genus of a surface defined
by the graph. Using his methods we will show that many similar issues appear in our theory for the SDW transition of a Fermi surface,
although subtle differences in RG properties imply that in the present case no genus expansion exists, and diagrams of increasingly higher
order in $N$ are generated as the number of loops is increased.

\begin{figure}[t]
\begin{center}
\includegraphics*[width=2in]{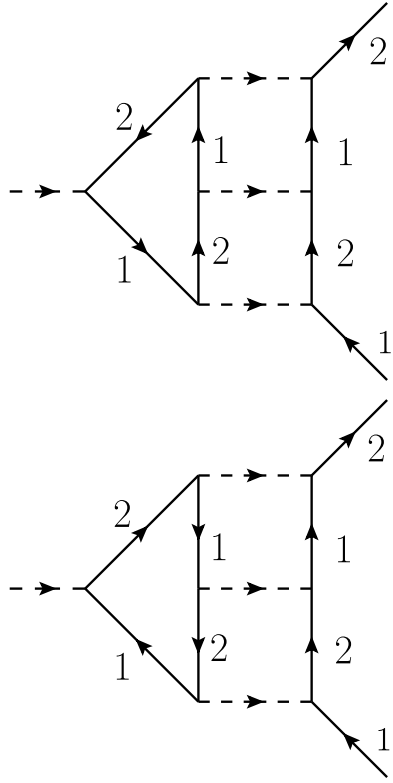}
\caption{Three loop corrections to the boson-fermion vertex that are enhanced in $N$, scaling as ${\cal O}(N^0)$.}
\label{fig:vertex2}
\end{center}
\end{figure}

In the absence of an external pairing vertex (see section \ref{sec:pairing}), the simplest diagrams exhibiting the above effect are the 
three-loop corrections to the boson-fermion vertex, see Fig.~\ref{fig:vertex2}.
In fact, the two diagrams are equal as they are related by particle-hole symmetry. 
The external fermions are taken to have hot spot index $\ell = 1$, while the fermions running in the loop can come 
from any hot spot $\ell'$, although we will see that the singular contributions will originate from $\ell' = 1$ and $\ell' = 3$. 
The diagram is given by,
\bea \delta \Gamma_{\phi \psi_2 \psi^{\dagger}_1}(p,q) \tau^{a}= - \tau^{a_1} \tau^{a_2} \tau^{a_3} \int \frac{d k_\tau d \vec{k} d k'_\tau d \vec{k'}}{(2 \pi)^6} f^{a a_1 a_2 a_3}(q, p-k', k'-k, k - p - q)\times\nn\\ 
G_1(k) G_2(k') D(k' - p) D(k - k') D(p+q-k)\nn\eea
Substituting the four-point boson vertex $f$, Eq.~(\ref{f}),
\bea \delta \Gamma_{\phi \psi_2 \psi^{\dagger}_1}(p,q) &=& -\frac{7 N}{2\pi v_x v_y} 
\sum_{\ell'} \int  \frac{d k_\tau d \vec{k} d k'_\tau d \vec{k}'}{(2 \pi)^6} (|q_\tau| - |p_\tau - k'_\tau| + |k'_\tau - k_\tau| - |k_\tau - p_\tau - q_\tau|)\nn\\
&\times&\frac{1}{(i \eta (p_\tau - k_\tau) - \vec{v}^{\ell'}_1 \cdot (\vec{p} - \vec{k})) (i \eta (q_\tau + p_\tau - k'_\tau) - \vec{v}^{\ell'}_2 \cdot 
(\vec{q} + \vec{p} - \vec{k}'))}\nn  \label{Vertex3l2}\\
&\times &\frac{1}{(i \eta k_\tau - \vec{v}_1 \cdot \vec{k}) (i \eta k'_\tau - \vec{v}_2 \cdot \vec{k}')} D(k' - p) D(k - k') D(p+q-k) \label{Vertex3l3}\eea
Observe that if $\ell' = 2$ or $\ell' = 4$ the four denominators in Eq.~(\ref{Vertex3l3}) involve four linearly independent combinations of internal momenta $\vec{k}$, $\vec{k}'$. As a result, the integral has a well defined limit when $\eta \to 0$. 
On the other hand, when $\ell' = 1$ or $\ell' = 3$ (which we will also denote as $\ell' = -1$), $\vec{v}^{\ell'}$ and $\vec{v}$ are parallel.
Keeping only these two hot spots, let us integrate over the momentum components $\vec{v}_1 \cdot \vec{k}$, $\vec{v}_2 \cdot \vec{k}'$. We focus on the contribution from the fermionic poles, which, as we will see, is infrared singular.
\bea \delta \Gamma_{\phi \psi_2 \psi^{\dagger}_1}(p,q) &\approx&   
\frac{7 N}{2\pi v_x v_y |\vec{v}|^2} 
\sum_{\ell' = \pm1} \int  \frac{d k_\tau dk_{\parallel} d k'_\tau dk'_{\parallel}}{(2 \pi)^4} (|q_\tau| - |p_\tau - k'_\tau| + |k'_\tau - k_\tau| - |k_\tau - p_\tau - q_\tau|)\nn\\
&&\times\frac{(\theta(k_\tau) - \theta(\ell' (k_\tau - p_\tau))) (\theta(k'_\tau) - \theta(\ell'(k'_\tau - p_\tau - q_\tau)))} {(i \eta ((1-\ell') k_\tau - p_\tau) + \ell' \vec{v}_1 \cdot \vec{p})(i \eta ((1- \ell') k'_\tau - p_\tau - q_\tau)+ \ell' \vec{v}_2 \cdot (\vec{p} + \vec{q}))}
\nn \\ &&~~~~~~D(k' - p) D(k - k') D(p+q-k). \nn\eea
Here $k_{\parallel}$, $k'_{\parallel}$ denote the components of $\vec{k}$, $\vec{k'}$ along  the Fermi surface of $\psi_1$ and $\psi_2$ respectively, and the arguments of boson propagators are evaluated at $\vec{v}_1 \cdot \vec{k} = \vec{v}_2 \cdot \vec{k}' = 0$. (Strictly speaking, only one pair of poles has $\vec{v}_1 \cdot \vec{k} = \vec{v}_2 \cdot \vec{k}' = 0$, while the other pair has $\vec{v}_1 \cdot \vec{k} = \vec{v}_1 \cdot \vec{p}$ and $\vec{v}_2 \cdot \vec{k}' = \vec{v}_2 \cdot (\vec{p} + \vec{q})$. However, in situations of interest to us discussed below the above difference may be neglected in the bosonic propagators).

Note that if we take the initial and final fermion momenta to lie on the Fermi surface, {\em i.e.\/} $\vec{v}_1 \cdot \vec{p} = 0$, $\vec{v}_2 \cdot (\vec{p} + \vec{q}) = 0$, then $\delta \Gamma$ diverges as $\eta^{-2}$. Since the dimension of $\eta$ is $\omega^{-1/2}$, this is synonymous to an infra-red divergence, 
\beq \delta \Gamma_{\phi \psi_2 \psi^{\dagger}_1} \sim \eta^{-2} N^{-2} \omega^{-1} \label{div1} .
\eeq
This behavior can be easily checked by, for instance, setting all the external momenta to zero ({\em i.e.\/} taking the external fermions to be at the hot spots). We also note that in the case when the external fermion momenta do not lie on the Fermi surface, the limit $\eta \to 0$ can be taken in the contribution of hot spot pair $\ell' = 1$, but not $\ell' = -1$, as the latter contains a non-local $UV$ divergence. Keeping $\eta$ finite, we obtain,
\beq \delta \Gamma_{\phi \psi_2 \psi^{\dagger}_1} \sim \eta^{-1} N^{-2} p_\perp^{-1} \label{div2}.\eeq 
where $p_\perp$ schematically denotes the distance of external fermion momenta to the Fermi surface.

The infra-red divergences in Eqs. (\ref{div1}), (\ref{div2}) are a product of the bare fermion propagator having $z = 1$ dynamics, whereas we expect that the full fermion propagator has the same dynamics as the spin-density wave excitations. We saw that this, indeed, holds at the one-loop level, where both the boson (\ref{bprop}) and fermion (\ref{Sigma}) propagators are invariant under scaling with $z = 2$ (up to logarithmic corrections in the latter case). As in Ref.~\onlinecite{SSLee}, the divergence can be cured by including the one-loop fermion self-energy within the fermion propagators, before taking the $\eta \rightarrow 0$
limit. This is the approach that will be adopted below. From Eq.~(\ref{Sigmak0}), we know that the self-energy is $\sim \sqrt{\omega}/N$. Therefore, mapping
$\eta \omega \rightarrow \sqrt{\omega}/N$, we find from Eq.~(\ref{div1}) that
\beq \delta \Gamma_{\phi \psi_2 \psi^{\dagger}_1} \sim \mathcal{O}(1) \label{Gammaest}\eeq 
Thus, the vertex correction is not suppressed relative to the bare value, and the naive $1/N$ expansion has broken down. In the appendix \ref{app:vert3l}, we compute the vertex correction in Fig.~\ref{fig:vertex2} with dressed fermion propagators and find to logarithmic accuracy,
\beq \delta \Gamma_{\phi \psi_2 \psi^{\dagger}_1} \sim X(\alpha) \log \frac{\Lambda}{|\vec{q}|} \label{dGammaX}\eeq
where $X$ is a finite negative function of $\alpha$. Note that the strong infra-red divergence of Eq.~(\ref{div1}) is now replaced by a mild logarithmic divergence that one may hope to treat with renormalization group. However, the price one has to pay for curing the strong infra-red divergence is the enhancement of the diagram with $N$, as anticipated in Eq.~(\ref{Gammaest}). This enhancement occurs for any external fermion momenta (not only for momenta on the Fermi surface). Finally, the presence of a logarithm implies that not only is the diagram itself unsuppressed relative to its bare value, but also that the anomalous dimensions are not expected to be suppressed with $N$. 

Having seen an explicit example of violation of naive large-$N$ counting, we would like to investigate the general scaling of diagrams with $N$ in our theory, when a one-loop dressed fermion propagator is used. Our procedure closely follows that of Ref.~\onlinecite{SSLee}. A general diagram can be schematically written as,
\beq {\cal D} = N^{L_f} \int \prod_{i = 1}^{L} 
{d^2 p_i d \omega_i}\prod_{j = 1}^{I_f} \frac{1}{\Sigma_{1loop}(l_j) + \vec{v} \cdot \vec{l}_j} \prod_{k = 1}^{I_b} D(q_k)\eeq
Here, $I_f$ and $I_b$ are numbers of fermion and boson propagators respectively, $L_f$ is the number of fermion loops and $L$ is the number of total loops. The momenta $l_j$ and $q_k$ are linear combinations of $p_i$ entering the fermion and boson propagators. The ``naive" scaling of the diagram with $N$ is given by ${\cal D} \sim N^{Q_0}$,
\beq Q_0 = L_f - I_b \label{Q0} \eeq
It is clear that the enhancement of diagrams with $N$ comes from the dangerous factor of $1/N$ in the fermion self-energy. However, in order to access this factor the fermion momentum must be on the Fermi surface. Given a diagram, let us call the phase-space for all internal fermion momenta to lie on the Fermi surface, the ``singular manifold." Having identified this manifold, one can divide the momentum integration variables into components parallel $p_\parallel$ and perpendicular $p_\perp$ to the manifold,
\beq  \prod_{i = 1}^{L} d^2 p_i = \prod_{a = 1}^{n} dp_{\parallel a} \prod_{b = 1}^{2 L - n} dp_{\perp b}\eeq
where $n$ is the dimension of the manifold. Linear combinations of $p_{\perp}$'s enter the fermion energy $\vec{v} \cdot \vec{l}_j$ and hence scale as $1/N$, making the fermion propagators scale as $N$. On the other hand, the components $p_\prl$ only enter the bosonic propagators and the one-loop fermion self-energy $\Sigma_{1loop}$ and scale as $N^0$. Hence, the diagram acquires an enhancement, ${\cal D} \sim N^{Q}$, $Q = Q_0 + \Delta Q$,
\beq \Delta Q = [I_f - 2 L + n] \label{DeltaQ} \eeq
where $[x] = x$ if $x \ge 0$ and $[x] = 0$ if $x < 0$. 

Thus, to find the degree of a diagram in $N$, one has to find the singular manifold and compute its dimension $n$. This can be done diagramatically by introducing a double-line representation, originally used in the study of electron-phonon interactions.\cite{Shankarphon} Below, we will consider diagrams involving opposite hot spot pairs $\ell = 1$ and $\ell = -1$ only. Subsitution of fermions from hot spots $\ell = 2$ and $\ell = -2$ into these diagrams is expected to reduce the dimension of the singular manifold. Moreover, we for simplicity consider diagrams without the quartic bosonic vertex $u$. Finally, we take all the exernal fermion momenta to be on the Fermi surface. 

\begin{figure}[t]
\begin{center}
\includegraphics*[width=3.5in]{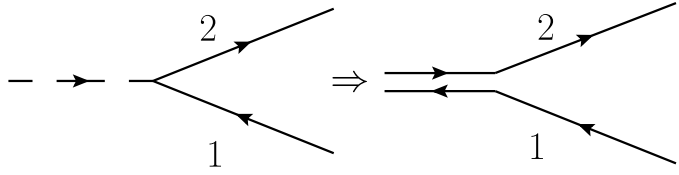}
\caption{Double line representation for the boson-fermion vertex.}
\label{fig:DblV}
\end{center}
\end{figure}

\begin{figure}[t]
\begin{center}
\includegraphics*[width=3.5in]{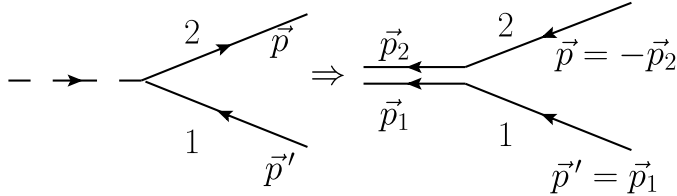}
\caption{Double line representation for the boson-fermion vertex. The direction of momentum and particle flow need not coincide.}
\label{fig:DblVinv}
\end{center}
\end{figure}

Now, we are ready to introduce the double-line representation. We would like to find under what conditions do all the fermions in a diagram go to the Fermi surface. Observe, that any momentum can be uniquely decomposed into components along the Fermi surface of fermion $1$ and fermion $2$. Thus, we fatten bosonic propagators into double lines, one carrying momentum along the Fermi surface of fermion $1$, and the other along the Fermi surface of fermion $2$. If a fermion is to absorb this bosonic momentum and stay on the Fermi surface, its incoming and outgoing momenta are fixed in terms of the components of the double line. Hence, the boson-fermion vertices can be redrawn as shown in Fig.~\ref{fig:DblV}. Note that if a certain momentum is along the Fermi surface of fermion $1$ from hot spot $\ell = 1$, it is also along the Fermi surface of fermion $1$ from hot spot $\ell = -1$. Thus, the fermion lines in our diagrams can come from either of these hot spots. Also, the direction of lines in the double-line representation is not fixed, and need not coincide with that in the single line representation. If the two are opposite, then it is understood that the physical fermion momentum $\vec{p}$ is the negative of the momentum carried by the fermion in the double-line representation, see Fig.~\ref{fig:DblVinv}. 
Because we are neglecting the Fermi surface curvature in the low-energy theory, a particle with momentum $\vec{p}$ is on the Fermi surface if and and only if a particle with momentum $-\vec{p}$ is on the Fermi surface, and the above representation is consistent. 
(We remind the reader that here all the fermion momenta are defined relative to hot spot locations). 

\begin{figure}[t]
\begin{center}
\includegraphics*[width=2in]{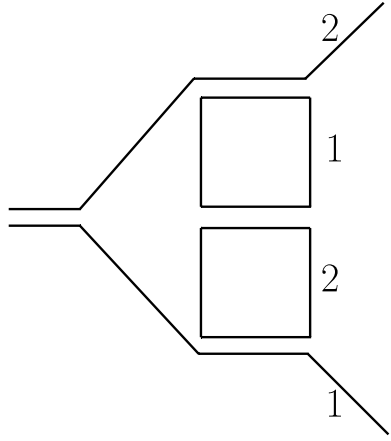}
\caption{Double line representation applied to the diagrams in 
Fig.~\ref{fig:vertex2}. The enhancement of the diagram in $N$ is related to the number of loops $n$ in the double line-representation via Eq.~\ref{DeltaQ}.}\label{fig:vertex4}
\end{center}
\end{figure}

\begin{figure}[t]
\begin{center}
\includegraphics*[width=2in]{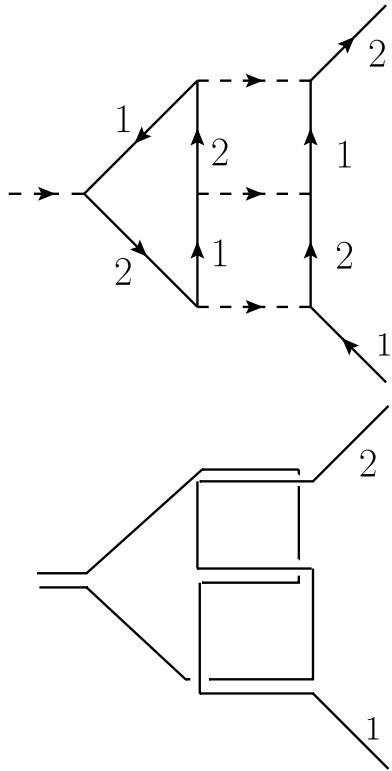}
\caption{A three loop vertex correction with no enhancement in $N$.}
\label{fig:vertex5}
\end{center}
\end{figure}

Thus, the double line representation completely specifies the singular manifold. In particular, the dimension of the manifold $n$ is just given by the number of loops in this representation. As an example, consider the double line represenation of the diagrams in 
Fig.~\ref{fig:vertex2} shown in Fig.~\ref{fig:vertex4}.  We see that Fig.~\ref{fig:vertex4} contains two closed loops, which implies that the
singular manifold is two-dimensional. From Eq.~(\ref{DeltaQ}), the enhancement of the diagram is $\Delta Q = 2$, which combined with the naive degree of the diagram, Eq.~(\ref{Q0}), $Q_0 = -2$, gives $Q = 0$, consistent with the explicit calculation in Eq.~(\ref{dGammaX}). In Fig.~\ref{fig:vertex5} we also give an example of a vertex correction which is not enhanced in $N$. Here, the double line representation contains no loops so the dimension of the singular manifold is zero, $\Delta Q = 0$ and the degree of the diagram is given by the naive $N$ counting, $Q = -2$.

\begin{figure}[t]
\begin{center}
\includegraphics*[width=3.5in]{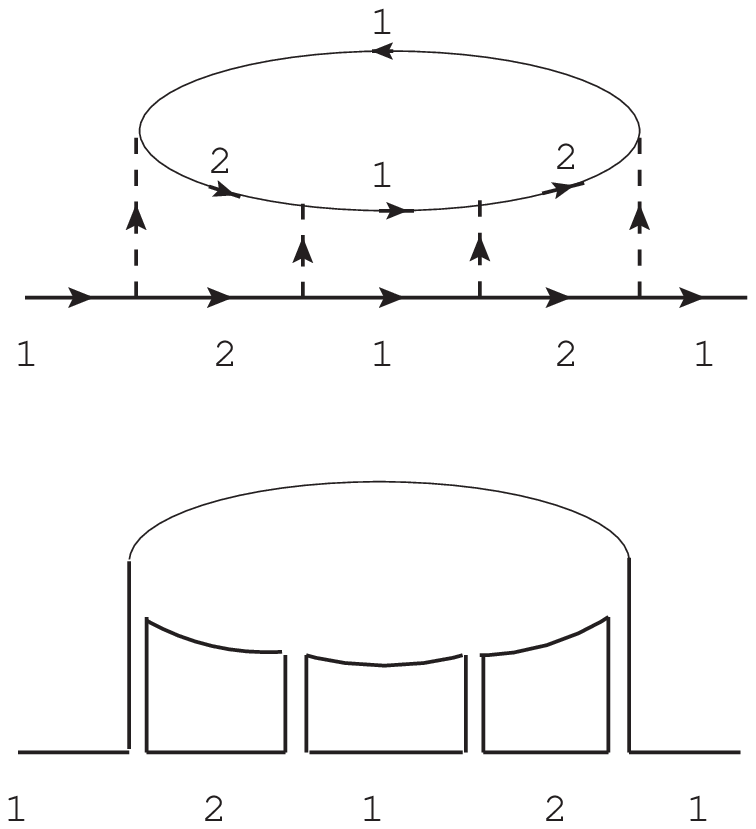}
\caption{A diagram for the fermion self-energy that is of ${\cal O}(1/N)$ as a result of enhancement.}
\label{fig:FSE4loop}
\end{center}
\end{figure}

\begin{figure}[t]
\begin{center}
\includegraphics*[width=3.5in]{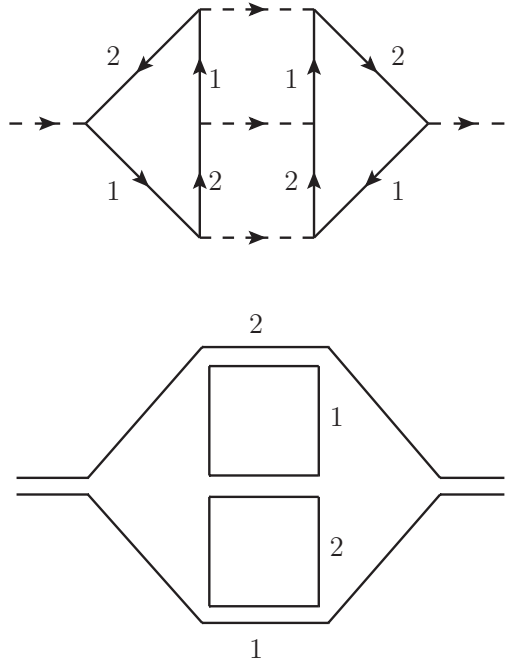}
\caption{A diagram for the boson self-energy that is of ${\cal O}(N)$ as a result of enhancement.}
\label{fig:Pi4loop}
\end{center}
\end{figure}

It is easy to see that the violations of naive large-$N$ counting are not confined to vertex corrections alone. In Fig.~\ref{fig:FSE4loop} we show a 
fermion self-energy diagram that acquires an enhancement. Indeed, the naive degree of the graph is $Q_0 = -3$. However, since the double line representation contains three loops, the graph receives an enhancement $\Delta Q = 2$, so that the total degree of the graph is $Q = -1$. Hence, the graph is of the same order $\mathcal{O}(1/N)$ as the one-loop fermion self-energy. Similarly, in Fig.~\ref{fig:Pi4loop} we show an enhanced diagram for the boson self-energy. In this case, $Q_0 = -1$, $\Delta Q = 2$, $Q = 1$. Hence, the diagram is of $\mathcal{O}(N)$, again the same as the tree level contribution. 

\begin{figure}[t]
\begin{center}
\includegraphics*[width=3in]{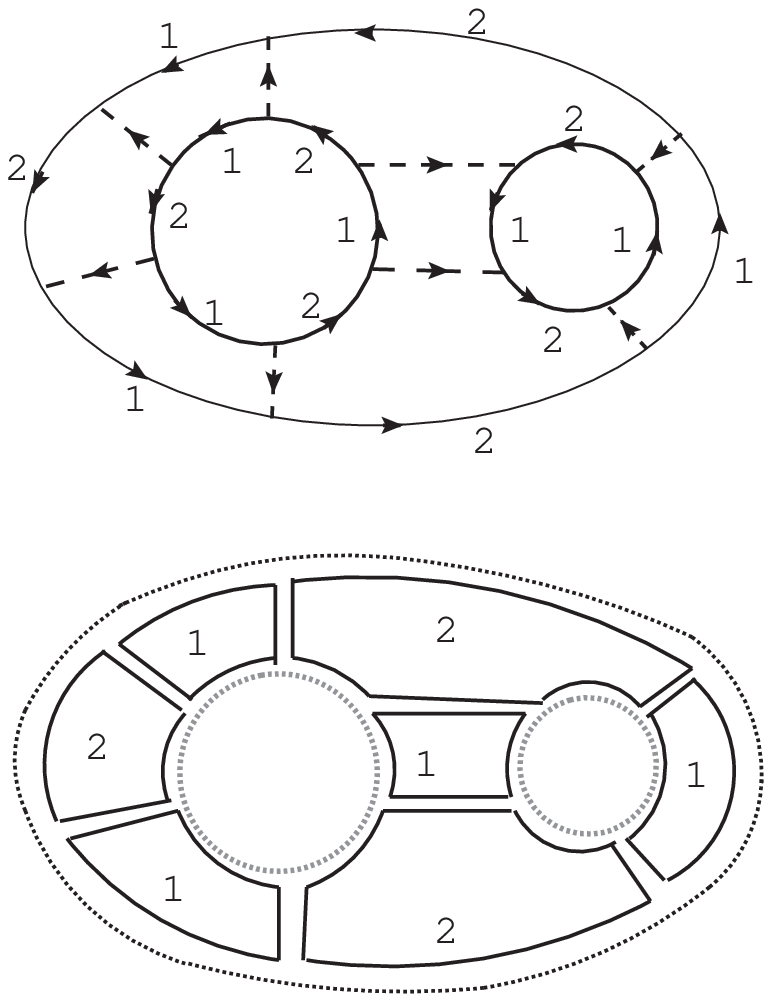}
\caption{Converting vacuum energy diagrams into surfaces: a face is attached to each solid and grey loop in the double-line representation (on the right). In the present case, the resulting surface is a sphere.}
\label{fig:surface}
\end{center}
\end{figure}

A remarkable feature of the large-$N$ counting in Eqs.~(\ref{Q0}), (\ref{DeltaQ}), pointed out in Ref. \onlinecite{SSLee}, is that the degree of a diagram is related to its topology. Let us first apply the topological classification to vacuum energy diagrams, {\em i.e.\/} 
graphs with no external lines. We can convert these diagrams into two-dimensional surfaces in the following way. First, let us introduce fermion loops back into the double line represenation (they will appear dotted in our diagrams, see Fig.~\ref{fig:surface}). Then attach  a face to each solid loop of the double-line representation and a face to each dotted loop ({\em i.e.\/} fermion loop). As a result, each boson propagator is shared by two faces with solid boundaries, while each fermion propagator is shared by a face with a solid boundary and a face with a dotted boundary. Therefore, if we glue the faces along propagators we obtain a closed surface. Now consider the Euler characteristic of this surface,
\beq \chi = F - E + V \eeq
where $F$ is the number of faces, $E$ is the number of edges and $V$ is the number of vertices of the surface. We have, $F = L_f + n$, $E = I_b + I_f$ and $V$ is just the number of vertices in the original Feynman graph. Now, using $V = 2 I_b$, $2 V = 2 I_f$ we obtain,
\beq \chi = L_f + n - \frac{V}{2} \eeq
However, using $L = I_b + I_f - V + 1$, we see that the degree of a diagram in $N$ given by Eqs. (\ref{Q0}), (\ref{DeltaQ}), is,
\beq Q = L_f - \frac{V}{2} + n - 2\eeq
where we've assumed that the argument of $[\, ]$ in Eq.~(\ref{DeltaQ}) is positive. Thus, we arrive at the relation,
\beq Q = \chi - 2 \eeq
This result means that at each order in $1/N$ one has to sum an infinite set of diagrams with a given Euler characteristic. In particular, at $N = \infty$ the theory is dominated by diagrams with $\chi = 2$, {\em i.e.\/} those whose double-line representation can be drawn on a sphere. Such graphs are often referred to as planar diagrams.

It is straightforward to extend the classification above to diagrams with external legs. For instance, fermion self-energy diagrams  can be obtained by cutting one fermion propagator in a vacuum graph. This results in $I_b \to I_b$, $L_f \to L_f - 1$, so $Q_0 \to Q_0 - 1$, and $I_f \to I_f - 1$, $L \to L - 1$, $n \to n - 1$, as cutting a fermion propagator destroys a solid loop in the double line representation. Hence, $\Delta Q \to \Delta Q$ and $Q \to Q - 1$, {\em i.e.\/}
\beq Q = \chi - 3\eeq
with $\chi$ the Euler characteristic of the initial vacuum diagram. In particular, planar vacuum graphs give rise to fermion self-energy diagrams of $\mathcal{O}(1/N)$. 

Similarly, to obtain a boson self-energy diagram, we cut a boson propagator in a vacuum bubble. This gives $I_b \to I_b - 1$, $L_f \to L_f$, so $Q_0 \to Q_0 + 1$, and $I_f \to I_f$, $L \to L - 1$, $n \to n - 2$, as we now destroy two solid loops in the double line representation. Hence, $\Delta Q \to \Delta Q$ and $Q \to Q + 1$, {\em i.e.\/}
\beq Q = \chi - 1\eeq
Hence, planar graphs give rise to boson self-energy diagrams of $\mathcal{O}(N)$.

Likewise, to obtain vertex correction diagrams, we remove a vertex in a vacuum bubble. As a result, $I_b \to I_b - 1$, $L_f \to L_f - 1$, so $Q_0 \to Q_0$, and $I_f \to I_f - 2$, $L \to L - 2$, $n \to n - 2$, as we again destroy two solid loops in the double line representation. Hence, $\Delta Q \to \Delta Q$ and $Q \to Q$, {\em i.e.\/}
\beq Q = \chi - 2\eeq
and all planar graphs give rise to vertex diagrams of $\mathcal{O}(1)$. 

At this point, we would like to make a remark about conditions on external momenta in diagrams needed for the enhancements to occur. Up to now we have been assuming that all the external fermion momenta in a diagram are on the Fermi surface. If all the diagrams in our theory were $UV$ finite then this condition would, indeed, be required. However, as we have seen, some of the diagrams actually contain logarithmic divergences, {\em i.e.\/} they receive contributions from momenta, which are much larger than the external momenta. For the purpose of computing the $UV$ divergent contribution to these diagrams and estimating its scaling with $N$, we can set the external momenta to zero (which certainly puts the external fermions on the Fermi surface). This explains why the vertex correction in Figs.~\ref{fig:vertex2},\ref{fig:vertex4} receives an enhancement for any external fermion momentum, as can be explicitly seen in Eq.~(\ref{dGammaX}). 

\begin{figure}[t]
\begin{center}
\includegraphics*[width=3.5in]{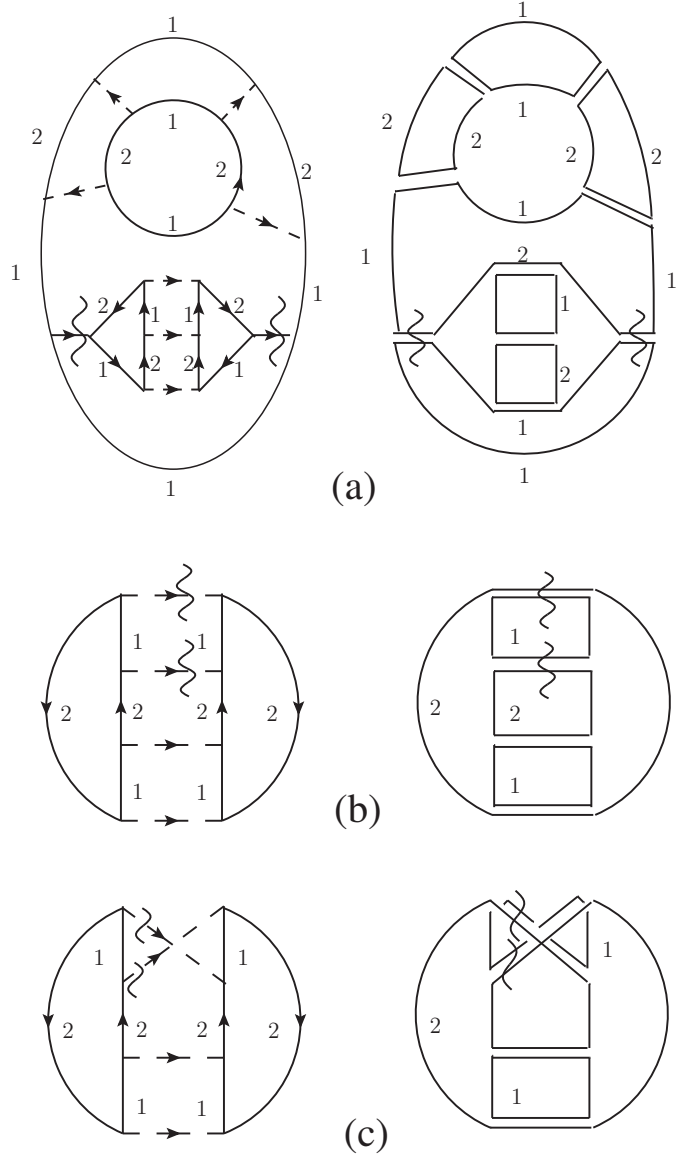}
\caption{Producing a boson four-point function from a vacuum bubble by cutting two boson propagators. If the initial diagram is planar and only two solid lines are cut in the double-line representation then the resulting diagram is disconnected, as in (a). Diagrams of highest degree are obtained by starting with a planar diagram and cutting three solid line loops, as in (b), or starting with a diagram with $\chi = 1$ and cutting two solid line loops, as in (c).}
\label{fig:Cut2}
\end{center}
\end{figure}
\begin{figure}[t]
\begin{center}
\includegraphics*[width=2.5in]{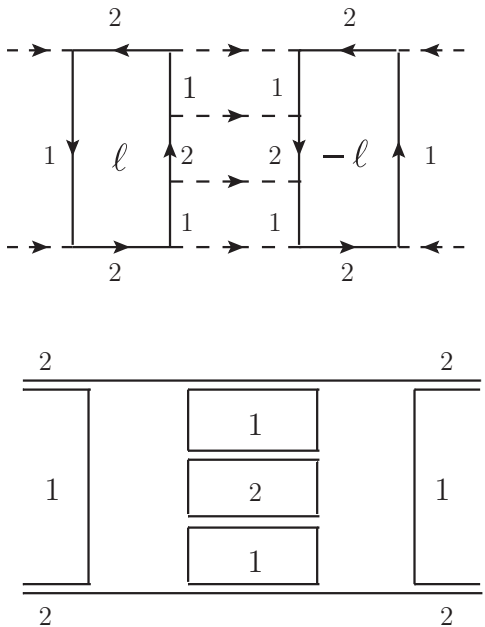}
\caption{A diagram for the boson four-point function that diverges logarithmically and scales as $N^3$.}
\label{fig:Quartic5l}
\end{center}
\end{figure}

So far, we have left out one type of diagram which is important from the point of view of RG properties of the theory, namely diagrams for the boson four-point function. Such diagrams can be obtained by cutting two boson propagators in a vacuum bubble. This results in $I_b \to I_b - 2$, $L_f \to L_f$, so $Q_0 \to Q_0 + 2$. Now let us discuss the change in the enhancement $\Delta Q$. We see that $I_f \to I_f$, $L \to L - 2$. The change in the dimension of the singular manifold $\delta n$ depends on how many loops in the double line representation the two propagators that we cut share. If both the components $1$ and $2$ of the two propagators are part of the same two solid loops, see Fig.~\ref{fig:Cut2}c, then the change in the dimension of the singular manifold $\delta n = -2$. If these two propagators share only one solid loop, see Fig.~\ref{fig:Cut2}b, then $\delta n = -3$. Finally, if the two propagators don't share any solid loops, then $\delta n =  -4$. Thus, we obtain, $\Delta Q \to \Delta Q + 4 + \delta n$ and $Q \to Q+ 6 + \delta n$, {\em i.e.\/}
\beq Q = \chi + 4 + \delta n \label{Qquartic}\eeq
It appears that the highest possible degree of the four-point vertex corresponds to starting with a planar graph and cutting two bosonic propagators, which are part of the same double-line loop, to obtain, $Q = 4$. However, it is easy to see that this always produces a diagram, which is disconnected, see Fig.~\ref{fig:Cut2}a. To obtain a connected diagram for the four-point function starting from a planar graph, we must cut at least three solid loops, such that the highest possible degree of a four-point function is $Q = 3$. The fact that the four-point vertex scales as $N^3$ could be anticipated from the simple one-loop result in Eq.~(\ref{f}). Indeed, for special kinematic conditions, $\vec{v}_1 \cdot (\vec{q}_2 + \vec{q}_3) = 0$, $\vec{v}_2 \cdot (\vec{q}_1 + \vec{q}_2) = 0$, Eq.~(\ref{f}) diverges as $N (\eta^2 \omega)^{-1}$, which after including the one-loop fermion self-energy is expected to become of order $N^3$. Such kinematic conditions are automatically assumed in our double line representation that led to the large-$N$ counting in Eq.~(\ref{Qquartic}). However, as was already noted, diagrams that have ultraviolet divergences are expected to receive the enhancement in Eq.~(\ref{DeltaQ}) independent of external momenta. The simplest diagram for the boson four-point vertex that is expected to scale as $N^3$ and exhibits such a divergence is shown in Fig.~\ref{fig:Quartic5l}. 
In the appendix, we explicitly evaluate this diagram obtaining to logarithmic accuracy,
\beq \delta\Gamma^4 = N^3 Y(\alpha) \gamma \log \frac{\Lambda}{|\vec{q}|}\eeq
with $Y$ a finite function of $\alpha$. 

The fact that there are diagrams for the four-point boson function that scale as $N^3$ for arbitrary external momenta has drastic consequences for the theory. Indeed, a diagram with just quartic internal vertices (which can themselves have a non-trivial internal structure), will scale as $N^Q$, with $Q = V_4 + \frac{E_b}{2}$, where $V_4$ is the number of quartic vertices and $E_b$ is the number of external bosons. Thus, the degree of the diagram in $N$ grows with the number of quartic vertices. This means that perturbation theory based on the one-loop dressed fermion propagator is not a good starting point for taking the large-$N$ limit, and no genus expansion similar to that of Ref.~\onlinecite{SSLee} exists in the present case. Note that this effect was not captured in our initial large-$N$ counting, as we have ignored the possible presence of $UV$ divergent subdiagrams.

\section{Pairing vertex}
\label{sec:pairing}

\begin{figure}[h]
\begin{center}
\includegraphics[width=2.5in]{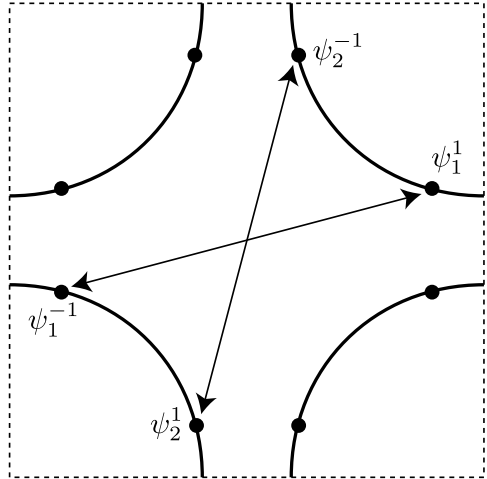}
\caption{Pairing of the electrons at the $\ell=\pm 1$ hotspots of Fig.~\ref{fig:hotspots}.
Electrons at opposite ends of the arrows form spin-singlet pairs.
The $\mu=+1$ ($\mu=-1$) pairing amplitude in Eq.~(\ref{Vmunu}) has the same (opposite) sign
on the two arrows. Only the $\mu=-1$ spin singlet pairing is enhanced near the SDW critical point.
}
\label{fig:hotspots2}
\end{center}
\end{figure}

In this section we will study the renormalization properties of the BCS order parameter to one loop order. We consider 
pairing in the spin singlet, parity even, momentum zero channel. There are four order parameters that one can form out of our four 
pairs of hot spots,
\beq V_{\mu \nu} = \epsilon_{\sigma \sigma'} (\psi^{\ell = -1}_{1\sigma} \psi^{\ell = 1}_{1 \sigma'} + \mu \psi^{\ell = -1}_{2 \sigma} \psi^{\ell = 1}_{2 \sigma'}) + \nu \epsilon_{\sigma \sigma'} (\psi^{\ell = -2}_{1\sigma} \psi^{\ell = 2}_{1 \sigma'} + \mu \psi^{\ell = -2}_{2 \sigma} \psi^{\ell = 2}_{2 \sigma'}) \label{Vmunu}\eeq
Here the minus sign in the hot spot labels $\ell = -1 \equiv 3$ and $\ell = -2 \equiv 4$ denotes the opposite hot spot pair. The geometry of the pairing operators for $\ell=1$ is illustrated in Fig.~\ref{fig:hotspots2}.
The coefficients $\mu = \pm 1$, $\nu = \pm 1$ determine the transformation properties of $V$ under 
the lattice rotation symmetry $R_{\pi/2}$ and the reflection symmetry $I_{(-1,1)}$ about the $(-1,1)$ axis:
\bea R_{\pi/2}: V_{\mu \nu} &\to& \nu V_{\mu \nu}\\
I_{(-1,1)}: V_{\mu \nu} &\to& \mu V_{\mu \nu}\eea
These properties are summarized in Table \ref{TblV}.
\begin{table}
\medskip
\begin{center}
\begin{tabular}{|c||c|c|c|}
\hline
& \multicolumn{3}{c|}{$\mu$}\\
\hline\hline
& &1 & -1\\
\cline{2-4}
$\nu$ & 1& $s$ & $g$\\
\cline{2-4}
&~~ -1~~& ~~$d_{xy}$~~ & ~~$d_{x^2-y^2}$~~\\
\hline
\end{tabular}
\end{center}
\caption{Symmetry properties of the pairing vertex.}\label{TblV}
\end{table}
Since the theory (\ref{Lmarg}) conserves the number of fermions at each hot spot pair $\ell$, the parts of the order parameter involving $\ell = \pm 1$ and $\ell = \pm 2$ renormalize independently. Hence, the scaling dimension of the pairing vertex in the low-energy theory is independent of $\nu$ and is sensitive only to $\mu$, i.e the operators with $s$ and $d_{xy}$, and $g$ and $d_{x^2-y^2}$ symmetries are degenerate.

\begin{figure}[t]
\begin{center}
\includegraphics*[width=1.5in]{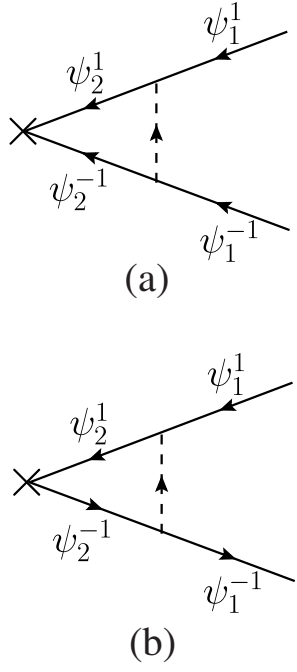}
\caption{The leading corrections to (a) BCS pairing vertex, (b) density-wave vertex.}
\label{FigV}
\end{center}
\end{figure}

The renormalization properties of the operator $V$ can be determined from its insertion into the correlation function, 
\beq \epsilon_{\sigma \sigma'} \Gamma_{V \psi^{\dagger} \psi^{\dagger}}(k_1, k_{-1}) = \int d^D x_{1} d^D x_{-1} \langle V(0) \psi^{\dagger\ell = -1}_{1 \sigma'}(x_{-1}) \psi^{\dagger\ell = 1}_{1 \sigma}(x_1) \rangle_{1PI} e^{i (k_1 x_1 + k_{-1} x_{-1})} \eeq  
At tree level, $\Gamma_{V \psi^{\dagger} \psi^{\dagger}} = 1$. Let us now consider the one-loop renormalization of $V$, shown in Fig. \ref{FigV} a). 
This diagram is given by
\beq \delta \Gamma_{V \psi^{\dagger} \psi^{\dagger}}(k_1, k_{-1}) = 
- 3 \mu \int \frac{d^3 l}{(2\pi)^{3}} D(l) G^1_{2}(k_1 - l) G^{-1}_{2}(k_{-1} + l). \label{GV1}
\eeq
Details of the evaluation of (\ref{GV1}) appear in Appendix~\ref{app:pairing}. Direct computation with bare fermion propagators gives rise to strong infra-red divergences, which are cured by using the one-loop dressed propagators. With this approach, we obtain
to logarithmic accuracy
\beq \delta \Gamma_{V \psi^{\dagger} \psi^{\dagger}}  = - \frac{\mu \alpha}{\pi (\alpha^2 + 1)} \log^2\left(\frac{\Lambda^2}{\gamma \omega}\right) \label{Vfin}\eeq

Note that the one loop renormalization of the pairing vertex (\ref{Vfin}) is of order unity, and is 
not suppressed in $1/N$. Thus the naive counting in powers of $1/N$ is violated, as was already noted in 
Ref.~\onlinecite{ChubukovFinkelstein}.
Moreover, the one-loop contribution gives a suppression of the vertex for $\mu = 1$ ($s$ and $d_{xy}$ channels) and an enhancement for $\mu = -1$ ($d_{x^2-y^2}$, $g$ channels) as expected. Finally,
we find that the one-loop result has a non-local $\log^2$ divergence. 
The origin of this non-local divergence is BCS pairing of the Fermi surface away from the hot spots. Indeed, as noted
in Appendix~\ref{app:pairing}, the divergence comes from the regime 
where $\gamma |l_\tau| \ll l^2_\parallel$, with $l_\parallel$ the component of $\vec{l}$ along the Fermi surface of $\psi_2$. This is precisely the regime in which one has good Landau-quasiparticles, suggesting
that it may be possible to obtain Eq.~(\ref{Vfin}) in a Fermi liquid computation.

We now show this is indeed the case, and obtain (\ref{Vfin}) in a physically transparent form.
Let us approximate the propagators in Eq.~(\ref{GV1}) by the Fermi-liquid form Eq.~(\ref{GFSscal}),
\beq \delta \Gamma_{V \psi^{\dagger} \psi^{\dagger}} = \frac{3 \mu}{N} \int \frac{dl_\parallel}{2 \pi} \int_{\gamma |l_\tau| \lesssim l^2_\parallel} \frac{d l_\tau}{2 \pi} \int \frac{dl_\perp}{2\pi} \frac{1}{\gamma |l_\tau| + l^2_\parallel} \frac{{\cal Z}(l_\parallel)}{i (l_\tau - \omega) - v_F(l_\parallel) l_\perp} \frac{{\cal Z}(l_\parallel)}{i (l_\tau+ \omega) + v_F(l_\parallel) l_\perp}\eeq
with the Fermi-liquid parameters given by Eq.~(\ref{ZVoneloop}).
Note that due to the restriction $\gamma |l_\tau| \ll l^2_\parallel$ the bosonic propagator is static. Changing variables to $\epsilon = v_F(l_\parallel) l_\perp$,
\beq \delta \Gamma_{V \psi^{\dagger} \psi^{\dagger}} = \frac{3 \mu}{N} \int \frac{dl_\parallel}{2 \pi} \frac{{\cal Z}^2(l_\parallel)}{v_F(l_\parallel)l^2_\parallel }\int_{\gamma |l_\tau| \lesssim l^2_\parallel} \frac{d l_\tau}{2 \pi} \int \frac{d\epsilon}{2\pi} \frac{1}{i (l_\tau - \omega) - \epsilon} \frac{1}{i (l_\tau+ \omega) + \epsilon}\eeq
The integral over $l_\tau$, $\epsilon$ has the form familiar from Fermi-liquid theory and gives the usual BCS logarithm,
\beq \int \frac{d l_\tau}{2 \pi} \int \frac{d\epsilon}{2\pi} \frac{1}{i (l_\tau - \omega) - \epsilon} \frac{1}{i (l_\tau+ \omega) + \epsilon} = - \frac{1}{2 \pi} \log \frac{\Lambda_{FL}}{\omega} \eeq
where $\Lambda_{FL}$ is the frequency/energy cut-off, which in the present case is $\Lambda_{FL} = l^2_\parallel/\gamma$. Of course, for the above form to hold, we need $\omega \ll \Lambda_{FL}$. Thus,
\beq \delta \Gamma_{V \psi^{\dagger} \psi^{\dagger}} = -\frac{3 \mu}{2 \pi^2 N} \int_{\sqrt{\gamma \omega}}^{\infty} dl_\parallel \frac{{\cal Z}^2(l_\parallel)}{v_F(l_\parallel)l^2_\parallel } \log \frac{l^2_\parallel}{\gamma \omega} = -\frac{\mu \alpha}{\pi (\alpha^2 + 1)} \log^2 \frac{\Lambda^2}{\gamma \omega}\eeq
which agrees with the result in Eq.~(\ref{Vfina}) obtained from a more complete computation. Note that the prefactor
of $1/N$ arising from the boson propagator has disappeared from the final result.
A similar log-squared term has been noted for the pairing vertex in a theory of a Fermi surface coupled to a gauge field
in three dimensions\cite{sonqcd,Schafer} and in a theory of a Fermi surface interacting via a Chern-Simons gauge field and a $1/r$ potential
in two dimensions.\cite{nayak}

The appearance of the log-squared term above indicates a breakdown of the present RG in analyzing
the renormalization of the pairing vertex. It is clearly a consequence of two different physical effects.
One is the familiar BCS logarithm of Fermi liquid theory, which appears here from the Fermi surface away
from the hot spots. The second logarithm is a critical singularity associated with SDW fluctuations at the hot
spot. Our RG approach, defined in terms of a cutoff $\Lambda$ which measures distance from the hot spot,
is unable to regulate the first logarithm: the Fermi surface is present at momenta all the way upto $\Lambda$.

An alternative RG is necessary to analyze the consequences of the log-squared term. One possible approach
is that of Son \cite{sonqcd}, who worked with an RG defined in terms of momentum shells a fixed distance 
from the Fermi surface of fermions coupled to a gauge field. We leave such investigations for future work.

\section{Density vertices}
\label{sec:cdw}

In this section we focus attention on one of the interesting consequences of the pseudospin symmetries
of the critical theory of the SDW transition, specified by Eq.~(\ref{SU2}). Note that the pseudospin rotations
can be performed independently on different pairs of hotspots.

Under the operation in Eq.~(\ref{SU2}), the pairing operator (\ref{Vmunu}) in the particle-particle channel
becomes exactly degenerate with certain operators in the particle-hole channel which connect opposite patches
of the Fermi surface. 
Indeed, consider spin-singlet operators that can be built out of fermions coming from hot spots $\ell$ and $-\ell$. Using the spinor representation (\ref{Psi}), we may write these as,
\beq V^\ell_{\alpha \beta} = M_{ij} \epsilon_{\sigma \sigma'} \Psi^{-\ell}_{i \alpha \sigma} \Psi^{\ell}_{j \beta \sigma'} \label{Mij}\eeq 
The indices $\alpha$, $\beta$ of $V_{\alpha \beta}$ carry spin $1/2$ under the independent $SU^{-\ell}(2)$ and $SU^{\ell}(2)$ particle-hole symmetries. Hence, we have a set of four degenerate operators. Choosing $\alpha = 1$, $\beta = 1$,
\beq V^\ell_{11} = M_{ij} \epsilon_{\sigma \sigma'} \psi^{-\ell}_{i \sigma} \psi^{\ell}_{j \sigma'}\eeq
The mixing matrix $M_{ij}$ is fixed by lattice symmetries to give operators,
\bea V^{\ell,\vec{Q} = (0,0)}_\mu = \epsilon_{\sigma \sigma'} \left(\psi^{-\ell}_{1 \sigma} \psi^{\ell}_{1 \sigma'}+\mu \psi^{-\ell}_{2 \sigma} \psi^{\ell}_{2 \sigma'}\right)\label{Q00}\\
V^{\ell,\vec{Q} = (\pi,\pi)}_\mu = \epsilon_{\sigma \sigma'} \left(\psi^{-\ell}_{1 \sigma} \psi^{\ell}_{2 \sigma'}+\mu \psi^{-\ell}_{2 \sigma} \psi^{\ell}_{1 \sigma'}\right)\label{Qpipi}\eea
which correspond to superconducting order parameters with momenta $(0,0)$ and $(\pi,\pi)$ respectively. The index $\mu = \pm 1$ determines the parity of the operator under a reflection about a lattice diagonal. The operator (\ref{Q00}) was considered above. We will not discuss the other operator (\ref{Qpipi}) below; due to kinematics, its renormalization at one-loop order contains neither the large-$N$ enhancement, nor the unusual powers of logarithm squared.

\begin{figure}[h]
\begin{center}
\includegraphics[width=3.3in]{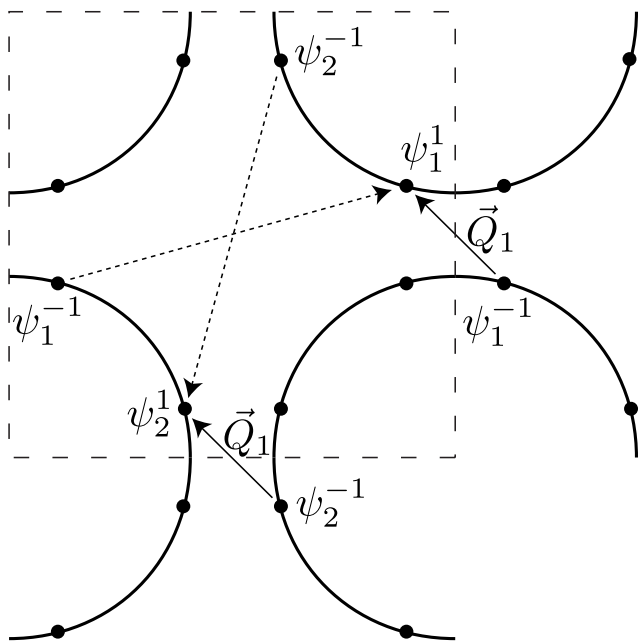}
\caption{Spin singlet density operators ($\sim \psi^\dagger \psi$)
of the electrons at the $\ell=\pm 1$ hotspots of Fig.~\ref{fig:hotspots} (see also Fig.~\ref{fig:hotspots2}), 
shown with an arrow pointing
from the Brillouin zone location of $\psi^\dagger$ to that of $\psi$. The dashed arrows are
the density operators in the first Brillouin zone. The full arrows are in an extended zone scheme
which shows that these operators have net momentum
$\vec{Q}_1 = 2 K_y (-1,1)$, where $(K_x, K_y)$ is the location of the $\ell=1$, $i=1$ hot spot.
The density operator with opposite signs ($\mu=-1$) on the two arrows is enhanced near 
the SDW critical point. Similarly the $\ell = \pm 2$ hot spots contribute density operators
at $\vec{Q}_2 = 2 K_y (1,1)$.
}
\label{fig:hotspots3}
\end{center}
\end{figure}

Now, let us discuss the particle-hole partners of (\ref{Q00}). Setting $\alpha = 2$, $\beta = 2$ in (\ref{Mij}) simply gives rise to the Hermitian conjugate of (\ref{Q00}). On the other hand $\alpha = 2$, $\beta = 1$ gives the operators,
\beq O^{\ell}_\mu = \psi^{-\ell\dagger}_{1\sigma} \psi^{\ell}_{1\sigma} + \mu \psi^{-\ell\dagger}_{2\sigma} \psi^{\ell}_{2\sigma} \label{Omu}\eeq
The other choice $\alpha = 1$, $\beta = 2$ generates the Hermitian conjugates of (\ref{Omu}). 
Following Fig.~\ref{fig:hotspots2}, the $O^{\ell}_\mu$ operators are illustrated in
Fig.~\ref{fig:hotspots3}.
To determine the wavevectors of these operators, let the $\ell=1$, $i=1$ hot spot be at $\vec{K}_1 = (K_{x}, K_{y})$. (Note that here we are using the principal axes of the square lattice
for the momentum co-ordinates, not the diagonal axes indicated in Fig.~\ref{fig:hotspots}.)
Then, from Fig.~\ref{fig:hotspots} we note that the $\ell=1$, $i=2$ hot spot is at
$(-K_y, -K_x)$, and so the value of the SDW wavevector $\vec {Q} = (\pi, \pi)$ implies that $K_x + K_y = \pi$.
Also from Fig.~\ref{fig:hotspots}, the $\ell=-1$, $i=1$ hot spot is at $(-K_x, -K_y)$, and so we conclude
that the ordering wavevector of the first term in $O^1_\mu$ is $(2 K_x, 2 K_y)$. Similarly, the ordering
wavevector of the second term in $O^1_\mu$ is seen to be $(-2K_y, -2K_x)$. Using $K_x + K_y = \pi$,
we observe that these two ordering wavevectors are actually equal, and take 
the common value $\vec{Q}_1 = 2K_y (-1,1)$,
which is therefore the momentum of the $O^1_\mu$ order parameters, as shown in Fig.~\ref{fig:hotspots2}. Similarly, the momentum
of the $O^2_\mu$ order parameters is seen to be $\vec{Q}_2 = 2 K_y (-1,-1)$. Thus the
$O^{\ell}_\mu$ represent density modulations along the diagonals of the square lattice.

For a clearer physical interpretation of the $O^{\ell}_\mu$ orders, it is useful to express them in terms
of the lattice fermions $c_{\vec{k} \sigma}$, where the momentum $\vec{k}$ ranges over the full square
lattice Brillouin zone. Then by looking at the transformations of Eq.~(\ref{Omu}) under all square lattice
space group operations, and under time-reversal, we find that the $O^{\ell}_+$ are orders are characterized
by
\begin{equation}
\left \langle c_{\vec{k} - \vec{Q}_\ell /2, \sigma}^\dagger c_{\vec{k} + \vec{Q}_ \ell /2, \sigma} \right\rangle = 
O^{\ell}_+ \,  f_0 ( \vec{k}), \label{CN1}
\end{equation}
where $f_0 (\vec{k})$ is any periodic function on the Brillouin zone 
that is invariant under the point group operations
which leave the wavevector $\vec{Q}_\ell$ invariant {\em i.e.\/} under the little group of $\vec{Q}_\ell$.
Also time-reversal and inversion symmetries imply $f_0 (\vec{k})$ is real and even.
The little group consists only of reflections along the diagonals, and so a simple
choice is  $f_0 (\vec{k}) = 1 + c_1 \left( \cos k_x + \cos k_y \right) + \ldots$, where $c_1$ is a constant.
By taking a Fourier transform of Eq.~(\ref{CN1}), it is clear that $O^{\ell}_1$ corresponds
to an ordinary charge density wave (CDW) on the sites of the square lattice:
\begin{equation}
\left \langle c^{\dagger}_{\vec{r} \sigma}  c_{\vec{r} \sigma} \right\rangle
= \sum_{\ell=1,2} \left( O^{\ell}_{+} e^{ i \vec{Q}_\ell \cdot \vec{r}} + \mbox{c.c.} \right) \label{cdw}
\end{equation}
As we saw in Section~\ref{sec:pairing}, SDW fluctuations suppress pairing with $\mu = +1$, and so
its particle-hole partner, the CDW order parameter $O^{\ell}_+$ will also be suppressed.
We will therefore not consider it further.

\begin{figure}[t]
\begin{center}
\includegraphics[width=4in]{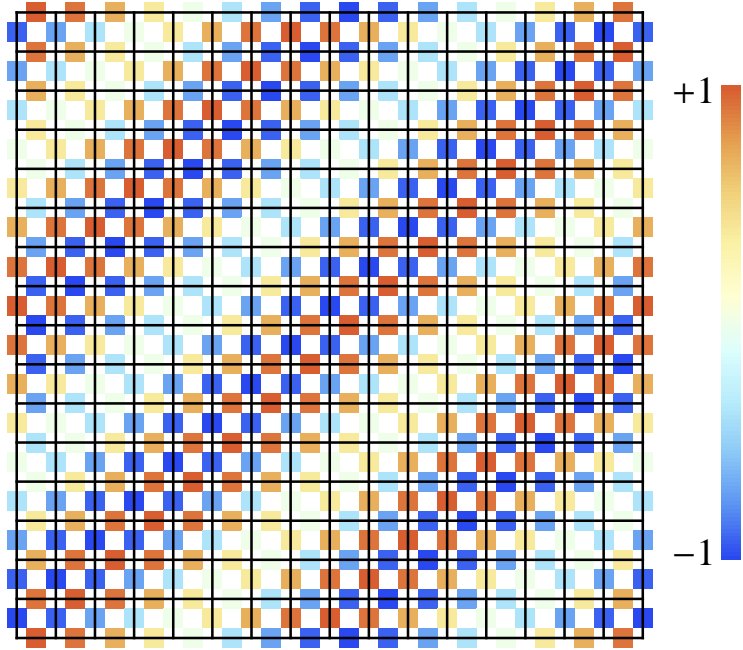}
\caption{(Color online) Plot of the bond density modulations in Eq.~(\ref{CN3}). The lines are the links of the underlying
square lattice. Each link contains a colored square representing the value of $\left \langle c^{\dagger}_{\vec{r} \sigma}  c_{\vec{s} \sigma} \right\rangle$, where $\vec{r}$ and $\vec{s}$ are the sites at the ends of the link.
We chose the ordering wavevector $\vec{Q}_1 = (2 \pi/16) (1,-1)$. Notice the local Ising-nematic ordering,
and the longer wavelength sinusoidal envelope along the diagonal.}
\label{fig:density}
\end{center}
\end{figure}
\begin{figure}[t]
\begin{center}
\includegraphics[width=4in]{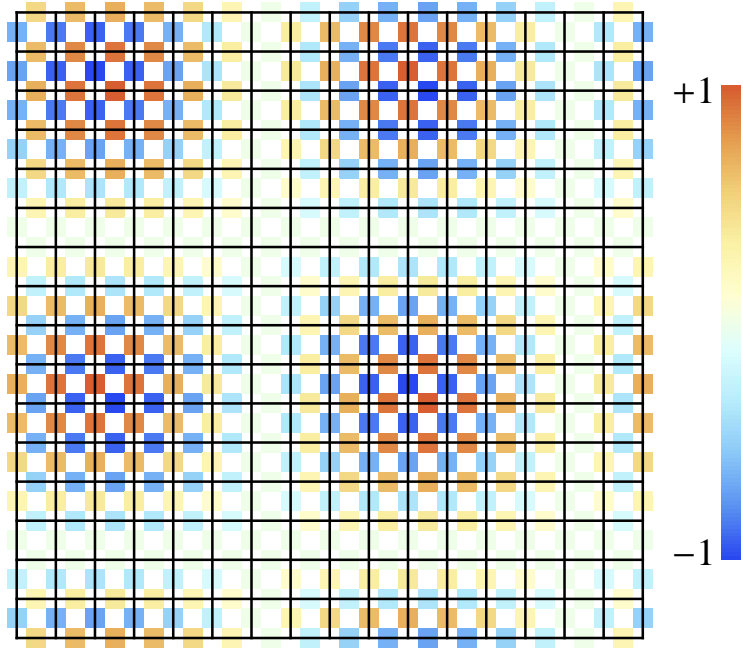}
\caption{(Color online) As in Fig.~\ref{fig:density}, but for orderings along both
$\vec{Q}_1 = (2 \pi/16) (1,-1)$ and $\vec{Q}_2 = (2 \pi/16) (1,1)$.}
\label{fig:density2}
\end{center}
\end{figure}
By the same reasoning, the order parameter $O^{\ell}_-$ should be enhanced by the SDW
fluctuations, and so it is of far greater interest. Following the steps leading to Eq.~(\ref{CN1}),
we now find 
\begin{equation}
\left \langle c_{\vec{k} - \vec{Q}_\ell /2, \sigma}^\dagger c_{\vec{k} + \vec{Q}_ \ell /2, \sigma} \right\rangle = 
O^{\ell}_-  \, \tilde{f}_0 ( \vec{k}) \left( \cos k_x - \cos k_y \right), \label{CN2}
\end{equation}
where $\tilde{f}_0 (\vec{k})$ has the same structure as $f_0 (\vec{k})$.
Time-reversal symmetry played an important role in constraining the rhs: it is easily verified
that Eq.~(\ref{CN2}) is invariant under time-reversal for general complex $O^{\ell}_-$.
The order in Eq.~(\ref{CN2}) is odd under reflections along the diagonals, and so it 
is a $p_{x \pm y}$-density wave, in the nomenclature of Ref.~\onlinecite{nayak2}. Despite the $d$-wave-like factor
on the rhs of Eq.~(\ref{CN2}), this order is not the popular $d$-density wave \cite{sudip}; the 
latter is odd
under time-reversal, and in the present notation takes the form
\begin{equation}
\left \langle c_{\vec{k} - \vec{Q}/2, \sigma}^\dagger c_{\vec{k} + \vec{Q} /2, \sigma} \right\rangle 
\sim i  \left( \sin k_x - \sin k_y \right), \label{dden}
\end{equation}
with $\vec{Q} = ( \pi, \pi)$. The order in Eq.~(\ref{dden}) is not enhanced near the SDW critical point, while
that in Eq.~(\ref{CN2}) is.
By taking the Fourier transform of Eq.~(\ref{CN2}),
it is easy to see that $O^{\ell}_-$ does not lead to any modulations in the site charge density
$\left \langle c^{\dagger}_{\vec{r} \sigma}  c_{\vec{r} \sigma} \right\rangle$, and so it is not 
a CDW. 
The non-zero modulations occur in the off-site correlations $\left \langle c^{\dagger}_{\vec{r} \sigma}  c_{\vec{s} \sigma} \right\rangle$ with $\vec{r} \neq \vec{s}$. For $\vec{r}$ and $\vec{s}$ nearest-neighbors, we have
\begin{equation}
\left \langle c^{\dagger}_{\vec{r} \sigma}  c_{\vec{s} \sigma} \right\rangle = 
\sum_{\ell =1,2}\left( O^{\ell}_{-} e^{ i \vec{Q}_\ell \cdot (\vec{r} + \vec{s})/2} + \mbox{c.c.} \right)
\left[ \delta_{\vec{r} - \vec{s}, \hat{x}} + \delta_{\vec{s} - \vec{r}, \hat{x}} - \delta_{\vec{r} - \vec{s}, \hat{y}} - \delta_{\vec{s} - \vec{r}, \hat{y}} \right], \label{CN3}
\end{equation}
where $\hat{x}$ and $\hat{y}$ are unit vectors corresponding to the sides of the square lattice unit cell.
The modulations in the nearest neighbor bond variables $\left \langle c^{\dagger}_{\vec{r} \sigma}  c_{\vec{r}+\hat{x}, \sigma} \right\rangle$ and $\left \langle c^{\dagger}_{\vec{r} \sigma}  c_{\vec{r}+\hat{y}, \sigma} \right\rangle$ are plotted in Figs.~\ref{fig:density} and~\ref{fig:density2}. 
These observables measure spin-singlet correlations
across a link: if there are 2 electrons on the 2 sites of a link, this observable takes different values
depending upon whether the electrons are in a spin singlet or a spin triplet state. Thus $O^{\ell}_-$ has
the character of a valence bond solid (VBS) order parameter. The first factor on the rhs of Eq.~(\ref{CN3})
shows that the VBS order has modulations at the wavevectors $\vec{Q}_\ell$ along the square
lattice diagonals. 
However, from our discussion above, note that $|\vec{Q}_\ell| = 2 \sqrt{2} K_y$, where
the magnitude of $K_y$ is quite small for the Fermi surface in Fig.~\ref{fig:hotspots}: the $\ell=1$, $i=1$ hot spot
is at $(K_x, K_y)$. Thus the first factor
in Eq.~(\ref{CN3}) contributes a relatively long-wavelength modulation, as is evident from Figs.~\ref{fig:density} and~\ref{fig:density2}. 
This long-wavelength modulation serves as an envelope to the oscillations given by
the second factor in Eq.~(\ref{CN3}). The latter indicates
indicates that the bond order has opposite signs on the $x$ and $y$ directed bonds: this short distance behavior
corresponds locally
to an {\em Ising-nematic \/} order, which is also evident in Figs.~\ref{fig:density} and~\ref{fig:density2}.
The ordering in Eq.~(\ref{CN3}) becomes global Ising-nematic order in the limit $\vec{Q}_\ell \rightarrow 0$. Non-linear terms
in the effective action for the bond order will lock in commensurate values of $\vec{Q}_\ell$, and so
it is possible that strong-coupling effects will prefer $\vec{Q}_\ell =0$.

As already remarked, the particle-hole symmetry of our theory guarantees a degeneracy between the $d$-wave superconducting vertex and the density-wave vertex. However, this degeneracy is lifted once effects which break the particle-hole symmetry are introduced. One such effect is the curvature of the Fermi surface at the hot spots. Nominally, the curvature is irrelevant under the scaling towards hot spots (\ref{scalferm}). However, we recall that the double-log structure in Eq.~(\ref{Vfin}) originates from an interplay between scaling in a Fermi-liquid and quantum critical scaling. Moreover, we know that the scaling of the superconducting vertex and the density-wave vertex in a Fermi liquid are very different: at one loop the corrections to  former are logarithmic, while corrections to latter are suppressed by $\omega^{1/2}$. Thus, one might expect that the Fermi surface curvature will play an important role in the renormalization of the density-wave vertex, reducing its enhancement compared to the BCS vertex and establishing superconductivity as the dominant instability of the SDW critical point. We check this by an explicit calculation below.

We introduce the Fermi-surface curvature into the theory via a perturbation,
\beq L_c = \frac{1}{2m} \sum_{\ell,i} |(\nabla \cdot \hat{n}^{\ell}_{\parallel,i}) \psi^{\ell}_i|^2\label{curv}\eeq
where $\hat{n}^{\ell}_{\parallel, i} = \hat{z} \times \hat{v}^{\ell}_i$ is the unit tangent to the Fermi surface of $\psi^{\ell}_i$. 

Let us define the insertion of the density-wave order parameter $O^{\ell}_\mu$ into the fermion correlation function,
\beq \Gamma_{O\psi\psi^{\dagger}}(k_1,k_{-1}) \delta_{\sigma \sigma'}= \int d^D x_{1} d^D x_{-1} \langle O^{\ell}_\mu(0) \psi^{-\ell}_{1 \sigma}(x_{-1}) \psi^{\dagger\ell}_{1 \sigma'}(x_1) \rangle_{1PI} e^{i (k_1 x_1 - k_{-1} x_{-1})} \eeq  
At tree level $\Gamma_{O\psi\psi^{\dagger}}(k_1,k_{-1}) = 1$. The one loop correction to the vertex is given by the diagram in Fig.~\ref{FigV}b). We perform the calculations with propagators dressed by the one-loop fermion self-energy and by the curvature (\ref{curv}). Details are presented in Appendix \ref{app:2kF}. To leading logarithmic accuracy we obtain,
\beq \delta \Gamma_{O \psi \psi^{\dagger}} = -\frac{\mu \alpha}{3\pi (\alpha^2 + 1)} \log^2 \frac{\Lambda^2}{\gamma \omega}\eeq
which is a factor of $3$ smaller than the corresponding expression for the superconducting vertex (\ref{Vfin}). 

Finally, we note the resemblance between our results and those obtained by Halboth and Metzner,\cite{halboth} and Honerkamp et.~al,\cite{honerkamp}
using a functional renormalization group treatment of the Hubbard model. They find dominant instabilities
to SDW order and $d$-wave pairing, along with a sub-dominant enhancement of Ising-nematic order. 
They assumed their Ising-nematic order was at $\vec{Q}_\ell =0$, but their results could be limited by
the finite resolution of Fermi surface points, and their specific Fermi surface configurations. It would be interesting
if higher resolution studies of more generic Fermi surfaces lead to ordering compatible with Eq.~(\ref{CN2}).

\section{Conclusions}
\label{sec:conc}

Quantum phase transitions involving symmetry breaking in the presence of a Fermi surface
can be associated with the appearance of a condensate of particle-hole pairs of the Fermi surface quasiparticles. 
Such transitions can be divided
into two broad classes: those in which the particle-hole condensate carries net momentum $\vec{Q} \neq 0$,
and those in which the particle-hole condensate is at $\vec{Q} = 0$. Both classes were considered by Hertz in his 1976 paper \cite{hertz},
using a self-consistent RPA approach, formulated in terms of a RG analysis of an effective action for
the condensate fluctuations. He argued that for both cases, and for all spatial dimensions $d \geq 2$, the condensate fluctuations were effectively Gaussian, and hence the leading critical behavior could be exactly calculated.

We have re-examined both classes of Fermi surface transitions in this and a previous paper \cite{max1}. While Hertz's conclusions
are expected to be largely correct in $d=3$, they break down \cite{ChubukovShort} in both classes 
for the physically important case of $d=2$.
Our previous paper \cite{max1} proposed and analyzed a critical theory in $d=2$ 
for a paradigm of the $\vec{Q}=0$
case: the onset of Ising-nematic order. This theory involved both the bosonic order parameter and the fermionic quasiparticles
as fundamental degrees of freedom, which interact strongly at the quantum critical point. 
The present paper has considered a typical case in $d=2$ with $\vec{Q} \neq 0$,
the onset of spin density wave (SDW) order, using a field theory for the bosonic order parameter and the fermions proposed by
Abanov and Chubukov \cite{ChubukovShort0}. 

Our analysis for $\vec{Q} \neq 0$ begins by focusing on the vicinity of the ``hot spots'' on the 
Fermi surface shown in Fig.~\ref{fig:hotspots}. Zooming in on a single pair of hot spots, and shifting one
of the hot spots by a momentum $\vec{Q}$, we obtain the situation shown in Fig.~\ref{fig:fermions}, where
we can approximate the two Fermi surfaces near the hot spots by two non-collinear
straight lines. The two Fermi surfaces are coupled at the hot spot by the SDW order parameter $\phi$, and the low
energy physics is then described by the field theory in Eq.~(\ref{L}). In the phase with SDW order with $\langle \phi \rangle \neq 0$,
the Fermi surfaces reconnect into the configuration shown in Fig.~\ref{fig:sdwfs2}, leading to electron and hole
pockets appearing from the original large Fermi surface in Fig.~\ref{fig:hotspots}.

Our RG analysis of Eq.~(\ref{L}) was performed using the $1/N$ expansion, where the fermions are endowed
with an additional flavor index which runs over $N$ values. Initially, it seems that the counting of powers
of $1/N$ is simple: each boson propagator comes with a factor of $1/N$, and each fermion loop yields
a factor $N$. Using this ``naive'' counting, all RG flow equations were computed to order $1/N$ in 
Section~\ref{naive}. We found a consistent renormalization of the couplings in the local field theory 
in Eq.~(\ref{L}); the damping parameter $\gamma$ appearing in the boson propagator was tied to the local
couplings via Eq.~(\ref{Pi}), and this relation was maintained under the RG. The flow of the spin-damping rate
under RG implies that the dynamical critical exponent $z$ renormalizes away from its RPA value $z = 2$. This is in stark
contrast to Hertz theory\cite{hertz} and previous studies of the present theory.\cite{ChubukovLong} One of the main consequences
of the RG flow in Section~\ref{naive} was a logarithmic divergence in the ratio of Fermi velocity components
with length scale: this implied that the Fermi surfaces at the quantum critical point took the shape
in Fig~\ref{fig:sdwfs3}. The effective dynamical nesting of the Fermi surfaces at low energies gives rise to a divergence
of anomalous dimensions, which may lead to a first order phase transition. 

Section~\ref{sec:genus} looked at higher loop effects which showed that the naive counting of
powers of $1/N$ was not correct. The enhancements in powers of $N$ arose from infrared singularties
appearing when internal fermion lines were restricted to momenta on the Fermi surface, similar
to the Fermi surface enhancements discovered by S.-S.~Lee for the problem of a Fermi surface
coupled to a U(1) gauge field. These enhancements distinguish the present problem from that considered
in Refs.~\onlinecite{eunah,yejin}: the Ising-nematic transition in a $d$-wave superconductor. Formally,
the latter problem is described by a field theory similar to that of the present paper: fermions with
linear dispersion coupled via a Yukawa interaction to a scalar field $\phi$. Also, in both problems we find
a logarithmic divergence of velocity ratios in the infrared at order $1/N$ for the RG flows. However, for
the $d$-wave superconductor, with Dirac fermions whose energy vanishes only at isolated
``hot spots'', the $1/N$ expansion was found to be stable at higher loops. In contrast, for the present
SDW problem, the fermion hot spots are connected to ``cold'' Fermi lines, and singularities associated
with these lines lead to a breakdown in the naive $1/N$ counting. Because of this breakdown, the nature
of the $N \rightarrow \infty$ limit of Eq.~(\ref{L}) remains unclear.

Next, we examined the instability of the SDW metal to the onset of superconductivity
near the quantum critical point in Section~\ref{sec:pairing}. We found a strong tendency towards
spin-singlet pairing, with pairing amplitude having opposite signs across a pair of hot spots. For the cuprate
Fermi surface in Fig.~\ref{fig:hotspots} this includes $d_{x^2-y^2}$ pairing, while for the pnictide Fermi surfaces
this includes $s_{+-}$ pairing. This pairing instability was
manifested in a log-squared divergence of the renormalization of the pairing vertex, arising from
an interplay of the infrared singularities associated with the Fermi surfaces and the hot spot. 
This log-squared singularity cannot be resolved by the present RG approach, and other methods are needed
to determine its consequences. An important problem for future research is to understand the feedback of the pairing
fluctuations on the non-Fermi liquid singularities at the metallic hot spot. 
Clearly, superconductivity appears near the quantum critical point as $T \rightarrow 0$.
The interesting question is the behavior above $T_c$, involving the interplay between
the metallic quantum criticality and the pairing fluctuations.

In our discussion of the critical theory for the SDW transition in Section~\ref{sec:model}, we noted
that the field theory had emergent pseudospin SU(2) symmetries (Eq.~(\ref{SU2})) containing the particle-hole
transformation; note that the pseudospin rotations can be carried out independently on different pairs of hot spots. 
Given the strong instability towards $d$-wave pairing near the SDW critical point
described in Section~\ref{sec:pairing}, it is natural to examine the action of the SU(2) pseudospin
symmetries on the $d$-wave pairing order parameter. This was described in Section~\ref{sec:cdw},
where we found a similar log-squared enhancement of the susceptibility to a modulated valence bond solid 
(VBS) order
parameter illustrated in Figs.~\ref{fig:density} and~\ref{fig:density2}.  
Notice that at short scales this ordering has an
Ising-nematic character: this corresponds to the breaking of a 90 degree rotation symmetry of the square
lattice by the values of the bond order parameter in Eq.~(\ref{CN3}). 
It would be interesting if future work supports a connection between 
the ordering instability of Section~\ref{sec:cdw}, and the bond and Ising-nematic
ordering observed in experiments \cite{ando02,kohsaka07,borzi07,hinkov08a,taill10b,lawler10}.
While the present analysis has focused exclusively on the vicinity of the hot spots, it is quite possible
that strong coupling physics away from the hot spot could lock in a preference for commensurate
values, such as $\vec{Q}_\ell = 0$, in Eq.~(\ref{CN3}), leading to global Ising-nematic order.
Also, it would be interesting to study the changes in the VBS ordering for the case of a SDW transition
at an incommensurate ordering wavevector, like that found in the hole-doped cuprates.

Finally, we note an interesting possibility for future theoretical work.
Given the breakdown of the $1/N$ expansion for the theory in Eq.~(\ref{L}) for the SDW critical point
in a two-dimensional metal, other systematic methods of analyzing this field theory are clearly needed.
Following Ref.~\onlinecite{nayak}, one possibility is to modify the $(\nabla \vec{\phi})^2$ term in Eq.~(\ref{L})
to $k^{1+x} \vec{\phi}^2$, where $k$ is the momentum carried by $\phi$. Then at the RPA level, we obtain
a theory with $z=1+x$, and an expansion in small $x$ appears possible. 

\acknowledgements

We thank A.~Chubukov, C.~Honerkamp, G.~Kotliar, S.-S.~Lee, W.~Metzner, and L.~Taillefer for useful discussions.
This research was supported by the National Science Foundation under grant DMR-0757145, by the FQXi
foundation, and by a MURI grant from AFOSR.

\appendix
\section{RG computations}
\label{app:rg}
In this appendix we give the details of our calculations in Sections~\ref{sec:model} and~\ref{naive}.

\subsection{RPA polarization}
\label{app:rpa}
 
We begin with the RPA polarization bubble,
\beq \Pi^{ab}(q) = 2 N \delta^{ab} \sum_\ell \int \frac{dl_\tau d^2 \vec{l}}{(2 \pi)^3} (G^{\ell}_{1}(l+q) G^{\ell}_{2} (l) + 
G^{\ell}_{2}(l+q) G^{\ell}_{1} (l))\eeq
The two terms in brackets come from the two graphs in Fig. \ref{FigPi} with different directions of the particle flow. As discussed in Section \ref{sec:model} such graphs are equal by the emergent particle-hole symmetry.
Thus, focusing on the contribution from $\ell = 1$,
\beq \Pi^{\ell = 1}(q) = 2 N \int \frac{dl_\tau d^2 \vec{l}}{(2 \pi)^3} \frac{1}{(i \eta (l_\tau + q_\tau) - \vec{v}_1 \cdot (\vec{l} + \vec{q})) (i \eta l_\tau - \vec{v}_2 \cdot \vec{l})} + (q \to -q)\eeq
We change variables to $l_1 = \hat{v}_1 \cdot (\vec{l} + \vec{q})$, $l_2 = \hat{v}_2 \cdot \vec{l}$, and take the limit $\eta \to 0$ using the relation,
\beq \frac{1}{x + i \eta} = \frac{P}{x} - \pi i \mbox{sgn}(\eta) \delta(x)\label{etazero}\eeq
which yields,
\beq \Pi^{\ell = 1}(q) = \frac{N}{v_x v_y} \int \frac{dl_\tau d^2 \vec{l}}{(2 \pi)^3} \left(\frac{P}{l_1} + \pi i \mbox{sgn}(l_\tau + q_\tau) \delta (l_1)\right)\left(\frac{P}{l_2} + \pi i \mbox{sgn} (l_\tau) \delta (l_2)\right) + (q\to -q)\eeq
Evaluating the integrals over $l_1$, $l_2$,
\beq \Pi^{\ell = 1}(q) = -  \frac{N}{8 \pi v_x v_y} \int d l_\tau\, \mbox{sgn}(l_\tau + q_\tau) \mbox{sgn}(l_\tau) + (q \to -q) \label{Pil0}\eeq
Here, we've taken the principal value integral to be zero, as it would be if we used a particle-hole symmetric regularization. Otherwise, one can check that any terms generated by the pv integral are of the form $i q_\tau$ and are cancelled by the $(q \to -q)$ term of Eq.~\ref{Pil0}.  Now, subtracting the value of the polarization bubble at $q = 0$, we obtain,
\beq \Pi^{\ell = 1}(q) - \Pi^{\ell = 1}(q=0)=  - \frac{N}{8 \pi v_x v_y} \int d l_\tau\, (\mbox{sgn}(l_\tau + q_\tau) \mbox{sgn}(l_\tau) -1 )+ (q \to -q) = \frac{N}{2 \pi v_x v_y} |q_\tau|\eeq
which, taking into account contributions from the other hot spots, gives,
\beq \Pi(q) = \Pi(q = 0) + \frac{N n}{2 \pi v_x v_y} |q_\tau|\eeq

\subsection{Fermion self energy}
\label{app:fself}

We next proceed to the self-energy of fermion $\psi^{\ell = 1}_1$, Fig. \ref{FigSigma},
\bea \Sigma_{1, \sigma \sigma'}(p) &=&  \tau^a_{\sigma \rho} \tau^a_{\rho \sigma'} \int \frac{dl_\tau d^2 \vec{l}}{(2\pi)^3} G_{2}(p-l) D(l) \nn \\ &=& 
\frac{3}{N} \delta_{\sigma \sigma'} \int \frac{dl_\tau d^2 \vec{l}}{(2\pi)^3} \frac{1}{ i \eta (p_\tau -l_\tau) -\vec{v}_2 \cdot (\vec{p} - \vec{l}) }\frac{1}{\gamma |l_\tau| + \vec{l}^2} \eea
We take the limit $\eta \to 0$ and use Eq.~(\ref{etazero}). Moreover, we change variables, so that $l_\perp = \hat{v}_2 \cdot \vec{l}$ and $l_{\parallel}$ is the momentum component along the Fermi surface of $\psi_2$ (i.e. perpendicular to $\hat{v}_2$). Then,
\beq \Sigma_1(p) = \frac{3}{N |\vec{v}|} \int \frac{dl_\tau dl_\perp d l_{\parallel}}{(2\pi)^3} \left(\frac{P}{l_\perp - \hat{v}_2 \cdot \vec{p}} + \pi i \mbox{sgn} (l_\tau - p_\tau) \delta(l_\perp - \hat{v}_2 \cdot \vec{p})\right) \frac{1}{\gamma|l_\tau| + l_\perp^2 + l_{\parallel}^2}\eeq
Thus, the imaginary part of $\Sigma$ is given by,
\beq \mbox{Im} \Sigma_1(p) =  \frac{3}{N |\vec{v}|} \int \frac{d l_\tau}{8 \pi} \mbox{sgn}(l_\tau - p_\tau) \frac{1}{\sqrt{\gamma |l_\tau| + |\hat{v_2} \cdot \vec{p}|^2}}\eeq
where we have performed the integral over $l_\perp$, $l_{\parallel}$. Since, $\mbox{Im} \Sigma(p_\tau = 0) = 0$, 
\bea \mbox{Im} \Sigma_1(p) &=& \frac{3}{N |\vec{v}|} \int \frac{d l_\tau}{8 \pi} (\mbox{sgn}(l_\tau - p_\tau) - \mbox{sgn}(l_\tau)) \frac{1}{\sqrt{\gamma |l_\tau| + |\hat{v_2} \cdot \vec{p}|^2}} \nn \\
&=& -\frac{3}{2 \pi N |\vec{v}| \gamma} \mbox{sgn}(p_\tau) \left(\sqrt{\gamma |p_\tau| + (\hat{v}_2 \cdot \vec{p})^2}-|\hat{v}_2 \cdot \vec{p}|\right)\label{ImSigma}\eea
On the other hand, the real part of $\Sigma$ is given by,
\beq \mbox{Re} \Sigma_1(p) = -\frac{3 \hat{v}_2 \cdot \vec{p}}{2 N |\vec{v}|} \int \frac{d l_\tau d l_{\parallel}}{(2 \pi)^2} \frac{1}{\sqrt{\gamma |l_\tau| + {l}_{\parallel}^2}} \frac{1}{\gamma |l_\tau| + {l}_{\parallel}^2 + (\hat{v}_2 \cdot \vec{p})^2}\eeq
Changing variables to $u = \sqrt{\gamma l_\tau + l^2_\parallel}$,
\bea \mbox{Re} \Sigma_1(p) &=& -\frac{3 (\hat{v}_2 \cdot \vec{p})}{2 \pi^2 N \gamma |\vec{v}|} \int dl_{\parallel} \int_{|l_{\parallel}|}^{\infty} du \frac{1}{u^2 + (\hat{v}_2 \cdot \vec{p})^2} \nn \\
&=& -\frac{3 (\hat{v}_2 \cdot \vec{p})}{2 \pi^2 N \gamma |\vec{v}|} \int \frac{ dl_{\parallel}}{|\hat{v}_2 \cdot \vec{p}|} \tan^{-1} \left(\frac{|\hat{v}_2 \cdot \vec{p}|}{|l_{\parallel}|}\right)\eea
The integral over $l_{\parallel}$ is ultra-violet divergent. Cutting off the integral at $|l_{\parallel}| = \Lambda$, we obtain to logarithmic accuracy,
\beq \mbox{Re} \Sigma_1(p) = -\frac{3 \hat{v}_2 \cdot \vec{p}}{\pi^2 N |\vec{v}| \gamma} \log \frac{\Lambda}{|\hat{v}_2 \cdot \vec{p}|}\label{ReSigma}\eeq
Combining eqs.~(\ref{ImSigma}), (\ref{ReSigma}) we obtain the self-energy (\ref{Sigma}). 


\subsection{Boson-fermion vertex}
\label{app:vertex}
Proceeding to the first correction in $1/N$ to the boson-fermion vertex, Fig. \ref{FigVertex},
\beq \delta \Gamma^{a}_{\sigma \sigma'}(p,q)  = (\tau^b \tau^a \tau^b)_{\sigma \sigma'} \int \frac{dl_\tau d^2 \vec{l}}{(2 \pi)^3} G_2(l+p) G_1(l + p + q)  D(l)\eeq
Evaluating the matrix product,
\beq \delta \Gamma(p,q)= - \frac{1}{N} \int \frac{dl_\tau d^2 \vec{l}}{(2 \pi)^3} \frac{1}{\vec{v}_2 \cdot (\vec{l} + \vec{p}) - i \eta (l_\tau + p_\tau)}\frac{1}{\vec{v}_1 \cdot (\vec{l} + \vec{p} + \vec{q}) - i \eta (l_\tau + p_\tau + q_\tau)}\frac{1}{\gamma |l_\tau| + \vec{l}^2}\label{dGammadet}\eeq
The integral (\ref{dGammadet}) is logarithmically divergent in the UV. To extract this divergence, we may set all external momenta to zero:
\beq \delta \Gamma(p,q) \stackrel{UV}{=} - \frac{1}{N} \int \frac{dl_\tau d^2 \vec{l}}{(2 \pi)^3} \frac{1}{(-v_x l_x + v_y l_y -i \eta l_\tau) (v_x l_x + v_y l_y -i \eta l_\tau)} \frac{1}{\gamma |l_\tau| + l^2_x + l^2_y}\label{dGamma2}\eeq
The poles in $l_y$ coming from the two fermion propagators in Eq.~(\ref{dGamma2}) are in the same half-plane; we may choose to close the $l_y$ integration contour in the opposite half-plane, picking up the pole from the bosonic propagator:
\bea  &&\delta \Gamma(p,q) \stackrel{UV}{=} -\frac{1}{N} \int \frac{dl_\tau dl_x}{(2 \pi)^2} \frac{1}{(-v_x l_x - i v_y \mbox{sgn} (l_\tau) \sqrt{\gamma|l_\tau| + l^2_x})
(v_x l_x - i v_y \mbox{sgn}(l_\tau) \sqrt{\gamma |l_\tau| + l^2_x})} \nn \\
&&~~~~~~~~~~~~~~~~~~~~~~~~~~~~~~~~~~~~~~\times \frac{1}{2 \sqrt{\gamma |l_\tau| + l^2_x}}\eea
Changing variables to $u = \sqrt{\gamma |l_\tau| + l^2_x}$,
\beq \delta \Gamma(p,q) \stackrel{UV}{=} \frac{2}{N \gamma} \int_{-\infty}^{\infty} \frac{dl_x}{2 \pi} \int_{|l_x|}^{\infty} \frac{du}{2\pi} \frac{1}{v^2_x l^2_x + v^2_y u^2}\eeq
We now go to polar coordinates, $v_x l_x + i v_y u = |\vec{v}| \rho e^{i \theta}$,
\beq \delta \Gamma(p,q) \stackrel{UV}{=} \frac{1}{N \pi (2 \pi v_x v_y \gamma)} \int_0^{\infty} \frac{d \rho}{\rho} \int_{\tan^{-1} \alpha}^{\pi - \tan^{-1} \alpha} d \theta \eeq
The integral over $\rho$ is logarithmically divergent in the $UV$; cutting off the integral at $\rho \sim \Lambda$,
\beq \delta \Gamma(p,q) \stackrel{UV}{=} \frac{2}{\pi n N} \tan^{-1}{\frac{1}{\alpha} \log \Lambda}\eeq

\subsection{Boson self energy}
\label{app:bself}

We now proceed to the $1/N$ corrections to the boson self-energy, Fig. \ref{FigPiCorr}. We first analyze the contribution of diagrams a),b) and c), which we label $\delta\Pi_I$. Utilizing the expression (\ref{f}) for the fermion induced quartic coupling, we obtain,
\bea \delta\Pi^{ab}_I(q) &=& \frac{1}{2} \int \frac{d l_\tau d^2 \vec{l}}{(2 \pi)^3} \Gamma^{ab c c}(q,-q,l,-l) D(l) \nn\\
 &=& \int \frac{d l_\tau d^2\vec{l}}{(2 \pi)^3} (f^{ab c c}(q,-q,l,-l)+ f^{a cc b}(q,l,-l, -q) + f^{a c b c}(q,l,-q,-l)) D(l) \nn\\\label{Pif}\eea
The first two terms in Eq.~(\ref{Pif}) vanish (these terms correspond to the diagrams in Fig. \ref{FigPiCorr} a),b) ). Thus, only the diagram in Fig. \ref{FigPiCorr} c) contributes,
\beq \delta\Pi_I(q_\tau, \vec{q}) = |q_\tau| A(q_\tau, \vec{q}) + B(q_\tau, \vec{q})  \label{PicorrAB} \eeq
with
\bea
A(q_\tau, \vec{q}) &=& - \frac{N}{\pi v_x v_y} \sum_{\ell} \int \frac{d l_\tau d^2 \vec{l}}{(2 \pi)^3} G^{\ell}_1(l-q) G^{\ell}_2(l+q)D(l)\\
 B(q_\tau, \vec{q}) &=& \frac{N}{\pi v_x v_y}  \sum_{\ell} \int \frac{d l_\tau d^2 \vec{l}}{(2 \pi)^3} |l_\tau| G^{\ell}_1(l-q) G^{\ell}_2(l+q)D(l)
 \eea
The quantity $A(q_\tau, \vec{q})$ is logarithmically divergent in the $UV$. The coefficient of the divergence may be extracted by setting the external momenta and $r$ to zero. Then, from Eq.~(\ref{dGammadet}), we recognize,
\beq A(q_\tau, \vec{q}) \stackrel{UV}{=} \frac{N}{\pi v_x v_y} \sum_\ell \delta \Gamma(p,q) = \frac{4 \gamma}{n \pi} \tan^{-1}\frac{1}{\alpha} \log \Lambda \label{Afinal}\eeq
Now, let us  evaluate $B$. We temporarily keep only the contribution from the hot spot pair $\ell = 1$.
\bea
B^{\ell = 1}(q_\tau, \vec{q}) &=& \frac{1}{\pi v_x v_y}   \int \frac{d l_\tau d^2 \vec{l}}{(2 \pi)^3} \frac{1}{(v_x l_x + v_y l_y - 
\vec{v}_1 \cdot \vec{q} - i \eta (l_\tau - q_\tau))} \nn \\ 
&~&~~~~\times \frac{1}{(-v_x l_x + v_y l_y +\vec{v}_2\cdot \vec{q} - i \eta (l_\tau + q_\tau))}
\frac{|l_\tau|}{(\gamma |l_\tau| + l^2_x + l^2_y+r)}. \label{B}\eea
Note that the region $|l_\tau| < |q_\tau|$ does not contain any UV divergences. Thus, to compute the UV divergent part, we can confine our attention
to  the region $|l_\tau| > |q_\tau|$. In this case, the two poles in $l_y$ coming from the fermion propagators in Eq.~(\ref{B}) lie in the same half-plane; we may choose to close the $l_y$ integration contour in the opposite half-plane, picking up the pole from the bosonic propagator:
\bea B^{\ell = 1}(q_\tau, \vec{q}) &\stackrel{UV}{=}& \frac{1}{\pi v_x v_y} \int_{|l_\tau| > |q_\tau|} \frac{dl_\tau dl_x}{(2 \pi)^2} \frac{1}{v_x l_x - i v_y \mbox{sgn}(l_\tau) \sqrt{\gamma|l_\tau| + l^2_x+r} - \vec{v}_1 \cdot \vec{q}} \nn\\
 &\times& \frac{1}{-v_x l_x - i v_y \mbox{sgn}(l_\tau) \sqrt{\gamma|l_\tau| + l^2_x +r} + \vec{v}_2 \cdot \vec{q}}\, \frac{|l_\tau|}{2 \sqrt{\gamma |l_\tau| + l^2_x+r}}\label{A2}\eea
Note that we may extend the integration over $l_\tau$ in Eq.~(\ref{A2}) back to the whole real line without influencing the $UV$ part of the result. Thus,
\bea B^{\ell = 1}(q_\tau, \vec{q}) &\stackrel{UV}{=}& - \frac{1}{\pi v_x v_y} \int_{0}^{\infty} \frac{dl_\tau}{2 \pi} \int_{-\infty}^{\infty} \frac{dl_x}{2 \pi} \frac{1}{(v_x l_x - i v_y \sqrt{\gamma l_\tau + l^2_x+r} - \vec{v}_1 \cdot \vec{q})} \nn \\
&~&~~~~\times \frac{1}{
(v_x l_x + i v_y \sqrt{\gamma l_0 + l^2_x + r} - \vec{v}_2 \cdot \vec{q})} \frac{l_\tau}{2 \sqrt{\gamma l_\tau + l^2_x+r}} + c.c.\eea
It is convenient to change variables to $u = \sqrt{\gamma |l_\tau| + l^2_x+r}$,
\beq B^{\ell = 1}(q_\tau, \vec{q}) \stackrel{UV}{=} - \frac{1}{\pi v_x v_y \gamma^2}  \int_{-\infty}^{\infty} \frac{dl_x}{2 \pi} \int_{\sqrt{l^2_x+r}}^{\infty} \frac{du}{2 \pi} \frac{u^2 - l^2_x-r}{(v_x l_x - i v_y u - \vec{v}_1 \cdot \vec{q})
(v_x l_x + i v_y u - \vec{v}_2 \cdot \vec{q})} + c.c.\eeq
The $r$ in the lower limit of the integral over $u$ may be dropped without influencing the $UV$ behaviour. We now go to polar coordinates, $v_x l_x + i v_y u = |\vec{v}| \rho e^{i \theta}$,
\bea B^{\ell = 1}(q_\tau, \vec{q}) \stackrel{UV}{=} - \frac{1}{\pi (2 \pi v_x v_y \gamma)^2} \frac{|\vec{v}|^2}{v_x v_y} \int \rho d\rho \int_{\tan^{-1} \alpha}^{\pi - \tan^{-1} \alpha} d\theta\, \frac{\rho^2(\frac{1}{\alpha} \sin^2 \theta - \alpha \cos^2 \theta) - \frac{v_x v_y}{|\vec{v}|^2} r}{(\rho e^{i \theta} - \hat{v}_2 \cdot \vec{q})
(\rho e^{-i \theta} - \hat{v}_1 \cdot \vec{q})} + c.c.\nn\\\eea
The integral over $\rho$ is quadratically divergent. Expanding the divergent part in $\vec{q}$ and $r$,
\bea B^{\ell = 1}(q_\tau, \vec{q}) &\stackrel{UV}{=}& - \frac{2}{\pi n^2} \frac{|\vec{v}|^2}{v_x v_y} \int \rho d \rho \int_{\tan^{-1} \alpha}^{\pi - \tan^{-1} \alpha} d \theta \, \Bigg[\left(\frac{1}{\alpha} \sin^2 \theta - \alpha \cos^2 \theta\right) \bigg(1 + \frac{1}{\rho} (\hat{v}_1 + \hat{v_2}) \cdot \vec{q} \cos \theta\nn\\ &+& \frac{1}{\rho^2} \big((\hat{v}_1 \cdot \vec{q})(\hat{v}_2 \cdot \vec{q}) + ((\hat{v}_1 \cdot \vec{q})^2 + (\hat{v}_2 \cdot \vec{q})^2) \cos 2 \theta\big)\bigg)- \frac{v_x v_y}{|\vec{v}|^2}\frac{r}{\rho^2}\Bigg]\eea
As usual, the term constant in $\vec{q}$ corresponds to a shift in the position of the critical point and will be dropped below. The term linear in $\vec{q}$ vanishes under $\theta \to \pi - \theta$, {\em i.e.\/} $l_x \to - l_x$ (more rigorously, this term must vanish by symmetry, once the contributions from all 4 pairs of hot spots are summed). Finally, the term quadratic in $\vec{q}$ and the term linear in $r$ give logarithmic divergences. Cutting off the integral over $\rho$ at $\rho \sim \Lambda$, 
\beq B^{\ell = 1}(q_\tau, \vec{q}) \stackrel{UV}{=} \frac{4}{\pi n^2} \log \Lambda \left[ \frac{q^2_x}{\alpha^2} \left(\tan^{-1} \frac{1}{\alpha} + \frac{\alpha}{1+\alpha^2}\right) + \alpha^2 q^2_y \left(\tan^{-1} \frac{1}{\alpha} - \frac{\alpha}{1+\alpha^2}\right) + r \tan^{-1} \frac{1}{\alpha}\right]\eeq
Now, summing over the four pairs of hot spots, we restore rotational invariance,
\bea B(q_\tau, \vec{q}) &=& \frac{2}{\pi n} \left[ \frac{1}{\alpha} - \alpha + \left(\frac{1}{\alpha^2} + \alpha^2\right) \tan^{-1} \frac{1}{\alpha}\right] \vec{q}^2 \log \Lambda + \frac{4}{\pi n} \tan^{-1}\frac{1}{\alpha}\,  r\log \Lambda \label{Bfinal} \eea

We now compute the diagram in Fig.~\ref{FigPiCorr} d), which we label $\delta \Pi_{II}$. This diagram is present already in the Hertz-Millis theory and, being momentum independent, leads only to a renormalization of $r$,
\bea \delta \Pi_{II}(q) &=& 5 u \int \frac{dl_\tau d^2 \vec{l}}{(2\pi)^3} D(l) \stackrel{UV}{=} - \frac{5}{N} u r \int \frac{dl_\tau d^2 \vec{l}}{(2\pi)^3} \frac{1}{(\gamma |l_\tau| + \vec{l}^2)^2} = - \frac{5 u r}{\pi N \gamma} \int \frac{d^2 \vec{l}}{(2\pi)^2}\frac{1}{\vec{l}^2}\nn\\
&=& - \frac{5}{2 \pi^2 N} \tilde{u} r \log \Lambda \label{deltaPi2}\eea

Now combining Eqs.~(\ref{PicorrAB}), (\ref{Afinal}), (\ref{Bfinal}), (\ref{deltaPi2}) we obtain the UV part of the correction to the boson propagator, Eq.~(\ref{Dcorr}).

\section{Violatations of large-$N$ counting}
\subsection{Boson-fermion vertex correction at three loops}
\label{app:vert3l}
In this section we compute the vertex correction in Fig.~\ref{fig:vertex2}. As shown in section \ref{sec:genus}, an attempt to evaluate this graph directly with bare fermion propagators results in infra-red divergences. To cure this problem, we dress the fermion propagators by the one-loop self-energy (\ref{Sigma}). For simplicity, we include only the imaginary part of the self-energy responsible for the dynamics. The frequency independent real part responsible for the logarithmic running of the velocity $v$ will be ignored here. Thus, we use,
\beq G^{\ell}_i(\omega, \vec{k}) = \frac{1}{-i \frac{c_f}{N} g(\omega, \hat{v}^{\ell}_{\bar{i}} \cdot \vec{k}) + \vec{v}^{\ell}_i \cdot \vec{k}} \label{G1loop}\eeq
where $\bar{1} = 2$, $\bar{2} = 1$ and
\beq g(\omega, k) = \mbox{sgn} (\omega) (\sqrt{\gamma |\omega| + k^2} - |k|),\quad c_f = \frac{3}{2 \pi |\vec{v}| \gamma} \label{g}\eeq
Then, the diagram in Fig. \ref{fig:vertex2} is given by,
\bea \delta \Gamma_{\phi\psi_2\psi^{\dagger}_1} &=& - 28 N \int \frac{d^3 k}{(2 \pi)^3} \frac{d^3 l_1}{(2 \pi)^3} \frac{d^3 l_2}{(2 \pi)^3}G^{-1}_1(k) G^{-1}_2(k-l_1) G^{-1}_1(k-l_2) G^{-1}_2(k) G^{1}_2(l_1) G^{1}_1(l_2)\nn \\ && ~~~~~~~~~\times D(l_1) D(l_2) D(l_2 - l_1) \eea
The external fermions are taken to have hot spot index $\ell = 1$, while the fermions in the loop are taken to have $\ell' = -1$. As discussed in section \ref{sec:genus}, the contributions from $\ell' = 2$ and $\ell' = 4$ are not enhanced in $N$, while $\ell' = 1$ contributes a $UV$ finite term of $O(1)$ when the external fermion momenta are chosen to lie on the Fermi surface. As we are mainly interested in corrections to mean-field scaling, we only retain $UV$ divergent contributions below. Hence, all the external momenta of the diagram have been set to 0. Substituting the one-loop corrected propagators (\ref{G1loop}), we obtain,
\bea \delta \Gamma_{\phi\psi_2\psi^{\dagger}_1} &=& - 28 N \int \frac{d^3 k}{(2 \pi)^3} \frac{d^3 l_1}{(2 \pi)^3} \frac{d^3 l_2}{(2 \pi)^3}  \frac{1}{-i \frac{c_f}{N} g(k_\tau, \hat{v}_1 \cdot \vec{k}) - \vec{v}_2 \cdot \vec{k}} \times \frac{1}{-i \frac{c_f}{N} g(k_\tau, \hat{v}_2 \cdot \vec{k}) - \vec{v}_1 \cdot \vec{k}}\nn\\
&& \frac{1}{-i \frac{c_f}{N} g(k_\tau - l_{1\tau}, \hat{v}_1 \cdot(\vec{k}-\vec{l}_1)) - \vec{v}_2 \cdot (\vec{k} - \vec{l}_1)} 
\times \frac{1}{-i \frac{c_f}{N} g(k_\tau - l_{2\tau}, \hat{v}_2 \cdot(\vec{k}-\vec{l}_2)) - \vec{v}_1 \cdot (\vec{k} - \vec{l}_2)}\nn\\
&&\frac{1}{-i \frac{c_f}{N} g(l_{1\tau}, \hat{v}_1 \cdot \vec{l}_1) + \vec{v}_2 \cdot \vec{l}_1}
\times \frac{1}{-i \frac{c_f}{N} g(l_{2\tau}, \hat{v}_2 \cdot \vec{l}_2) + \vec{v}_1 \cdot \vec{l}_2} D(l_1)D(l_2) D(l_1-l_2)\eea
We may divide the spatial momenta into two groups: $\hat{v}_1 \cdot \vec{k},\, \hat{v}_2 \cdot \vec{k},\, \hat{v}_2 \cdot \vec{l}_1,\, \hat{v}_1 \cdot \vec{l}_2$ and $\hat{v}_1 \cdot \vec{l}_1$, $\hat{v}_2 \cdot \vec{l}_2$. The singular manifold of the diagram is given by setting the momenta in the first group to zero and can be parameterized by the two variables in the second group. We begin by integrating over the first set of variables, picking up the contribution from the poles of the fermion propagators. As this integration is saturated at momenta of ${\cal O}(1/N)$, we can neglect the dependence of the boson propagators and fermion self-energies on these momenta. We then obtain the result in terms of an integral over the singular manifold.

Due to the symmetry, $G(l) = - G(-l)$, the contributions to the integral from $k_\tau > 0$ and $k_\tau < 0$ are equal. Now, changing momentum variables to $\hat{v}_1 \cdot \vec{p}$, $\hat{v}_2 \cdot \vec{p}$, and integrating over $\hat{v}_2 \cdot \vec{l}_1$, $\hat{v}_1 \cdot \vec{l}_2$,
\bea \delta \Gamma_{\phi\psi_2\psi^{\dagger}_1} &=& -7 N \frac{|\vec{v}|^4}{(v_x v_y)^3} \int_0^{\infty} \frac{d k_\tau}{2\pi} \int \frac{d (\hat{v}_1 \cdot \vec{k}) d (\hat{v}_2 \cdot \vec{k}) d (\hat{v}_1 \cdot \vec{l}_1) d(\hat{v}_2 \cdot \vec{l}_2)}{(2 \pi)^4}\nn
\\&&\left[\int_{k_\tau}^{\infty} - \int_{-\infty}^{0}\right] \frac{dl_{1\tau}}{2 \pi}  \left[\int_{k_\tau}^{\infty} - \int_{-\infty}^{0}\right] \frac{dl_{2\tau}}{2 \pi} D(l_1) D(l_2) D(l_1 - l_2)\bigg|_{\hat{v}_1 \cdot \vec{l}_2 = \hat{v}_2 \cdot \vec{l}_1 = 0} \nn\\&&\frac{1}{-i \frac{c_f}{N}(g(l_{1\tau}, \hat{v}_1 \cdot \vec{l}_1) - g(k_\tau - l_{1\tau},\hat{v}_1 \cdot (\vec{k} - \vec{l}_1))) + \vec{v}_2 \cdot \vec{k}} \times \frac{1}{i\frac{c_f}{N}g(k_\tau, \hat{v}_1 \cdot \vec{k}) + \vec{v}_2 \cdot \vec{k}}\nn
\\ &&\frac{1}{-i \frac{c_f}{N}(g(l_{2\tau}, \hat{v}_2 \cdot \vec{l}_2) - g(k_\tau - l_{2\tau},\hat{v}_2 \cdot (\vec{k} - \vec{l}_2))) + \vec{v}_1 \cdot \vec{k}} \times \frac{1}{i \frac{c_f}{N} g(k_\tau, \hat{v}_2 \cdot \vec{k}) + \vec{v}_1 \cdot \vec{k}} \nn\\
\eea
Now, performing the integral over $\hat{v}_1 \cdot \vec{k}$, $\hat{v}_2 \cdot \vec{k}$,
\bea \delta \Gamma_{\phi\psi_2\psi^{\dagger}_1} &=& -7 N^3 \frac{|\vec{v}|^2}{(v_x v_y)^3 c^2_f} \int_0^{\infty} \frac{d k_\tau}{2\pi} \int_{k \tau}^{\infty} \frac{dl_{1 \tau}}{2\pi}  \int_{k_\tau}^{\infty} \frac{dl_{2 \tau}}{2\pi} \int  \frac{d (\hat{v}_1 \cdot \vec{l}_1) d(\hat{v}_2 \cdot \vec{l}_2)}{(2 \pi)^2}\nn\\
&& \frac{1}{g(k_\tau, 0) + g(l_{1 \tau}) + g(l_{1 \tau} - k_\tau, \hat{v}_1 \cdot \vec{l}_1)} \times \frac{1}{g(k_\tau, 0) + g(l_{2 \tau}) + g(l_{2 \tau} - k_\tau, \hat{v}_2 \cdot \vec{l}_2)}\nn\\
&& \times D(l_1) D(l_2) D(l_1 - l_2)\bigg|_{\hat{v}_1 \cdot \vec{l}_2 = \hat{v}_2 \cdot \vec{l}_1 = 0}\nn \eea
Changing variables to $l_{1,2\tau} = k_\tau x_{1,2}$, $l_{1,2y} = \sqrt{\gamma k_\tau} y_{1,2}$,
\beq \delta \Gamma_{\phi\psi_2\psi^{\dagger}_1} = \frac{1}{2} X(\alpha) \int_0^{\infty} \frac{d k_\tau}{k_\tau} = X(\alpha) \log \Lambda_y \eeq
with
\bea X(\alpha) &=& -\frac{7}{18 \pi^2 n} \left(\frac{1}{\alpha} + \alpha\right)^2 \int_1^{\infty} dx_1 \int_1^{\infty} dx_2 \int_{-\infty}^{\infty} dy_1 \int_{-\infty}^{\infty} dy_2 \frac{1}{\sqrt{x_1 + y^2_1} + \sqrt{x_1 - 1 + y^2_1} - 2 |y|_1 +1}\nn\\
&&\times \frac{1}{\sqrt{x_2 + y^2_2} + \sqrt{x_2 - 1 + y^2_2} - 2 |y|_2 +1}\times \frac{1}{x_1 + \frac{1}{4} (\frac{1}{\alpha} + \alpha)^2 y^2_1} \times
\frac{1}{x_2 + \frac{1}{4} (\frac{1}{\alpha} + \alpha)^2 y^2_2}\nn \\
&&\times \frac{1}{|x_1-x_2| + \frac{1}{4} (\frac{1}{\alpha} + \alpha)^2 (y^2_1 + y^2_2) - \frac{1}{2} (\frac{1}{\alpha^2} - \alpha^2) y_1 y_2}\eea

\subsection{Quartic vertex}
\label{app:u}
In this section we evaluate the five loop correcton to the boson four-point function shown in Fig.~\ref{fig:Quartic5l}. We recall that by the particle-hole symmetry of our theory, diagrams with a reversed direction of the two fermion loops have the same value. We focus only on the diagrams where the fermions in the two loops come from opposite hot spots as these give a result, which is of ${\cal O}(N^3)$ and logarithmically divergent. To identify the coefficient of the logarithmic divergence we may set all the external momenta to zero. Then by rotational invariance each hot spot pair gives the same contribution. Moreover, we can also consider the diagram as in Fig. \ref{fig:Quartic5l} but with fermions $1$ and $2$ interchanged. By reflection symmetry, this has the same $UV$ divergence. Finally, we should be able to absorb the $UV$ divergence into the coefficient of the quartic vertex $\vec{\phi^2}^2$, which specifies the spin structure,
\beq \delta\Gamma^{a_1 a_2 a_3 a_4}_4 \stackrel{UV}{=} \frac{1}{3}(\delta^{a_1 a_2} \delta^{a_3a_4} + \delta^{a_1 a_3} \delta^{a_2 a_4} + \delta^{a_1 a_4} \delta^{a_2 a_3})\delta\Gamma^{3333}_4\eeq
and
\bea  \delta\Gamma^{3333}_4 &=& -4 \cdot 6 \cdot 2 \cdot n \cdot S \cdot N^2\int \frac{d^3 p_1 d^3 p_2 d^3 l_1 d^3 l_2 d^3 l_3}{(2 \pi)^{15}} D(l_1) D(l_3) D(l_1 - l_2) D(l_2 - l_3)\nn\\
 &\times&G^{1}_1(p_1) G^{1}_2(p_1)^2 G^{1}_1(p_1 - l_1) G^{1}_2(p_1-l_2) G^{1}_1(p_1 - l_3) \nn\\
 &\times&G^{-1}_1(p_2) G^{-1}_2(p_2)^2 G^{-1}_1(p_2 - l_1) G^{-1}_2(p_2-l_2) G^{-1}(p_2 - l_3) \label{Gamma45l1}\eea
with
\beq S = tr(\tau^3 \tau^3 \tau^{a} \tau^b \tau^c \tau^d) tr(\tau^3 \tau^3 \tau^{a} \tau^b \tau^c \tau^d) = 84 \eeq

We will used the same strategy for evaluating the integral (\ref{Gamma45l1}) as for computing the vertex correction in section \ref{app:vert3l}. The singular manifold in the present case is specified by vanishing $\vec{p}_1$, $\vec{p}_2$, $\hat{v}_1 \cdot \vec{l}_1$, $\hat{v}_2 \cdot \vec{l}_2$, $\hat{v}_1 \cdot \vec{l}_3$ and can be parameterized by the three momenta $\hat{v}_2 \cdot \vec{l}_1$, $\hat{v}_1 \cdot \vec{l}_2$, $\hat{v}_2 \cdot \vec{l}_3$. We will integrate explicitly over the first set of momenta and leave the result as an integral over the later three momenta.

Let us call $I(p_{1\tau}, p_{2\tau})$ the result of integrating over all momenta and frequencies in Eq. (\ref{Gamma45l1}), except $p_{1\tau}$ and $p_{2 \tau}$. Then, using the particle-hole symmetry, $G(p) = - G(-p)$, and the inversion symmetry, $G^{-1}(p_\tau, \vec{p}) = G^1(p_\tau, -\vec{p})$, we obtain $I(p_{1\tau}, p_{2 \tau}) = I(-p_{1\tau}, -p_{2 \tau})$ and $I(p_{1\tau}, p_{2\tau}) = I(p_{2 \tau}, p_{1 \tau})$. Thus,
\bea \delta\Gamma^{3333}_4 &=& - 2^{10}\cdot 3^2\cdot 7 \cdot N^2 \left(\frac{|\vec{v}|^2}{2 v_x v_y}\right)^5 \int_0^{\infty}\frac{dp_{1 \tau}}{2 \pi}\int_{-p_{1 \tau}}^{p_{1 \tau}}\frac{dp_{2 \tau}}{2 \pi} \int \frac{dl_{1\tau} dl_{2\tau} dl_{3 \tau}}{(2 \pi)^3} 
\nn\\
&&\int \frac{d(\hat{v}_1 \cdot \vec{p}_1) d(\hat{v}_2 \cdot \vec{p}_1) d(\hat{v}_1 \cdot \vec{p}_2) d(\hat{v}_2 \cdot \vec{p}_2) d(\hat{v}_1 \cdot \vec{l}_1) d(\hat{v}_2 \cdot \vec{l}_1) d(\hat{v}_1 \cdot \vec{l}_2) d(\hat{v}_2 \cdot \vec{l}_2) d(\hat{v}_1 \cdot \vec{l}_3) d(\hat{v}_2 \cdot \vec{l}_3)}{(2 \pi)^{10}}\nn\\
&& \frac{1}{-i \frac{c_f}{N} g(p_{1 \tau}, 0) + \vec{v}_1 \cdot \vec{p}_1}\times\frac{1}{(-i \frac{c_f}{N} g(p_{1 \tau}, 0) + \vec{v}_2 \cdot \vec{p}_1)^2}\times\frac{1}{-i \frac{c_f}{N} g(p_{1 \tau} - l_{1 \tau}, \hat{v}_2\cdot \vec{l}_1) + \vec{v}_1 \cdot (\vec{p}_1-\vec{l}_1)}\nn\\
&\times&\frac{1}{-i \frac{c_f}{N} g(p_{1 \tau} - l_{2 \tau}, \hat{v}_1\cdot \vec{l}_2) + \vec{v}_2 \cdot (\vec{p}_1-\vec{l}_2)}\times\frac{1}{-i \frac{c_f}{N} g(p_{1 \tau} - l_{3 \tau}, \hat{v}_2\cdot \vec{l}_3) + \vec{v}_1 \cdot (\vec{p}_1-\vec{l}_3)}\nn\\
&\times& \frac{1}{-i \frac{c_f}{N} g(p_{2 \tau}, 0) - \vec{v}_1 \cdot \vec{p}_2}\times\frac{1}{(-i \frac{c_f}{N} g(p_{2 \tau}, 0) - \vec{v}_2 \cdot \vec{p}_2)^2}\times\frac{1}{-i \frac{c_f}{N} g(p_{2 \tau} - l_{1 \tau}, \hat{v}_2\cdot \vec{l}_1) - \vec{v}_1 \cdot (\vec{p}_2-\vec{l}_1)}\nn\\
&\times&\frac{1}{-i \frac{c_f}{N} g(p_{2 \tau} - l_{2 \tau}, \hat{v}_1\cdot \vec{l}_2) - \vec{v}_2 \cdot (\vec{p}_2-\vec{l}_2)}\times\frac{1}{-i \frac{c_f}{N} g(p_{2 \tau} - l_{3 \tau}, \hat{v}_2\cdot \vec{l}_3) - \vec{v}_1 \cdot (\vec{p}_2-\vec{l}_3)}\nn
\\&\times& D(l_1) D(l_3) D(l_1 - l_2) D(l_2 - l_3)\eea
Integrating over $\hat{v}_1\cdot \vec{l}_1$, $\hat{v}_2 \cdot \vec{l}_2$, $\hat{v}_1 \cdot \vec{l}_3$,
\bea \delta\Gamma^{3333}_4 &=& - i 2^{10}\cdot 3^2\cdot 7 \cdot N^2 \frac{|\vec{v}|^7}{(2 v_x v_y)^5} \int_0^{\infty}\frac{dp_{1 \tau}}{2 \pi}\int_{-p_{1 \tau}}^{p_{1 \tau}}\frac{dp_{2 \tau}}{2 \pi} \nn\\
&&\left[\int_{p_{1\tau}}^{\infty} - \int_{-\infty}^{p_{2 \tau}}\right]\frac{dl_{1\tau}}{2\pi} \left[\int_{p_{1\tau}}^{\infty} - \int_{-\infty}^{p_{2 \tau}}\right]\frac{dl_{2\tau}}{2\pi}\left[\int_{p_{1\tau}}^{\infty} - \int_{-\infty}^{p_{2 \tau}}\right]\frac{dl_{3\tau}}{2\pi}\nn\\
&&\int \frac{d(\hat{v}_1 \cdot \vec{p}_1) d(\hat{v}_2 \cdot \vec{p}_1) d(\hat{v}_1 \cdot \vec{p}_2) d(\hat{v}_2 \cdot \vec{p}_2) d(\hat{v}_2 \cdot \vec{l}_1) d(\hat{v}_1 \cdot \vec{l}_2) d(\hat{v}_2 \cdot \vec{l}_3)}{(2 \pi)^{7}}\nn\\
&& \frac{1}{-i \frac{c_f}{N} g(p_{1 \tau}, 0) + \vec{v}_1 \cdot \vec{p}_1}\times\frac{1}{(-i \frac{c_f}{N} g(p_{1 \tau}, 0) + \vec{v}_2 \cdot \vec{p}_1)^2}\nn\\&\times& \frac{1}{-i \frac{c_f}{N} g(p_{2 \tau}, 0) - \vec{v}_1 \cdot \vec{p}_2}\times\frac{1}{(-i \frac{c_f}{N} g(p_{2 \tau}, 0) - \vec{v}_2 \cdot \vec{p}_2)^2}\nn\\
&\times& \frac{1}{-i \frac{c_f}{N}(g(p_{1\tau}-l_{1\tau}, \hat{v}_2 \cdot \vec{l}_1) + g(p_{2\tau}-l_{1\tau}, \hat{v}_2 \cdot \vec{l}_1)) + \vec{v}_1 \cdot (\vec{p}_1 - \vec{p}_2)} \nn\\
&\times& \frac{1}{-i \frac{c_f}{N}(g(p_{1\tau}-l_{2\tau}, \hat{v}_1 \cdot \vec{l}_2) + g(p_{2\tau}-l_{2\tau}, \hat{v}_1 \cdot \vec{l}_2)) + \vec{v}_2 \cdot (\vec{p}_1 - \vec{p}_2)} \nn\\
&\times& \frac{1}{-i \frac{c_f}{N}(g(p_{1\tau}-l_{3\tau}, \hat{v}_2 \cdot \vec{l}_3) + g(p_{2\tau}-l_{3\tau}, \hat{v}_2 \cdot \vec{l}_3)) + \vec{v}_1 \cdot (\vec{p}_1 - \vec{p}_2)} \nn\\
&\times& D(l_1) D(l_3) D(l_1 - l_2) D(l_2 - l_3)|_{\hat{v}_1 \cdot \vec{l}_1 = \hat{v}_2\cdot \vec{l}_2 = \hat{v}_1 \cdot \vec{l}_3 = 0}\eea
Now, integrating over $\hat{v}_1 \cdot \vec{p}_1$, $\hat{v}_2 \cdot \vec{p}_1$,
\bea &&\delta\Gamma^{3333}_4 = - i 2^{10}\cdot 3^2\cdot 7 \cdot N^2 \frac{|\vec{v}|^5}{(2 v_x v_y)^5} \int_0^{\infty}\frac{dp_{1 \tau}}{2 \pi}\int_{-p_{1 \tau}}^{p_{1 \tau}}\frac{dp_{2 \tau}}{2 \pi} \int_{p_{1\tau}}^{\infty} \frac{d l_{2 \tau}}{2\pi} \nn\\
&&\int \frac{d(\hat{v}_1 \cdot \vec{p}_2) d(\hat{v}_2 \cdot \vec{p}_2) d(\hat{v}_2 \cdot \vec{l}_1) d(\hat{v}_1 \cdot \vec{l}_2) d(\hat{v}_2 \cdot \vec{l}_3)}{(2 \pi)^{5}} \frac{1}{-i \frac{c_f}{N} g(p_{2 \tau}, 0) - \vec{v}_1 \cdot \vec{p}_2}\times\frac{1}{(-i \frac{c_f}{N} g(p_{2 \tau}, 0) - \vec{v}_2 \cdot \vec{p}_2)^2}\nn\\
&\times& \frac{1}{(-i \frac{c_f}{N}(g(p_{1\tau},0)+ g(l_{2\tau}-p_{1\tau}, \hat{v}_1 \cdot \vec{l}_2) + g(l_{2\tau}-p_{2\tau}, \hat{v}_1 \cdot \vec{l}_2)) + \vec{v}_2 \cdot \vec{p}_2)^2} \nn\\
&\times& \Bigg[\int_{p_{1\tau}}^{\infty}\frac{dl_{1\tau}}{2\pi} \int_{p_{1\tau}}^{\infty} \frac{dl_{3\tau}}{2\pi} \frac{1}{i\frac{c_f}{N}(g(p_{1\tau},0) + g(l_{1\tau}-p_{1\tau},\hat{v}_2 \cdot \vec{l}_1) + g(l_{1\tau} - p_{2\tau}, \hat{v}_2 \cdot \vec{l}_1)) - \vec{v}_1 \cdot \vec{p}_2}\nn\\
&\times&\frac{1}{i\frac{c_f}{N}(g(p_{1\tau},0) + g(l_{3\tau}-p_{1\tau},\hat{v}_2 \cdot \vec{l}_3) + g(l_{3\tau} - p_{2\tau}, \hat{v}_2 \cdot \vec{l}_3)) - \vec{v}_1 \cdot \vec{p}_2}\nn\\
&+& \int_{p_{1\tau}}^{\infty}\frac{dl_{1\tau}}{2\pi} \int_{-\infty}^{p_{2\tau}} \frac{dl_{3\tau}}{2\pi}  \frac{1}{-i\frac{c_f}{N}(g(p_{1\tau},0) + g(l_{1\tau}-p_{1\tau},\hat{v}_2 \cdot \vec{l}_1) + g(l_{1\tau} - p_{2\tau}, \hat{v}_2 \cdot \vec{l}_1)) + \vec{v}_1 \cdot \vec{p}_2}\nn\\
&\times&\frac{1}{-i\frac{c_f}{N}(g(l_{1\tau}-p_{1\tau},\hat{v}_2 \cdot \vec{l}_1) + g(l_{1\tau} - p_{2\tau}, \hat{v}_2 \cdot \vec{l}_1)+ g(p_{1\tau}-l_{3\tau},\hat{v}_2 \cdot \vec{l}_3) + g(p_{2\tau} -l_{3\tau}, \hat{v}_2 \cdot \vec{l}_3))} \nn\\
&+&  \int_{-\infty}^{p_{2\tau}} \frac{dl_{1\tau}}{2\pi} \int_{p_{1\tau}}^{\infty}\frac{dl_{3\tau}}{2\pi} \frac{1}{-i\frac{c_f}{N}(g(p_{1\tau},0) + g(l_{3\tau}-p_{1\tau},\hat{v}_2 \cdot \vec{l}_3) + g(l_{3\tau} - p_{2\tau}, \hat{v}_2 \cdot \vec{l}_3)) + \vec{v}_1 \cdot \vec{p}_2}\nn\\
&\times&\frac{1}{-i\frac{c_f}{N}(g(p_{1\tau}-l_{1\tau},\hat{v}_2 \cdot \vec{l}_1) + g(p_{2\tau}-l_{1\tau}, \hat{v}_2 \cdot \vec{l}_1)+ g(l_{3\tau}-p_{1\tau},\hat{v}_2 \cdot \vec{l}_3) + g(l_{3\tau}-p_{2\tau}, \hat{v}_2 \cdot \vec{l}_3))}\Bigg]\nn\\
&&D(l_1)D(l_3)D(l_1-l_2)D(l_2-l_3)|_{\hat{v}_1 \cdot \vec{l}_1 = \hat{v}_2\cdot \vec{l}_2 = \hat{v}_1 \cdot \vec{l}_3 = 0}  \eea
Observe that under $l_1 \leftrightarrow l_3$ the first term in the square brackets is invariant, while the second and third terms map into each other. Utilizing this fact and integrating over $\hat{v}_1 \cdot \vec{p}_2$, $\hat{v}_2 \cdot \vec{p}_2$,
\bea \delta\Gamma^{3333}_4 &=& -  2^{12}\cdot 3^2\cdot 7 \cdot N^7 \frac{|\vec{v}|^3}{(2 v_x v_y c_f)^5} \int_0^{\infty}\frac{dp_{1 \tau}}{2 \pi}\int_{0}^{p_{1 \tau}}\frac{dp_{2 \tau}}{2 \pi} \int_{p_{1\tau}}^{\infty} \frac{d l_{1 \tau}}{2\pi}\int_{p_{1\tau}}^{\infty} \frac{d l_{2 \tau}}{2\pi} \nn\\
&&\int \frac{d(\hat{v}_2 \cdot \vec{l}_1) d(\hat{v}_1 \cdot \vec{l}_2) d(\hat{v}_2 \cdot \vec{l}_3)}{(2 \pi)^{3}} \frac{1}{g(p_{1\tau},0)+ g(p_{2 \tau}, 0) + g(l_{1\tau}-p_{1\tau},\hat{v}_2 \cdot \vec{l}_1) + g(l_{1\tau} - p_{2\tau}, \hat{v}_2 \cdot \vec{l}_1)}\nn\\
&\times& \frac{1}{(g(p_{1\tau},0)+ g(p_{2 \tau}, 0) + g(l_{2\tau}-p_{1\tau},\hat{v}_1 \cdot \vec{l}_2) + g(l_{2\tau} - p_{2\tau}, \hat{v}_1 \cdot \vec{l}_2))^3}\nn\\
&\times& \Bigg[\int_{p_{1\tau}}^{l_{1\tau}} \frac{dl_{3\tau}}{2\pi} \frac{1}{g(p_{1\tau},0)+ g(p_{2 \tau}, 0) + g(l_{3\tau}-p_{1\tau},\hat{v}_2 \cdot \vec{l}_3) + g(l_{3\tau} - p_{2\tau}, \hat{v}_2 \cdot \vec{l}_3)} \nn\\
&+& \int_{-\infty}^{p_{2\tau}} \frac{dl_{3\tau}}{2\pi} \frac{1}{g(l_{1\tau} - p_{1\tau},\hat{v}_2 \cdot \vec{l}_1)+ g(l_{1\tau} - p_{2\tau},\hat{v}_2 \cdot \vec{l}_1) + g(p_{1\tau}-l_{3\tau},\hat{v}_2 \cdot \vec{l}_3) + g(p_{2\tau}-l_{3\tau} , \hat{v}_2 \cdot \vec{l}_3)}\Bigg]\nn\\
&&D(l_1)D(l_3)D(l_1-l_2)D(l_2-l_3)|_{\hat{v}_1 \cdot \vec{l}_1 = \hat{v}_2\cdot \vec{l}_2 = \hat{v}_1 \cdot \vec{l}_3 = 0}\eea
We now introduce dimensionless variables, $p_{2 \tau} = x p_{1\tau}$, $l_{i \tau} = y_i p_{1\tau}$, $\hat{v}_2 \cdot \vec{l}_1 = \sqrt{\gamma p_{1\tau}} z_1$, $\hat{v}_1 \cdot \vec{l}_2 = \sqrt{\gamma p_{1\tau}} z_2$, $\hat{v}_2 \cdot \vec{l}_3 = \sqrt{\gamma p_{1\tau}} z_3$. Then,
\beq \delta\Gamma^{3333}_4 = \frac{1}{2} N^3 Y(\alpha) \gamma \int_0^{\infty} \frac{dp_{1\tau}}{p_{1\tau}} = N^3 Y(\alpha) \gamma \log \Lambda\eeq
with
\bea &&Y(\alpha) = -\frac{56}{27 \pi^2} \left(\frac{1}{\alpha} + \alpha\right)^4 \int_0^1 dx \int_1^{\infty} dy_1 \int_1^{\infty} dy_2 \int_{-\infty}^{\infty} dz_1 \int_{-\infty}^{\infty} dz_2 \int_{-\infty}^{\infty} dz_3 \nn\\
&&\frac{1}{1 + \sqrt{x} + \sqrt{y_1 - 1 + z^2_1} + \sqrt{y_1 -x + z^2_1} - 2|z_1|}\nn\\&\times& \frac{1}{(1 + \sqrt{x} + \sqrt{y_2 - 1 + z^2_2} + \sqrt{y_2 -x + z^2_2} - 2|z_2|)^3}\nn\\
&\times&\Bigg[\int_1^{y_1} dy_3 \frac{1}{1 + \sqrt{x} + \sqrt{y_3 - 1 + z^2_3} + \sqrt{y_3 -x + z^2_3} - 2|z_3|} \nn\\&+& \int_{-\infty}^{x} dy_3 \frac{1}{\sqrt{y_1 - 1 + z^2_1} + \sqrt{y_1 -x + z^2_1} + \sqrt{1-y_3 + z^2_3} + \sqrt{x-y_3 + z^2_3} - 2|z_1| -2 |z_3|}\Bigg]\nn\\
&\times&\frac{1}{y_1+\frac{1}{4}(\frac{1}{\alpha}+\alpha)^2 z^2_1}\frac{1}{|y_3|+\frac{1}{4}(\frac{1}{\alpha}+\alpha)^2 z^2_3}\frac{1}{|y_1-y_2|+\frac{1}{4}(\frac{1}{\alpha}+\alpha)^2 (z^2_1+z^2_2) + \frac{1}{2}(\alpha^2 - \frac{1}{\alpha^2}) z_1 z_2}\nn\\&\times&\frac{1}{|y_2-y_3|+\frac{1}{4}(\frac{1}{\alpha}+\alpha)^2 (z^2_2+z^2_3) + \frac{1}{2}(\alpha^2 - \frac{1}{\alpha^2}) z_2 z_3}\eea

\subsection{Pairing vertex}
\label{app:pairing}

This appendix will describe the direct evaluation of the pairing vertex correction in 
Eq.~(\ref{GV1}).
We first attempt to perform the calculation using bare fermion propagators,
\bea \delta \Gamma_{V \psi^{\dagger} \psi^{\dagger}}&=& \frac{-3 \mu}{N |\vec{v}|^2} \int \frac{d l_\tau dl_\perp dl_\parallel}{(2 \pi)^3} \frac{1}{\gamma |l_\tau| + l^2_\perp + l^2_\parallel}\times \frac{1}{l_\perp - \hat{v}_2 \cdot \vec{k}_1 - i \displaystyle \frac{\eta}{|\vec{v}|} (l_\tau - k_{1 \tau}) } \nn \\
&~&~~~~~~~~~~~~~~~~\times \frac{1}{l_\perp + \hat{v}_2 \cdot \vec{k}_{-1} + i \displaystyle \frac{\eta}{|\vec{v}|}(l_\tau + k_{-1 \tau}) }\nn \eea
where we've introduced variables $l_\perp = \hat{v}_2 \cdot \vec{l}$, $l_\parallel = \epsilon_{ij} (\hat{v}_2)_i l_j$.
For simplicity, let us choose $k_{1\tau} = k_{-1 \tau} = \omega>0$. We now perform the integral over $l_\perp$. For $|l_\tau| < \omega$ both poles in the fermion propagators are in the same half-plane and we can pick up just the pole from the bosonic propagator. In the opposite regime, $|l_\tau| > \omega$, we get contributions from both the bosonic and fermionic poles. Thus,
\bea && \delta \Gamma_{V \psi^{\dagger} \psi^{\dagger}} = \nn \\
& - & \frac{3 \mu}{N |\vec{v}|^2} \Bigg[ - \int_0^{\infty} \frac{dl_\tau}{2\pi} \int \frac{dl_\parallel}{2\pi}  \frac{1}{\sqrt{\gamma l_\tau + l^2_\parallel}} \frac{1}{\sqrt{\gamma l_\tau +l^2_\parallel} + i \hat{v}_2 \cdot \vec{k}_1}\frac{1}{\sqrt{\gamma l_\tau +l^2_\parallel} - i \hat{v}_2 \cdot \vec{k}_{-1}}\label{dV11}\\
&+& \frac{|\vec{v}|}{2 \eta} \int_\omega^{\infty} \frac{d l_\tau}{2 \pi} \int \frac{dl_\parallel}{2\pi} \bigg(\frac{1}{l_\tau - i \frac{\vec{v}_2}{\eta} \cdot (\vec{k}_1 + \vec{k}_{-1})} \frac{1}{\gamma l_\tau + l^2_\parallel + (\hat{v}_2 \cdot \vec{k}_1)^2} \label{dV12}\\ &+& \frac{1}{l_\tau + i \frac{\vec{v}_2}{\eta} \cdot (\vec{k}_1 + \vec{k}_{-1})} \frac{1}{\gamma l_\tau + l^2_\parallel + (\hat{v}_2 \cdot \vec{k}_{-1})^2}\bigg) \Bigg]\label{dV13}\eea
The contribution from the bosonic pole in Eq.~(\ref{dV11}) gives an expected logarithmic divergence,
\beq \delta^{bos} \Gamma_{V \psi^{\dagger} \psi^{\dagger}} \sim \frac{3 \mu}{N \pi^2 \gamma |\vec{v}|^2} \log\frac{\Lambda}{|\hat{v}_2 \cdot \vec{k}|}\eeq
On the other hand, the contribution from the fermionic poles in Eqs.~(\ref{dV12}),(\ref{dV13}) gives a much stronger infra-red singularity. If we set the total momentum of the fermion pair $\vec{k}_1 + \vec{k}_{-1}$ to zero, then 
\beq  \delta^{fer} \Gamma_{V \psi^{\dagger} \psi^{\dagger}} \sim 
 -\frac{3 \mu}{4 \pi N \eta |\vec{v}_2 \cdot \vec{k}_1|} f\left(\frac{\gamma|\omega|}{|\hat{v}_2 \cdot \vec{k}_1|^2}\right)\eeq
with
\beq f(a) = \int_a^{\infty} dx \frac{1}{x} \frac{1}{\sqrt{x+1}}\eeq
If the total pair momentum is non-vanishing, in particular, if $\frac{\gamma}{\eta} |\vec{v}_2 \cdot (\vec{k}_1 +\vec{k}_{-1})| \gg (\hat{v_2} \cdot \vec{k}_1)^2, \gamma \omega$, then,
\beq \delta^{fer} \Gamma_{V \psi^{\dagger} \psi^{\dagger}} = - \frac{3\mu}{4 N |\vec{v}| \sqrt{2 \gamma \eta}} \frac{1}{\sqrt{|\vec{v}_2 \cdot(\vec{k}_1 + \vec{k}_{-1})|}}\eeq

As usual, we cure the strong infra-red divergences by using a one-loop dressed fermion propagator (\ref{G1loop}). Then, 
\bea \delta \Gamma_{V \psi^{\dagger} \psi^{\dagger}}(k_1, k_{-1}) &=& - \frac{3 \mu}{N|\vec{v}|^2} \int \frac{d^3 l}{(2\pi)^{3}} \frac{1}{\gamma |l_\tau| + \vec{l}^2} \times \frac{1}{\hat{v}_2\cdot(\vec{l} - \vec{k}_1) - i \displaystyle \frac{c_f}{N|\vec{v}|} g(l_\tau - k_{1 \tau}, \hat{v}_1 \cdot (\vec{l} - \vec{k}_1))}\nn\\
%
%
&\times& \frac{1}{\hat{v}_2\cdot(\vec{l} + \vec{k}_{-1}) + i \displaystyle \frac{c_f}{N|\vec{v}|} g(l_\tau + k_{-1\tau}, \hat{v}_1 \cdot (\vec{l} + \vec{k}_{-1}))}\nn\eea
For simplicity, we take the external fermion momenta to lie at the hot spots, $\vec{k}_1 = \vec{k}_{-1} = 0$. Moreover, as before, we choose the external frequencies, $k_{1\tau} = k_{-1 \tau} = \omega > 0$. Switching to variables, $l_\perp$, $l_\parallel$, we perform the integral over $l_\perp$. As we saw above, the contribution from the pole in the bosonic propagator could be calculated without dressing the fermion Green's function and was of $\mathcal{O}(1/N)$ - we drop this piece below. On the other hand, as we will see the contribution from the poles in fermionic propagators is of $\mathcal{O}(1)$ in $N$. Moreover, since $l_\perp \sim \mathcal{O}(1/N)$ at these poles, we may ignore the dependence of the fermion self-energy on $l_\perp$, which gives, $\hat{v}_1 \cdot \vec{l} = \frac{2 \alpha}{\alpha^2 + 1} l_\parallel$. In this manner, we obtain,
\bea &&\delta \Gamma_{V \psi^{\dagger} \psi^{\dagger}} = - \frac{6 \mu}{c_f |\vec{v}|} \int^{\infty}_\omega \frac{d l_\tau}{2 \pi} \int \frac{dl_\parallel}{2\pi} \frac{1}{\gamma l_\tau + l^2_\parallel} \nn \\
&&~~~~~~~~~\times \frac{1}{g(l_\tau - \omega, \frac{2 \alpha}{\alpha^2+1} l_\parallel)+g(l_\tau + \omega, \frac{2 \alpha}{\alpha^2+1} l_\parallel)} \eea
We now perform the integral over $l_\tau$. This integral is convergent in the ultra-violet. However, when $\omega \to 0$, it is logarithmically divergent in the infra-red. This infra-red divergence comes from the region $\gamma l_\tau \ll l^2_\parallel$. Changing variables to $\gamma l_\tau =  x l^2_\parallel$, we obtain,
\beq \delta \Gamma_{V \psi^{\dagger} \psi^{\dagger}} = - \frac{3 \mu}{\pi^2 \gamma |\vec{v}| c_f}\int_0^{\infty} \frac{dl_\parallel}{l_\parallel} \int_{\frac{\gamma \omega}{l^2_\parallel}} \frac{dx}{x+1} \frac{1}{\sqrt{x+ (\frac{2 \alpha}{\alpha^2 + 1})^2 - \frac{\gamma \omega}{l^2_\parallel}} + \sqrt{x+ (\frac{2 \alpha}{\alpha^2 + 1})^2 +  \frac{\gamma \omega}{l^2_\parallel}} - \frac{4 \alpha}{\alpha^2 + 1}} \eeq
For $l^2_\parallel \gg \gamma \omega$, performing the integral over $x$ to logarithmic accuracy,
\beq \delta \Gamma_{V \psi^{\dagger} \psi^{\dagger}} \approx - \frac{6 \mu \alpha}{\pi^2 \gamma |\vec{v}| c_f (\alpha^2 + 1)} \int_{\sqrt{\gamma \omega}}^\infty \frac{dl_\parallel}{l_\parallel} \log\left(\frac{l^2_\parallel}{\gamma \omega}\right) = - \frac{\mu \alpha}{\pi (\alpha^2 + 1)} \log^2\left(\frac{\Lambda^2}{\gamma \omega}\right) \label{Vfina}\eeq

\subsection{Density vertex}
\label{app:2kF}
In this appendix, we compute the one-loop renormalization of the density-wave vertex, shown in Fig.~\ref{FigV}b),
\beq \delta \Gamma_{O \psi \psi^{\dagger}}(k_1, k_{-1}) = 
3 \mu \int \frac{d^3 l}{(2\pi)^{3}} D(l) G^1_{2}(k_1 - l) G^{-1}_{2}(k_{-1} - l). \label{GO1}
\eeq
If we ignore the effects of Fermi-surface curvature, $G(l)=-G(-l)$, and Eq.~(\ref{GO1}) reduces to its counterpart in the superconducting channel
with $k_{-1} \to - k_{-1}$. In the present calculation, we will keep the effects of the Fermi-surface curvature using a propagator,
\beq G^{\ell}_i(l) = \frac{1}{-\frac{i c_f}{N} g(l_\tau, \hat{v}^{\ell}_{\bar{i}}\cdot\vec{l}) + \vec{v}^{\ell}_i \cdot \vec{l} +  (\hat{n}^{\ell}_{\parallel,i} \cdot \vec{l})^2}\eeq
Here, we ignore any dressing of the curvature by the interactions.

For simplicity, we set external momenta to zero and  choose $k_{1\tau} =- k_{-1 \tau} = \omega>0$. As in Appendix \ref{app:pairing}, we introduce variables  $l_\perp = \hat{v}_2 \cdot \vec{l}$, $l_\parallel = \epsilon_{ij} (\hat{v}_2)_i l_j$.
Proceeding as in Section \ref{sec:pairing}, we keep only the contribution to the integral (\ref{GO1}) from the Fermi liquid regime, $\gamma l_\tau \ll l^2_\parallel$. Then,
\bea\delta \Gamma_{O \psi \psi^{\dagger}} &=&
\frac{3 \mu}{N} \int \frac{dl_\parallel}{2 \pi} \int_{\gamma |l_\tau| \lesssim l^2_\parallel} \frac{d l_\tau}{2 \pi} \int \frac{dl_\perp}{2\pi} \frac{1}{l^2_\parallel} \nn\\
&&\frac{1}{i {\cal Z}^{-1}(l_\parallel)(l_\tau - \omega) - |\vec{v}| l_\perp - \frac{1}{2m} l^2_\parallel }\frac{1}{i {\cal Z}^{-1}(l_\parallel)(l_\tau + \omega) + |\vec{v}| l_\perp - \frac{1}{2m} l^2_\parallel }\eea
Performing the integral over $l_\perp$,
\beq \delta \Gamma_{O \psi \psi^{\dagger}} = - \frac{3 \mu}{N |\vec{v}|} \int \frac{d l_\parallel}{2 \pi}  \int^{l^2_\parallel/\gamma}_\omega \frac{d l_\tau}{2\pi}
\frac{{\cal Z}(l_\parallel)}{l^2_\parallel} \frac{l_\tau}{l^2_\tau + \left(\frac{{\cal Z}(l_\parallel) l^2_\parallel}{2 m}\right)^2}\label{GO2}\eeq
Notice that the Fermi-surface curvature is present in the denominator of Eq.~(\ref{GO2}). This is in contrast to the corresponding calculation in the superconducting channel, where the Fermi-surface curvature drops out. Performing the integral over $l_\tau$,
\beq \delta \Gamma_{O \psi \psi^{\dagger}} = - \frac{3 \mu}{2 \pi N |\vec{v}|} \int_{\sqrt{\gamma \omega}}^{\infty} \frac{d l_\parallel}{2 \pi} \frac{{\cal Z}(l_\parallel)}{l^2_\parallel}\log \frac{l^4_\parallel}{(\gamma \omega)^2 + \left(\frac{\gamma{\cal Z}(l_\parallel) l^2_\parallel}{2 m}\right)^2}\label{GO3}\eeq
where we have ignored terms subleading in $l_\parallel$ in the numerator of the logarithm. Recall, $Z(l_\parallel) \sim N |\vec{v}| l_\parallel$. Hence, for
$l_\parallel \ll \left({m \omega}/{N |\vec{v}|}\right)^{1/3}$ the $l_\tau$ integral is cut-off in the infrared by the external frequency and the Fermi surface curvature may be neglected. On the other hand, for $l_\parallel \gg \left({m \omega}/{N |\vec{v}|}\right)^{1/3}$ the integral is cut-off by the curvature. By comparison, in the superconducting channel the integral is cut-off by the external frequency in both regimes resulting in a stronger enhancement. Notice that the cross-over scale $\left({m \omega}/{N |\vec{v}|}\right)^{1/3}$ is much larger than the infra-red cut-off of the $l_\parallel$ integral $\sqrt{\gamma \omega}$. Evaluating the integral over $l_\parallel$ to leading logarithmic accuracy,
\beq  \delta \Gamma_{O \psi \psi^{\dagger}} = - \frac{\mu \alpha}{3\pi (\alpha^2 + 1)} \log^2\left(\frac{\Lambda^2}{\gamma \omega}\right)\eeq

\end{document}